\documentclass{aa}
\usepackage{graphicx}
\usepackage{amsmath}
\newlength{\dlugskr}

\newcommand{\cmcub}{~cm$^{-3}$}
\newcommand{\kms}{~km~s$^{-1}$}

\newcommand{\Ms}{~M$_{\odot}$}

\newcommand{\msun}{\ifmmode M_{\odot} \else M$_{\odot}$\fi}
\newcommand{\rsun}{\ifmmode R_{\odot} \else R$_{\odot}$\fi}
\newcommand{\lsun}{\ifmmode L_{\odot} \else L$_{\odot}$\fi}
\newcommand{\zsun}{\ifmmode Z_{\odot} \else Z$_{\odot}$\fi}

\newcommand{\Hb}{\ifmmode {\rm H}\beta \else H$\beta$\fi}

\newcommand{\rOiii}{[O~{\sc iii}] $\lambda$4363/5007}
\newcommand{\rNii}{[N~{\sc ii}] $\lambda$5755/6584}

\newcommand{\rSii}{[S~{\sc ii}] $\lambda$6717/6731}

\newcommand{\Np}{N$^{+}$}
\newcommand{\Npp}{N$^{++}$}

\newcommand{\Op}{O$^{+}$}
\newcommand{\Opp}{O$^{++}$}

\newcommand{\Nepp}{Ne$^{++}$}

\newcommand{\Spp}{S$^{++}$}
\newcommand{\Sppp}{S$^{+++}$}
\newcommand{\Arpp}{Ar$^{++}$}

\begin{document}

\title{Abundances in the Galactic bulge: results from planetary nebulae
  and giant stars
\thanks{Based on observations made at the Cerro Tololo Interamerican
Observatory and the European Southern Observatory}
\thanks{Full Table~\ref{table_abundances} is only available in electronic form
at http://www.aanda.org}
}

\titlerunning{Abundance patterns in the Galactic bulge}

\author{Chiappini, C.
         \inst{1,2}
\and
        G\'{o}rny, S. K.
        \inst{3}
\and
        Stasi\'nska, G.
        \inst{4}
\and
        Barbuy, B.
        \inst{5}
}

\offprints{C. Chiappini}

\institute{
        Observatoire de Gen\`eve, Universit\'e de Gen\`eve,
        51 Chemin des Maillettes,
        CH-1290 Sauverny, Switzerland\\
        \email{Cristina.Chiappini@unige.ch}
\and
       Osservatorio Astronomico di Trieste - OAT/INAF,
       Via G. B. Tiepolo 11, 34131 Trieste, TS, Italy\\
       \email{chiappini@oats.inaf.it}
\and
        Copernicus Astronomical center, Rabia\'nska 8,
        PL-87-100 Toru\'n, Poland \\
        \email{skg@ncac.torun.pl}
\and
        LUTH, Observatoire de Paris, CNRS, Universit\'e Paris Diderot;
Place Jules Janssen 92190 Meudon, France\\
\email{grazyna.stasinska@obspm.fr}
\and
       Universidade de S\~ao Paulo, IAG, Rua do Mat\~ao 1226,
Cidade Universit\'aria, S\~ao Paulo 05508-900, Brazil;\\
\email{barbuy@astro.iag.usp.br}
}

\date{Received ???; accepted ???}

\abstract
{Our understanding of the chemical evolution (CE) of the Galactic bulge requires the determination of  abundances in large samples of giant stars and planetary nebulae (PNe). Studies based on high resolution spectroscopy of giant stars in several fields of the Galactic bulge obtained with very large telescopes have allowed important progress.} 
{We discuss PNe abundances in the Galactic bulge and compare these results with those presented in the literature for giant stars.} 
{We present the largest, high-quality data-set available for PNe in the direction of the Galactic bulge
(inner-disk/bulge).
For comparison purposes, we also consider a sample of PNe in the Large Magellanic Cloud (LMC). We derive the element abundances in a consistent way for all the PNe studied. By comparing the abundances for the bulge, inner-disk, and LMC, we identify elements that have not been modified during the evolution of the PN progenitor and can be used to trace the bulge chemical enrichment history. We then compare the PN abundances with abundances of bulge field giant.}
{At the metallicity of the bulge, we find that the abundances of O and Ne are close to the values for the interstellar medium at the time of the PN progenitor formation, and
hence these elements can be used as tracers of the bulge CE, in the same way as S and Ar, which are not expected to be affected by nucleosynthetic processes during the evolution of the PN progenitors.
The PN oxygen abundance distribution is shifted to
lower values by 0.3~dex with respect to the distribution given by giants. A similar shift appears to occur for Ne and S. We discuss possible reasons for this PNe-giant discrepancy and conclude that this is probably due to systematic errors in the abundance derivations in either giants or PNe (or both). We issue an important warning concerning the use of \emph{absolute} abundances in CE studies.} 
{}

\maketitle

\section{Introduction}\label{sec:intro}

The Galactic bulge is old, distant, and highly obscured and studies of its chemical composition include two types of stars that represent evolved stages of intermediate mass stars:  planetary nebulae (PNe) (e.g. G\'orny et al. 2004), for which abundances are derived using the intensities of conspicuous emission lines, and giant stars (e.g. Lecureur et al. (2007) and Zoccali et al. (2008)), whose luminosities enable spectra of sufficient quality suitable for abundance determination to be obtained, using very large telescopes.

However, the use of evolved stars as test particles for probing the chemical evolution of the Galactic bulge requires some caution, since it is known that some elements have their abundances modified during stellar evolution. For example, intermediate-mass stars compete with massive stars in the enrichment of  the interstellar medium (ISM) in N and C (e.g. Chiappini et al. 2003, Henry 2004), and contribute to the enrichment of He. However, other elements (such as S and Ar) are unaffected by nucleosynthetic processes during the evolution of intermediate-mass stars and probe the chemical composition of the ISM at the time when the stars were born. The status of O and Ne is less clear, since their abundances can be affected by nucleosynthesis and mixing during stellar evolution, in amounts that  depend on metallicity and other properties (Charbonnel 2005,  Leisy \& Dennefeld 2006, Pe\~na et al. 2007).

In this paper, we use data sets both for PNe and giant stars to discuss the chemical evolution of the Galactic bulge. For PNe, we use a high quality sample obtained by merging the data from G\'{o}rny et al. (2008) with those from G\'{o}rny et al. (2004) and Wang \& Liu (2007). All PN abundances were recomputed in a consistent way. A detailed comparison of our
sample with other bulge PN samples in the literature was presented by G\'orny et al. (2008). Our bulge PN sample constitutes the largest high-quality data-set of abundance measurements for bulge PNe available in the literature.

We complete a
detailed analysis of the chemical abundances of these PNe with two main goals. First, we compare the properties of the PN population in an old, metal-rich\footnote{The bulge appears to be metal-rich with respect to the halo and thick disk (the other two old components of the Milky Way). Within the central region, the bulge contains cold gas from which stars form until today.}
environment (the Galactic bulge) with those in both metal-poor (LMC) and metal-rich (inner-disk) environments with ongoing star formation. We are then able to gain insights into processes occurring during the evolution of low- and intermediate-mass stars that can affect the abundances observed in the PN stage (such as hot-bottom burning and dredge-up), and infer their dependence on stellar mass and metallicity. Second, we attempt to improve our
understanding of the formation and evolution of the Galactic bulge
by studying elements that have not been modified during the evolution of the PN progenitor and, hence, can be compared with chemical evolution model predictions. After these elements have been safely identified, we compare the bulge PNe abundances with those of bulge giant stars quoted in the literature.

This paper is organized as follows.
In Sects.~\ref{sec:PN} and ~\ref{sec:Stars}, we describe the PN and giant star samples on which we based our study, and
discuss the main uncertainties in their abundance determinations. In Sect.~\ref{sec:mixing}, we review current ideas about the main mixing processes occurring in low- and intermediate mass stars. In Sect.~\ref{sec:PNeCE} we compare the properties of PNe in different environments. We present compelling evidence that both O and Ne are reliable chemical-evolution tracers for metal-rich, old populations, in contrast to their unreliability for metal-poor systems. In Sect.~\ref{sec:StarsPNeCE}, bulge PN abundances are compared with field-giant bulge
abundances. We start 
this section by recalling the recent results on the bulge 
formation obtained from the study of stars. We approach the question of whether PNe and giant stars provide similar answers concerning the bulge chemical evolution and discuss the biases involved.
Section~\ref{sec:conclusions} contains a summary of our main results.

\section{Planetary nebulae: samples and abundances}\label{sec:PN}

\subsection{The adopted samples}\label{sec:samples}

\subsubsection{Bulge and inner-disk}\label{sec:BDsamples}

Our PN sample is obtained by merging the following data sets: 90 PNe
observed with 4m-class telescopes from G\' {o}rny et al. (2008, their
sample C), 164 PNe from G\' {o}rny et al. (2004), and 29 PNe from Wang
\& Liu (2007), the latter two data sets have been acquired with 2m-class
telescopes. The data set for 164 PNe from G\' {o}rny et al. (2004) come from the merging of data for 44 PNe observed by G\' {o}rny in 2000 with those for PNe
observed by Cuisinier et al. (2000), Escudero \& Costa (2001), and Escudero et al. (2004).
In this way, we obtain the largest sample of PNe with high-quality data
observed in the direction of the Galactic bulge (for a detailed description, see
G\'{o}rny et al. 2008), with 245 objects (for the 39 objects belonging to
more than one sample, the best data were chosen). This sample
contains both bulge and inner-disk objects.

\begin{figure}
\center
\includegraphics[height=6cm,width=8cm]{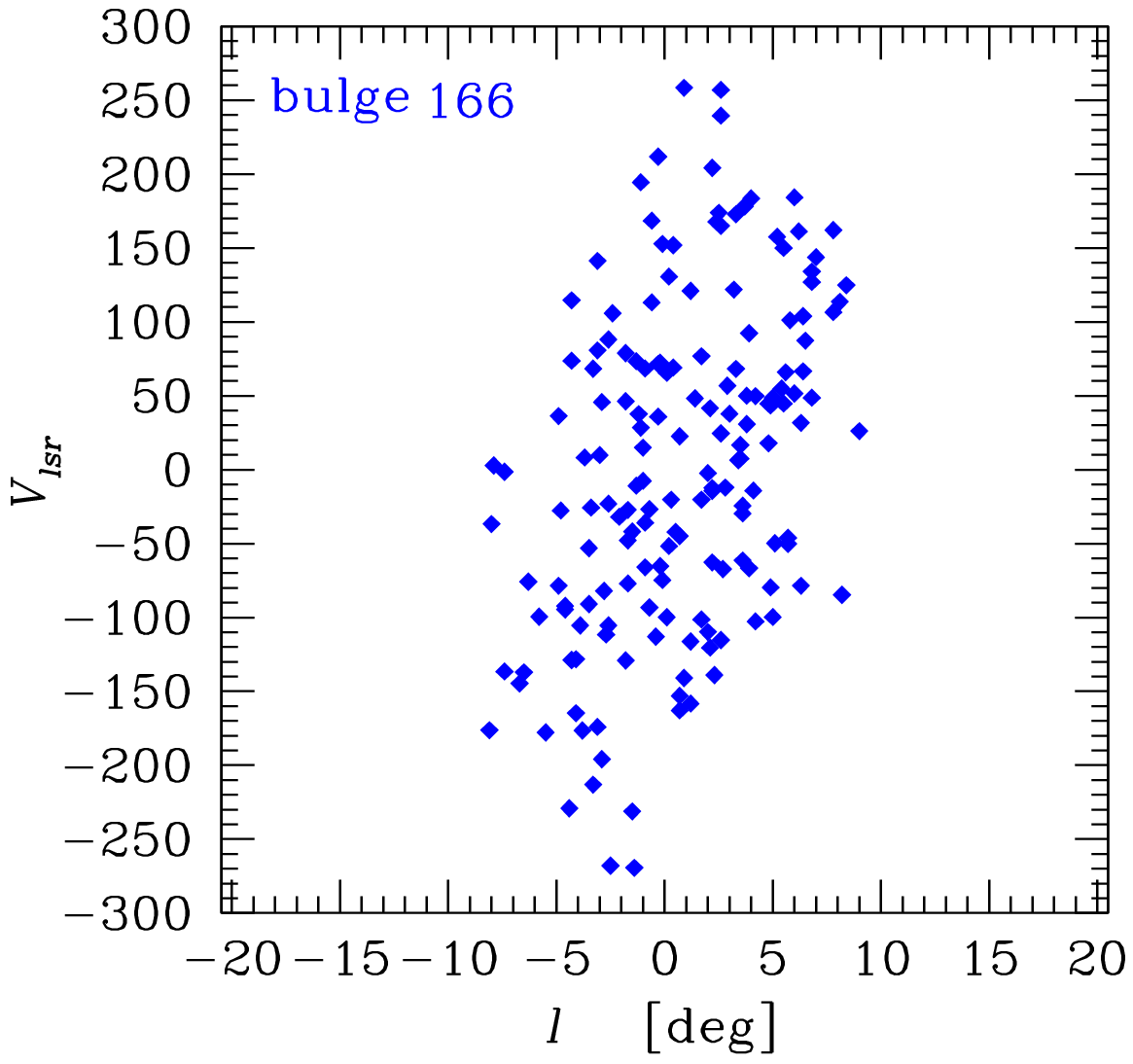}
\includegraphics[height=6cm,width=8cm]{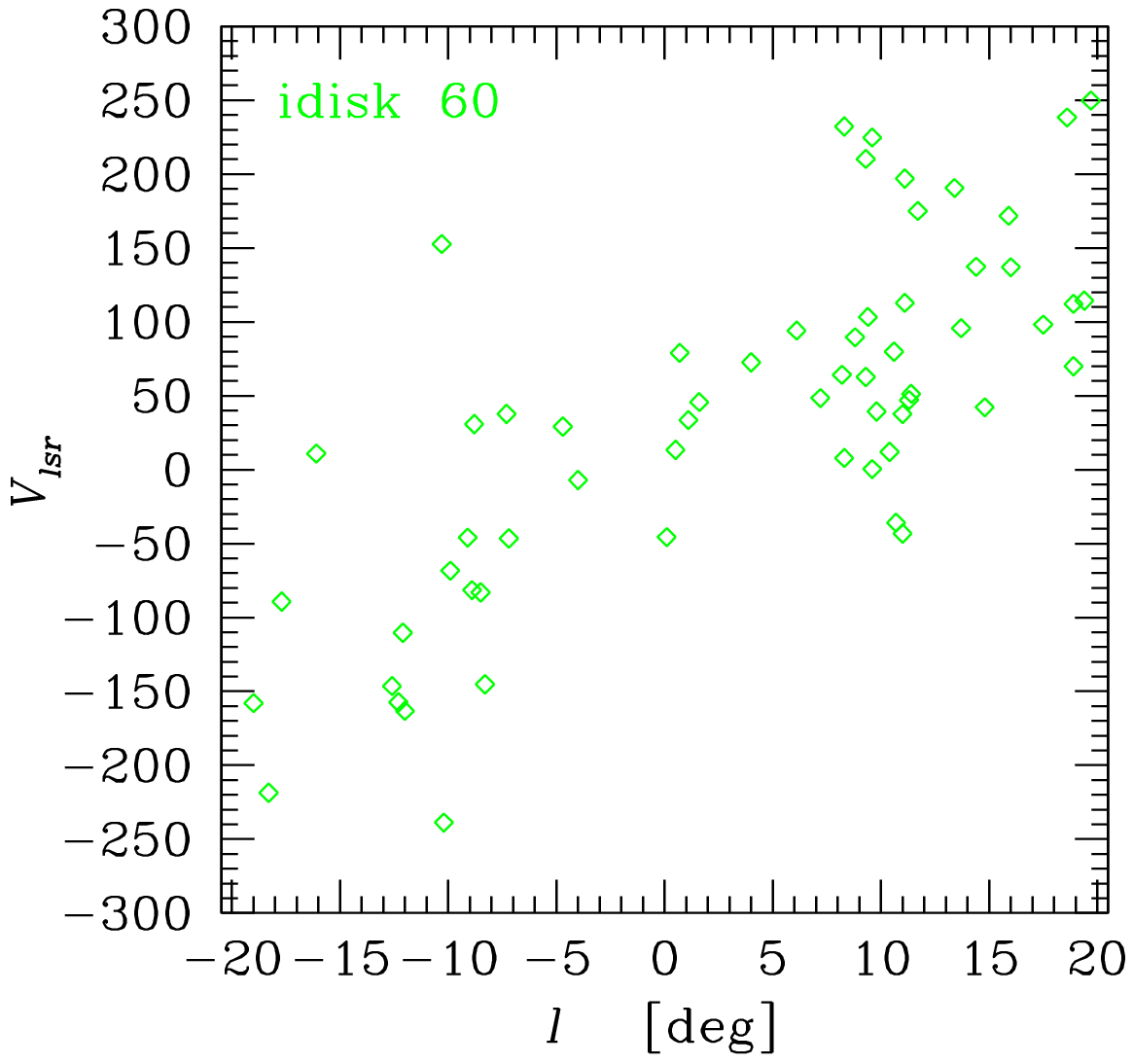}
\caption[]{Distributions of bulge (upper panel) and inner-disk (lower
panel) PNe in a radial velocity (corrected for the solar motion)
versus Galactic longitude
diagram. All the objects of the initial
sample with known velocities are represented (this excludes 11 bulge
and 5 inner-disk PNe for which the velocities were unavailable.).}
\label{vslr}
\end{figure}

The bulge sample was assembled following the standard criteria
(e.g. Stasi\'nska \& Tylenda 1994): they have locations within 10$^{\circ}$ 
of the Galactic centre, diameters smaller than 20 arcsec, and radio fluxes at
5~GHz of less than 100\,mJy. The contamination of a bulge PN
sample defined in this way by disk PNe is estimated to be less than 
10\% and most probably around 5\% (Stasi\'nska et al. 1991). Objects
observed in the direction of the Galactic centre, but rejected
according to the standard criteria described before, most probably
belong to the inner-disk\footnote{PNe seen in
the direction of the bulge but known to
belong to the Sgr B2 galaxy were removed from the sample.}.
We note that the two samples have different
distributions in a radial-velocity versus Galactic-longitude diagram
(see Fig.~\ref{vslr}), which are consistent with the bulge/inner-disk
classification. PNe have indeed the advantage over stars that one can
use their angular diameter as an additional constraint to decide
the Galactic component to which they belong.
Practically all of the 90 PNe of G\' {o}rny et al. (2008, sample C) can be
regarded as belonging to the bulge, given that most of the objects
were selected according to the above criteria.
Only 4 PNe were classified as inner-disk objects because they are
located just beyond 10 degrees from the Galactic centre.

We note that the inner-disk sample is less well defined than the bulge one,
and possibly includes both thick and thin disk objects.
However, we expect most of them to be
located at small Galactocentric distances due to the increasing 
density of stars towards the inner regions
of the Galaxy.

\begin{figure}
\center
\includegraphics[width=8cm]{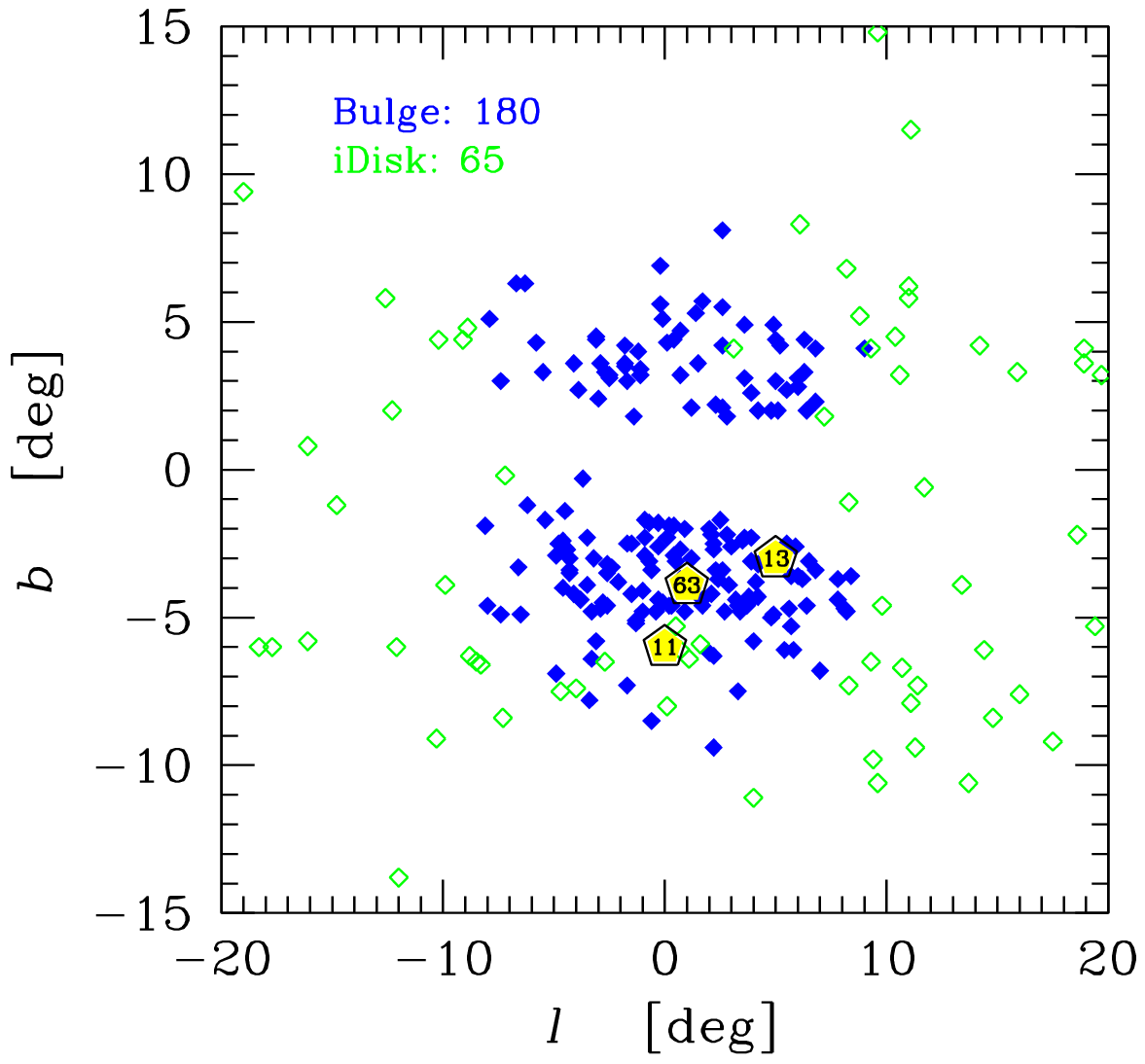}
\caption[]{Distribution of our bulge and inner-disk samples in a
Galactic latitude versus Galactic longitude diagram. The larger symbols
mark the location of the observed
fields in the case of bulge stars (the number of
stars in each field is also indicated).} \label{lb}
\end{figure}

Figure~\ref{lb} shows the distribution of the  bulge and inner-disk
PN samples in a Galactic latitude versus Galactic longitude diagram. We also
indicate the fields in which the bulge stars discussed in
Sect.~\ref{sec:Stars} are found.

From the above PN samples, we rejected objects for which abundances
could not be computed
from the available data or were too uncertain (see
Sect.~\ref{sec:biases}). The rejection criteria adopted in this work
were far more restrictive than in G\' {o}rny et al. (2004). This
explains why, despite adding data for many additional PN ($\sim$80), we have
increased the final bulge PN sample with
respect to G\'{o}rny et al. (2004) by only $\sim$40 objects. 
We considered it to be of greater
importance to analyse reliable abundances than increase the
number of objects.

\subsubsection{The Large Magellanic Cloud sample}\label{sec:LMCsample}

A data set of PNe with low metallicities is critical to a
comparison with the two samples described before.
Both the Small Magellanic Cloud (SMC) and the LMC are suitable candidates for this type of study
because both galaxies are close to the Milky Way and sufficiently
massive to contain a large number of PNe.
We chose to use data for PNe in the LMC in our comparison because of the larger
number of objects with good  quality spectra available.

The LMC PN sample originates in a compilation by Leisy \& Dennefeld
(2006) and contains 138 objects. We derived 
abundances in the same way as for the other two PN samples, and
applied the same restriction criteria as above concerning the 
abundance quality. Our final sample contains $\sim$100 objects.

\subsection{Abundance determinations in the PN samples}
\label{sec:PNabundances}

In all of our samples, the abundances were determined in exactly the
same way with an identical set of atomic data, using a classical
temperature-based empirical method described in detail by G\'orny et
al. (2008). Briefly, interstellar reddening was obtained from the
observed hydrogen Balmer decrement by comparison with the case B
theoretical decrement at a given temperature (first assumed to be
equal to 10$^4$K). All the line intensities were then corrected for
interstellar extinction. The electron temperature was then derived
from the \rOiii\ and/or \rNii\  ratios, and the electron density from
the \rSii\ ratio\footnote{If the density could
not be obtained from the \rSii\ ratio, we
adopted a value of 2000 {\cmcub} (the average of the values of the
measured densities).}. An iteration was performed over these first steps.
From the intensities of observed recombination lines of H and He and
forbidden lines of N, O, Ne, S, and Ar ions, ionic abundances were
then derived with respect to H. The elemental abundances were finally
obtained by using the ionization correction factors (ICFs) proposed
by Kingsburgh \& Barlow (1994) (except for Cl, in which case we used
the ICF of Liu et al. 2000).

In some cases, the observational data did not allow us to estimate the
electron temperature. We rejected the corresponding PNe from further
consideration because trustworthy abundances are unattainable
in such cases, even when using tailored photoionization modelling
instead of empirical methods (Stasi\'nska 2002). This involved discarding 12
objects out of 180 from our bulge sample, 4 objects out of 65 from our
inner-disk sample, and 27 out of 138 from the LMC
sample. 

The [N~{\sc ii}] $\lambda$5755 and [O~{\sc ii}]
$\lambda$$\lambda$7320,7330 lines can be affected by recombination from
\Npp\ and \Opp\ ions. We estimate the effects of the latter process by using the
expressions given in Liu et al. (2000), the \rNii\ temperature, and
by assuming that \Npp/H=\Opp/H $\times$ \Np/\Op. We  find that the final effect on the
computed abundances is negligible for our objects\footnote{
This effect could be larger if, as suggested by
Liu (2006 and references therein),
the recombination lines originate from
a much cooler zone. However, even
in the most extreme cases shown in Wang \& Liu
(2007), where the observational data
allow a more accurate correction for the effect of recombination, the resulting
abundances are modified by a few percent at most.}.

As explained in G\'orny et al. (2008), the uncertainties in abundance
ratios were estimated by propagating uncertainties in the observed
emission line intensities using Monte Carlo simulations.
In the case of the LMC sample, the  uncertainties in
the observed line intensities were estimated using information
provided in the tables of Leisy \& Dennefeld (2006), considering that 
the uncertainty is a factor of 0.4 the intensity of lines marked as
upper limits or a factor of 0.3 the intensity of the weakest line measured.
The accuracies in the final abundances in the bulge, inner-disk, and
LMC samples were found to be similar.

To avoid dealing with values that
are too uncertain, we remove from further consideration any abundance
ratio for which the two-sigma error from the Monte Carlo
simulation is larger than 0.3~dex.
The median uncertainty in the
abundance ratios for the data remaining in the sample is about
$\pm$0.1~dex for $\log\epsilon$(O), $\log\epsilon$(N), $\log\epsilon$(Ne), 
$\log\epsilon$(Ar), $\log\epsilon$(S), and $\log$(N/O),
where $\log\epsilon$(X) = log(X/H)+12.

We note that our Monte Carlo procedure does not take account
of possible variations in the reddening law, the effect of unknown
temperature and density structure of the nebulae, and the uncertainty
in the ionization correction factors (see below).

\subsection{Biases and uncertainties in PN abundances}
\label{sec:biases}

We now discuss in more detail the additional sources of errors mentioned above.\\

\noindent{\it Atomic data}

\noindent
The atomic data used in abundance determinations from optical
lines are generally believed to be quite accurate (to within 5 - 10\%).
Atomic data is presently not a major issue for abundance 
determinations in PNe from optical data.\\

\noindent{\it Extinction}

\noindent
It is known that the Galactic extinction curve is not identical in all
directions, and can be, to a first approximation, characterized by the
ratio of the total to selective extinction, $R_V$ (Cardelli et al.
1989, Fitzpatrick 1999). In the analysis of their sample of Galactic
bulge PNe, Wang \& Liu (2007) were able to derive the extinction
curve corresponding to each object by combining UV and optical H I
and He II lines, and found values of $R_V$ ranging from 1.8 to 4.2.
An inapropriated extinction law has its most significant effect on the
N/O ratio, when the extinction is high. We tested various
extinction laws and found that, in our samples, the N/O ratios 
obtained using $R_V$ =2.1 and 4.2, differed on average by less than 0.1~dex. 
Therefore, we assumed the same extinction law (Seaton 1979) for all the
samples, which corresponds to the standard case, $R_V$ = 3.1.\\

\noindent
{\it Temperature gradients and fluctuations}

\noindent
As known, the temperature in PNe is not uniform. Radial gradients are
expected, due to variations in heating and cooling processes
across the nebula. This is partly taken into account by our procedure
if temperatures from both  \rOiii\ and \rNii\ are measured with good
accuracy. However, some authors (Peimbert \& Peimbert 2006) argue
that significant temperature fluctuations occur within PNe, and that
abundances derived from collisionally excited lines may be
significantly underestimated if those temperature fluctuations are
not properly taken into account. In particular, they argue that the
correct abundances of the heavy elements are those obtained by using
\emph {recombination} lines. A more widespread view, however, is that
abundances from collisionally excited lines reflect the average
abundances in the nebulae, while the recombination lines are mostly
affected by cold, hydrogen-poor inclusions, which represent only a
small fraction of the total mass (see Wang \& Liu 2007 and references
therein). The origin of such inclusions is not yet fully understood.
In this paper, we adopt the view that the abundances derived from
collisionally excited lines are not biased in any significant way.\\

\noindent
{\it Ionization correction factors}

\noindent
The ionization correction factors (ICFs) are another source of
uncertainty. One could overcome this problem by constructing tailored
photoionization models for each object. However, this would require:
1) taking into account the geometrical structure of the nebula (rarely
known in detail at the distance of the bulge); 2) knowing the exact
spectral energy distribution of the ionizing radiation field (but the
model atmospheres of hot stars computed by various
state-of-the art codes for model atmospheres do not
agree on the energy distribution in the H Lyman continuum).

Another option could be to complement ionic abundances measured 
from optical spectra
with those from UV and IR spectra, so that most of -- if not all -- 
relevant ions are observed.
This procedure was employed by
Pottasch \& Bernard-Salas and their coworkers (see Pottasch \&
Bernard-Salas 2006; see also Gutenkunst et al. 2008). However, these
results can be affected by aperture and
calibration problems. In fact, without a detailed photoionization
modelling, it is impossible to take into account the effect of ionization
stratification in the different observing apertures. In addition, when
abundances of ions such as \Nepp, \Spp or \Arpp are derived simultaneously
from optical and infrared data, the latter
are generally higher by factors of 2--3.
The reason for this is presently not understood.
So, while the method is in principle
appealing, its results are not necessarily more
trustworthy than those from optical data alone.
We are thus left with the simple ICF method which, in
addition, has the advantage of being easily applicable to a large
number of objects.

Unfortunately, the uncertainty in
abundance ratios arising from the ICFs is difficult to evaluate. It
probably dominates the other sources of uncertainty in our final
samples. From our experiments with large grids of photoionization
models, we estimate a typical uncertainty of
$\pm$0.1~dex for $\log\epsilon$(O) (except for PNe of rather low
excitation, for which most of the oxygen is in the
optically visible forms \Op\ and \Opp and the ICF uncertainty is
almost zero). For $\log$(N/O), uncertainties could
exceed these values at the high excitation end, whereas for $\log$(Ne/O) this would 
happen at the low excitation end. The $\log$(Ar/O) ratio
should be accurate to within $\pm$~0.1~dex, except at the very low and
very high excitation ends. The $\log$(S/O) ratio, on the other hand, is probably
uncertain by more than $\pm$~0.2~dex in the entire excitation range
\footnote{
The comparison of our abundance determinations
for LMC PNe with those of Leisy \& Dennefeld
(2006) indicates differences. The most important one is for
the S/H values. The latter authors obtained a S/H larger than solar
for more than half of their sample. 
In contrast, we obtain a median log(S/H)+12
value of 6.45, below the solar value of 7.14.
This difference is due to the different
sulphur ICFs used in both studies.
By comparing HII regions and PNe, Henry et al.
(2006) noted that the latter tended to have much lower S/H ratios for
a given O/H ratio, a fact which they called the ``sulfur anomaly''
and attributed to an inadequate correction for \Sppp\ in PNe.}.

A practical way of testing whether the ICF
method may generate biases is to check whether the derived abundance
ratios show any
trend with respect to the ionization level of the nebula. We 
checked the values of He/H, O/H, N/O, Ne/O, S/O, and Ar/O as a
function of O$^+$/(O$^+$+ O$^{++}$) for the merged sample of all
PNe considered in this paper (as done in G\'orny et al. 2008, appendix). 
We found that the ICFs of Kingsburgh \&
Barlow (1994) do not lead to any artificial trend, provided that
one removes PNe with O$^+$/(O$^+$+ O$^{++}$)~$<$~0.4 in the
determination of He/H, Ar/O, and S/O. Hence, when considering the latter
abundance ratios, we removed from our samples the objects with
O$^+$/(O$^+$+ O$^{++}$)~$<$~0.4 (i.e. 23 out of 168 for the bulge
sample, 5 out of 61 from the inner-disk sample, and 7 out of 111
for the LMC sample).

The plasma parameters and chemical abundances on which the present work is based are listed in
Table~\ref{table_abundances}, where we show
separately bulge, inner-disk, and LMC PNe. There are three rows of data for
each object, and a fourth row used to separate them. The first row indicates the
values of parameters computed from the nominal values of the observational
data. 
The second row gives the one sigma upward deviation obtained from our Monte Carlo
simulations, while the third row gives the one sigma downward deviation.

\begin{figure}
\resizebox{\hsize}{!}{\includegraphics{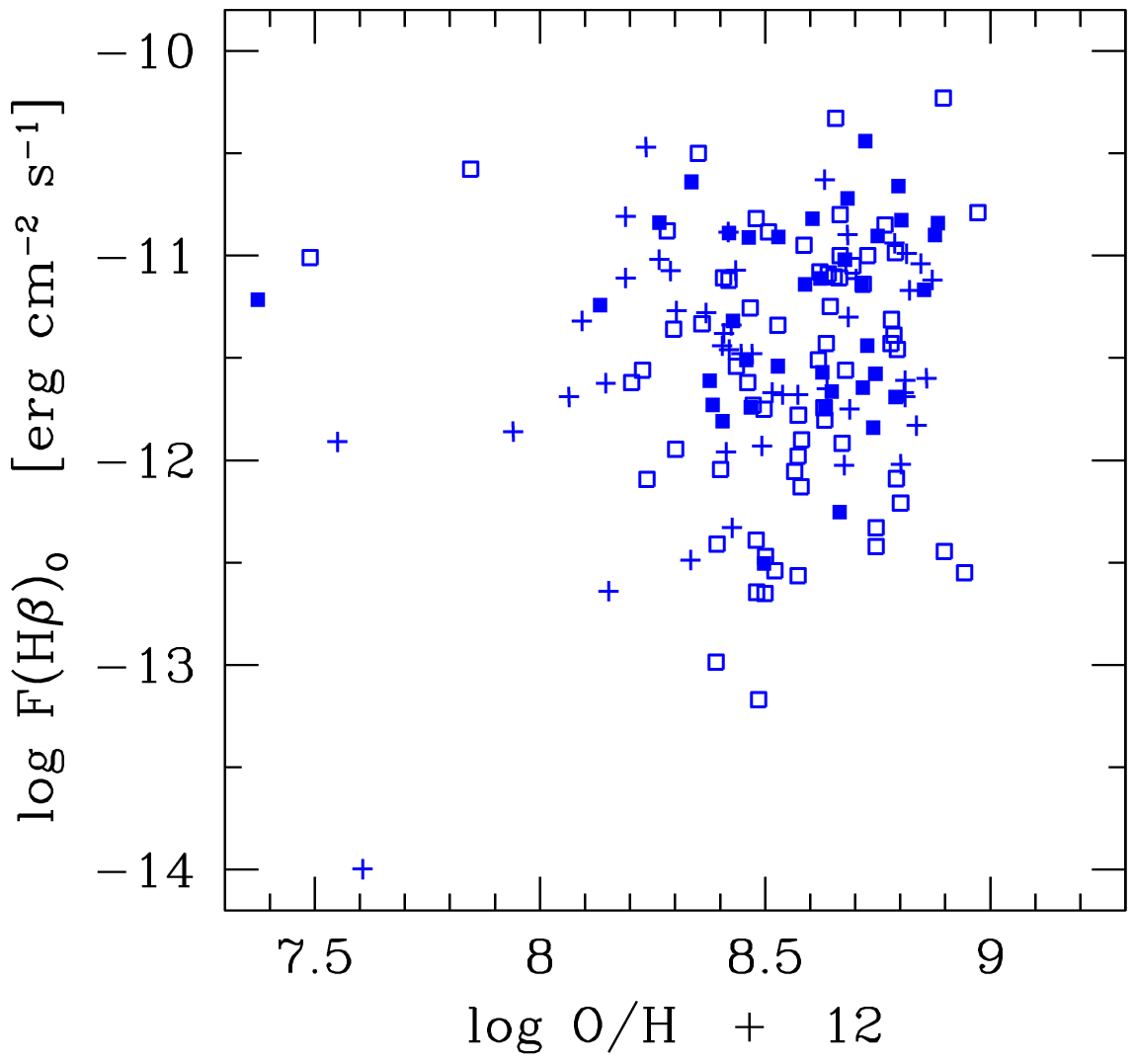}}
\caption{The oxygen abundance as a function of the total nebular
    H$\beta$ flux, corrected for extinction, for our final sample of
    bulge PNe. The filled symbols represent oxygen abundances of higher
quality (see Sect.~\ref{sec:biases} and ~\ref{sec:PNedistributions}).
The crosses represents objects with
errors in $\log\epsilon$(O) larger than 0.3~dex that
were excluded from our analysis.}
\label{O-FHB}\end{figure}

In Table~\ref{table_abundances} we also report the quality of the derived parameters, marked by one of the following
symbols: ``+'' marks data of the highest quality for which the derived individual
error is smaller than the computed median error of
that parameter for the entire PNe sample\footnote{
 The PNe marked with a ``+'' are represented by filled symbols in the 2D diagrams (Figs.~\ref{O-FHB},~\ref{ONO}, and \ref{OArNe}) and by filled bar histograms (Figs.~\ref{Distributions1},~\ref{Distributions2}, ~\ref{Distributions3}, and \ref{SARD}).
 }; 
``:'' indicates data of larger uncertainties but still not of errors exceeding the 0.3~dex limit; 
``;'' labels data rejected from further consideration due to either an error superior
to 0.3~dex or due to large uncertainties arising from the ICFs; an ``-'' marks objects
for which the considered parameter could not be derived. 
\\

\begin{table*}
\caption{Plasma parameters and chemical abundances obtained in the present work$^*$. The quantities
reported in each column are: 
Column (1) -- the PNG number; Column (2) -- the usual name of the PN; Column (3) --
the electron density deduced from the [S~{\sc ii}] $\lambda$6717/6731 ratio;
Column (4) -- the electron temperature deduced from the [O~{\sc iii}]
$\lambda$4363/5007 ratio; Column (5) -- the electron temperature deduced from the [N~{\sc ii}]
$\lambda$5755/6584  ratio (the value of T$_e$(N~{\sc ii}) is in        
parenthesis if T$_e$(O ~{\sc iii}) was chosen for all ions); Columns        
(6) to (12) -- the He/H, N/H, O/H, Ne/H, S/H, Ar/H, Cl/H abundance ratios,
respectively. }\label{table_abundances}
\tiny{
\begin{tabular}{
                    r
                    r @{\hspace{0.20cm}}
                    r @{\hspace{0.05cm}}
                    r @{\hspace{0.40cm}}
                    r @{\hspace{0.05cm}}
                    r @{\hspace{0.30cm}}
                    r @{\hspace{0.05cm}}
                    r @{\hspace{0.30cm}}
                    r @{\hspace{0.05cm}}
                    c @{\hspace{0.35cm}}
                    r @{\hspace{0.05cm}}
                    c @{\hspace{0.35cm}}
                    r @{\hspace{0.05cm}}
                    c @{\hspace{0.35cm}}
                    r @{\hspace{0.05cm}}
                    c @{\hspace{0.35cm}}
                    r @{\hspace{0.05cm}}
                    c @{\hspace{0.35cm}}
                    r @{\hspace{0.05cm}}
                    c @{\hspace{0.35cm}}
                    r @{\hspace{0.05cm}}
                    c}
\hline
  PN\,G &
  name &
  N$_{e}$ &  &
  T$_e$(OIII) &  &
  T$_e$(NII) &  &
                                                                                        He/H &  &            N/H &  &            O/H &  &           Ne/H &  &            S/H &  &           Ar/H &  &          Cl/H  &    \\
\hline
 \vspace{0.1cm}
 bulge:~~~~~ &            &             &  &             &  &               &  &             &  &                &  &                &  &                &  &                &  &                &  &                &    \\
  000.1-02.3 &   Bl\,3-10 & {\bf ~~446} &; & {\bf 12750} &: & {\bf    --~~} &--& {\bf 0.220} &: & {\bf     --~~} &--& {\bf 4.41E-04} &: & {\bf 6.38E-05} &: & {\bf     --~ } &--& {\bf 1.84E-06} &: & {\bf     --~ } & -- \\
             &            &        846  &  &      13440  &  &               &  &      0.272  &  &                &  &      5.16E-04  &  &      9.41E-05  &  &                &  &      1.95E-06  &  &                &    \\
             &            &        136  &  &      12410  &  &               &  &      0.149  &  &                &  &      3.60E-04  &  &      6.01E-05  &  &                &  &      1.51E-06  &  &                &    \\
  000.1+04.3 &    H\,1-16 & {\bf ~5890} &; & {\bf 10430} &: & {\bf (17000)} &: & {\bf 0.113} &: & {\bf 5.83E-05} &: & {\bf 6.76E-04} &; & {\bf     --~~} &--& {\bf 6.08E-06} &; & {\bf 2.94E-06} &; & {\bf     --~ } & -- \\
             &            &      15100  &  &      11000  &  &      (20570)  &  &      0.211  &  &      8.48E-05  &  &      1.12E-03  &  &                &  &      1.15E-05  &  &      4.58E-06  &  &                &    \\
             &            &       3820  &  &       9200  &  &      (10260)  &  &      0.105  &  &      4.86E-05  &  &      5.63E-04  &  &                &  &      4.55E-06  &  &      2.37E-06  &  &                &    \\
\hline
\end{tabular}
}\\
$^*$ The full table is available in electronic form at http://www.aanda.org.
\end{table*}

\noindent
{\it The relation between O/H and the PN luminosities}

\noindent
Since PN progenitors span a wide range of ages, one expects that the
O/H abundances observed in PNe would be correlated with their
progenitor ages, reflecting the chemical evolution of their environment.
Stasi\'nska et al.
(1998) found that, in both the Large and Small
Magellanic Clouds, more luminous PNe tend to be
more oxygen-rich but that for the Galactic
bulge, the picture was not clear. In
Fig.~\ref{O-FHB}, we display the value of the oxygen
abundance as a function of F(H$\beta$), which is the
extinction-corrected value of the total nebular
flux in the H$\beta$ line for our bulge PN sample. No trend is visible.
We note that, by rejecting objects with oxygen-abundance
uncertainties larger than 0.3 dex, we do not generate any
significant bias in the abundance distribution since these objects span the 
entire metallicity range shown in Fig.~\ref{O-FHB}.

\subsection{Abundance Distributions}
\label{sec:PNedistributions}

Table~\ref{table_PN} reports the median values of
$\log\epsilon$(He), $\log\epsilon$(O), $\log\epsilon$(Ar), 
$\log\epsilon$(Ne), $\log\epsilon$(S), $\log\epsilon$(N),
$\log\epsilon$(Cl), log(S/O), log(Ne/O), log(Ar/O), log(S/Ar),
log(N/O), and log(N/Ar)
in bulge, LMC, and inner-disk PNe. We show
in brackets the 25 and 75 percentiles, and in parenthesis the number of
objects with relevant data in each sample. Also shown, for
comparison, are the values found for the bulge
giants (see Section~\ref{sec:Stars}),
the solar abundances of Asplund et al. (2005), as well as
values of log(S/O), log(Ar/O), and log(Ne/O) at $\log\epsilon$(O)~$=$~7 and 8.66 for
a sample of 109 blue compact dwarf galaxies
with high quality spectra studied by Izotov et al. (2006). We note that,
in HII regions, errors in abundance ratios due to ICFs are expected
to be far smaller than in PNe. The ICFs used by Izotov et al. (2006)
in their study are based on a grid of photoionization models relevant
to giant HII regions, and are different from those used for our
PN samples. Izotov et al. (2006)
noticed a slight increase in the Ne/O
ratio with increasing O/H of $\sim$0.1~dex over the entire metallicity range considered. 
This was interpreted as an
indication that $\sim$20\% of oxygen is locked in dust grains in the
highest-metallicity HII regions of their sample.

\begin{figure*}
\resizebox{0.33\hsize}{!}{\includegraphics{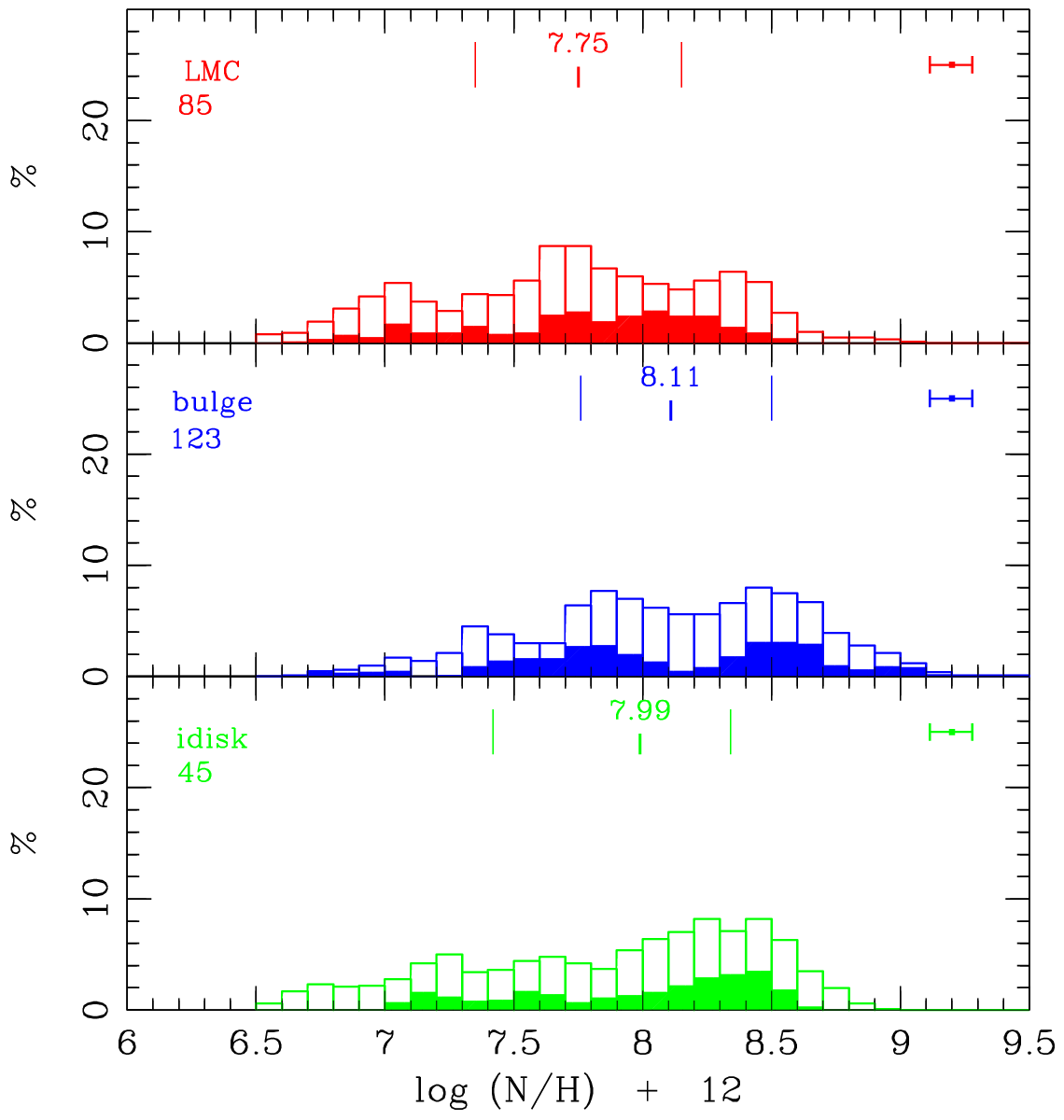}}
\resizebox{0.33\hsize}{!}{\includegraphics{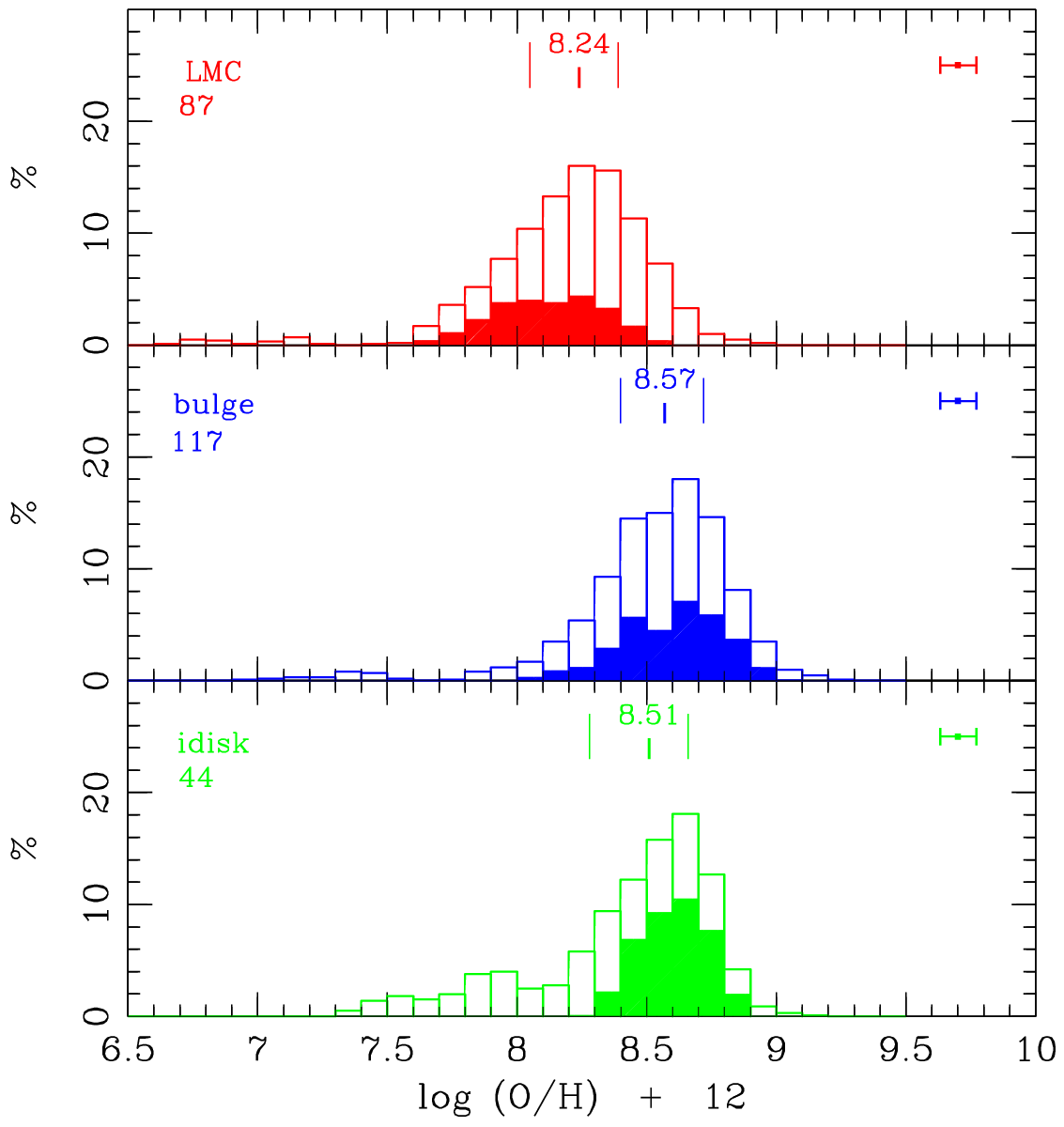}}
\resizebox{0.33\hsize}{!}{\includegraphics{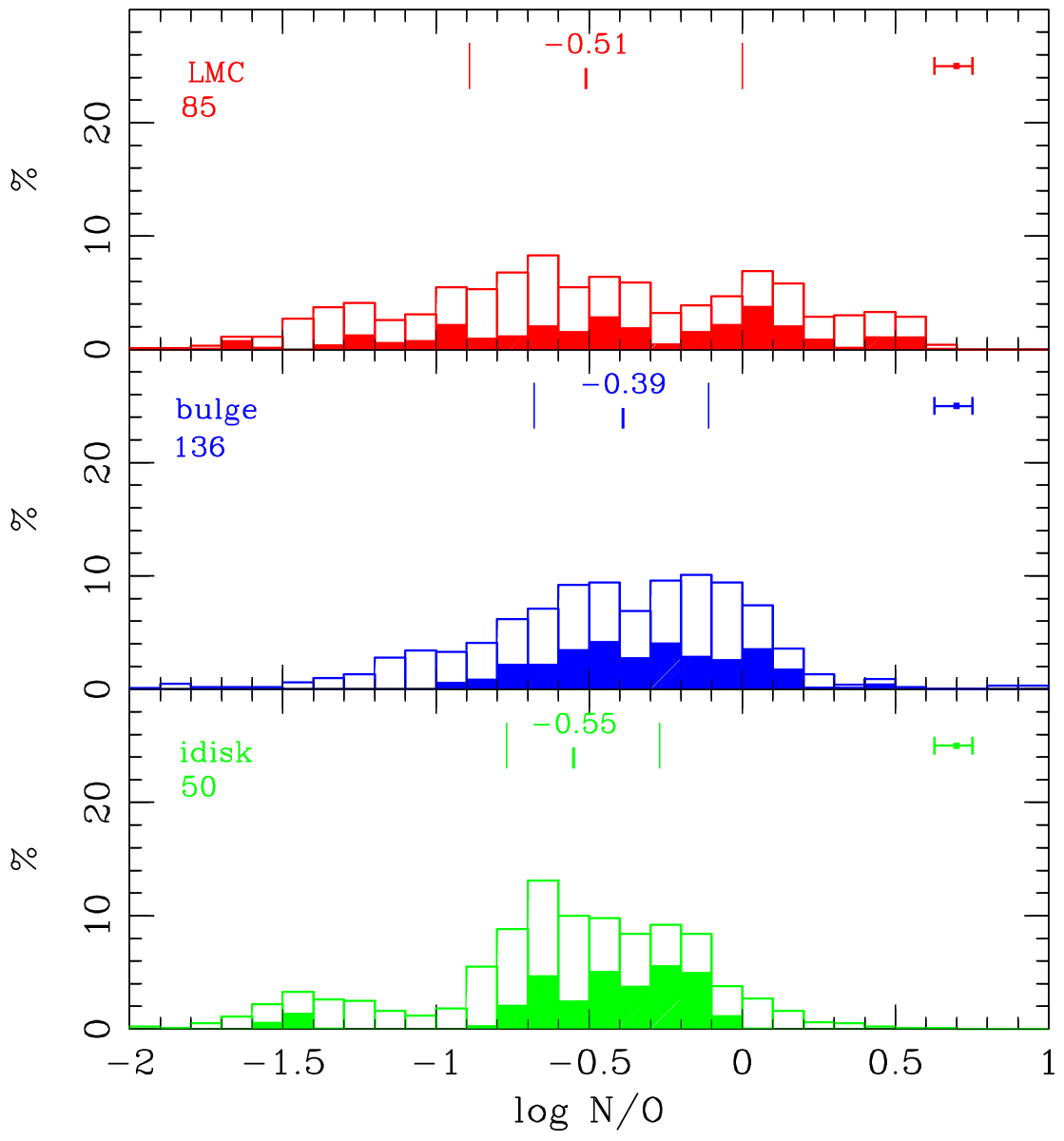}}
\resizebox{0.33\hsize}{!}{\includegraphics{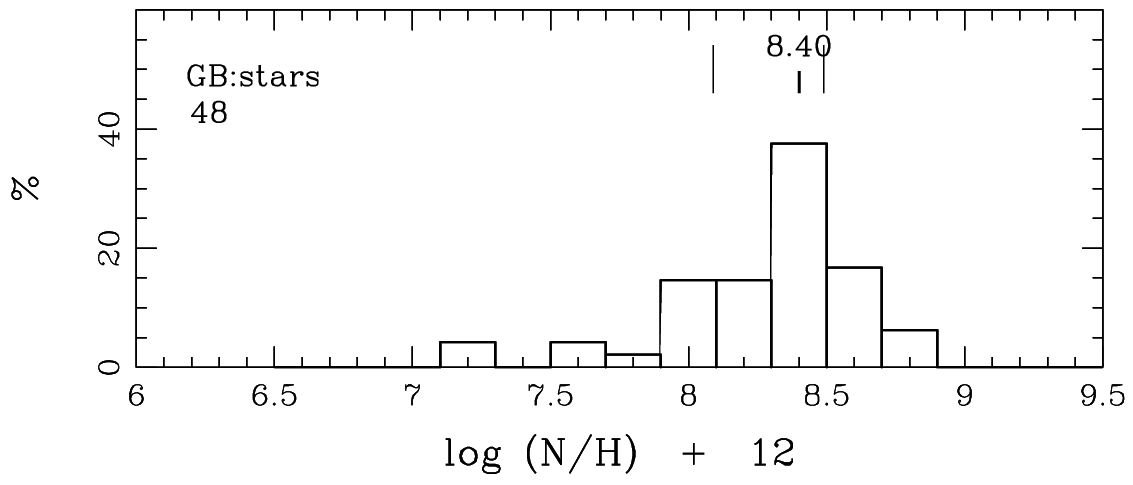}}
\resizebox{0.33\hsize}{!}{\includegraphics{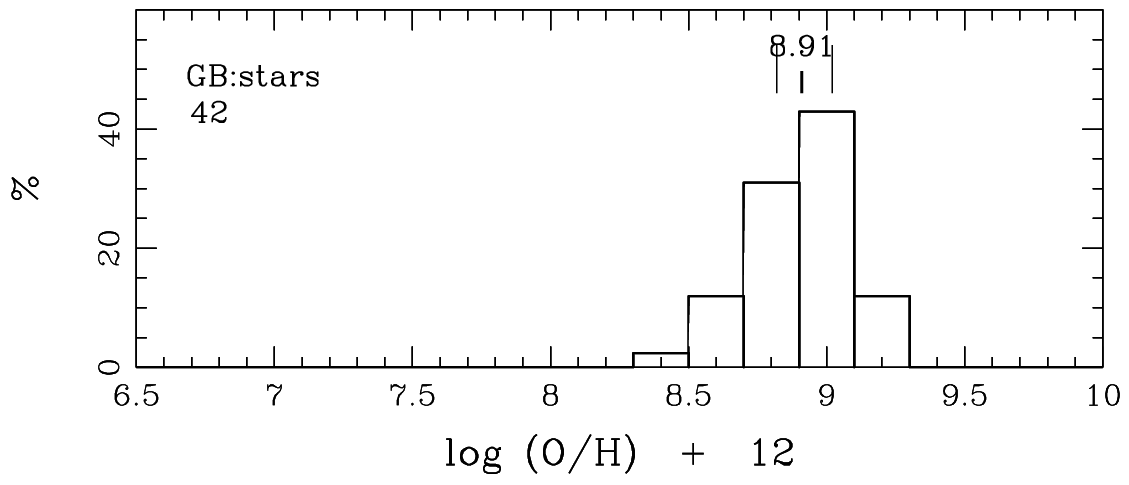}}
\resizebox{0.33\hsize}{!}{\includegraphics{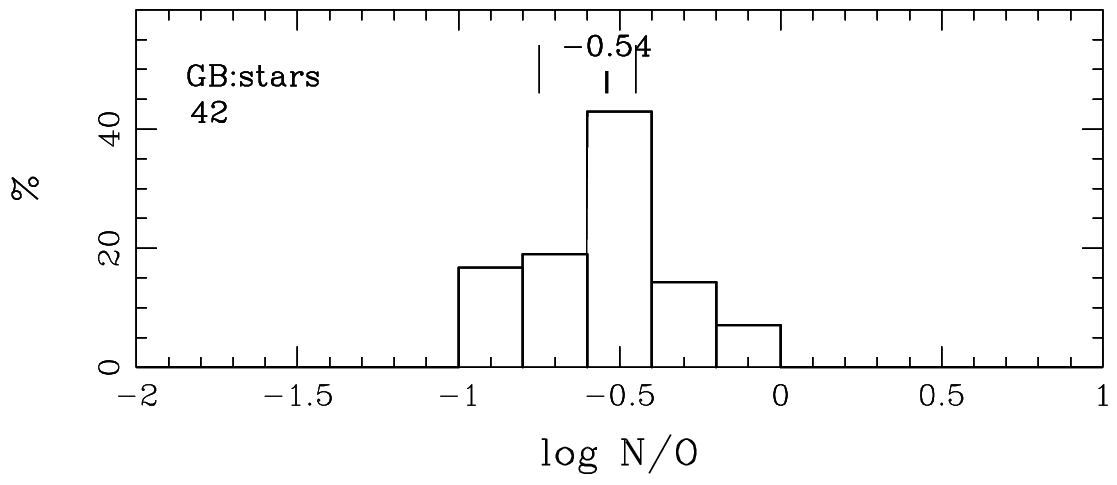}}
\caption[]{In the first three rows, we present the PN abundance distributions
in the LMC, bulge, and inner-disk for N (1st
column), O (2nd column), and N/O (3rd column).
The median values of the distributions, the 25
and 75 percentiles, the number of objects
used in each panel, and the typical error bar are also shown. 
In the three first rows, the filled-bar histograms show the location of our highest quality data namely that have errors in the abundance ratio smaller than
the median error for the entire PN sample (see text). In the last row, we show the
abundance distributions of N, O, and N/O obtained for field-bulge
giants.
}
\label{Distributions1}
\end{figure*}

\begin{figure*}
\resizebox{0.33\hsize}{!}{\includegraphics{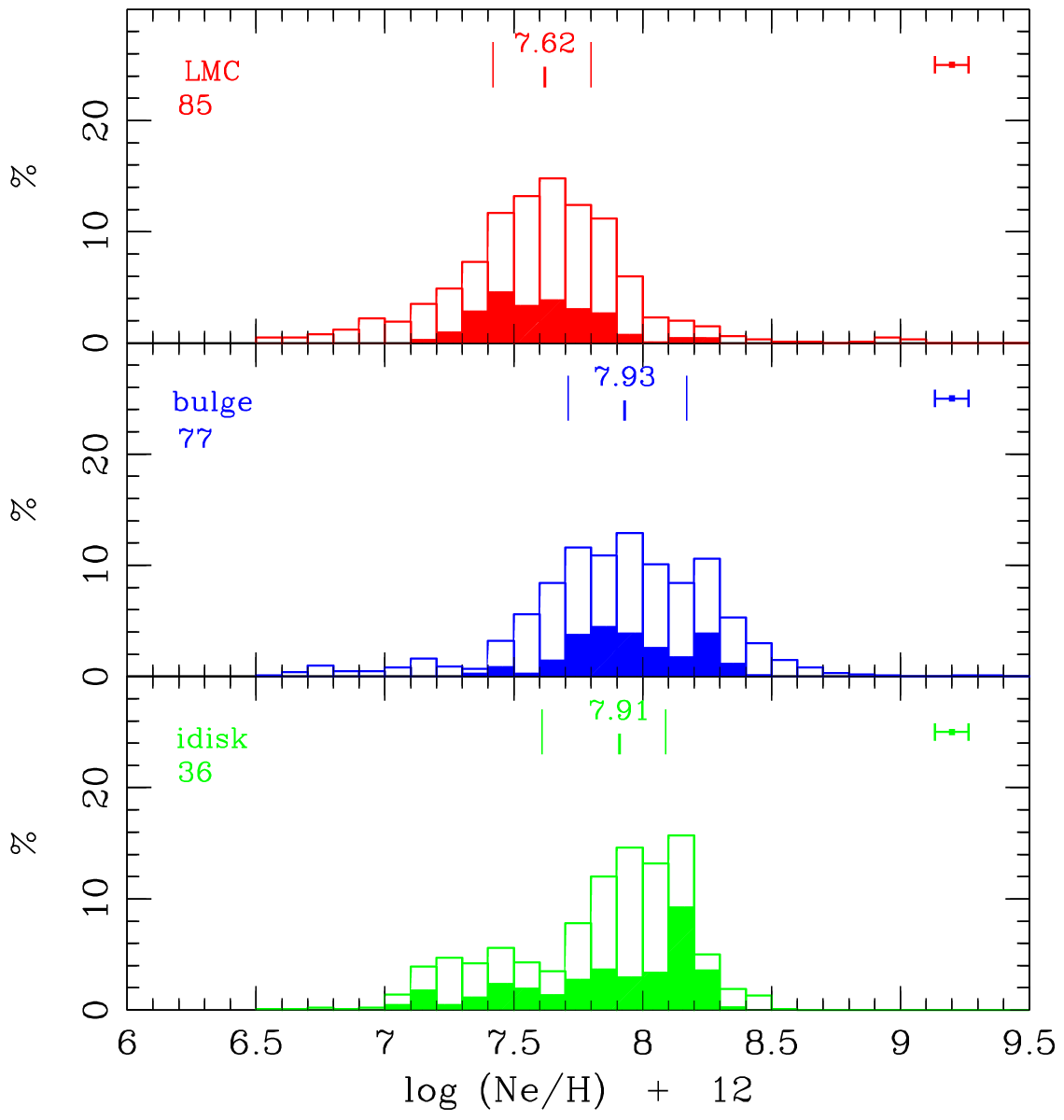}}
\resizebox{0.33\hsize}{!}{\includegraphics{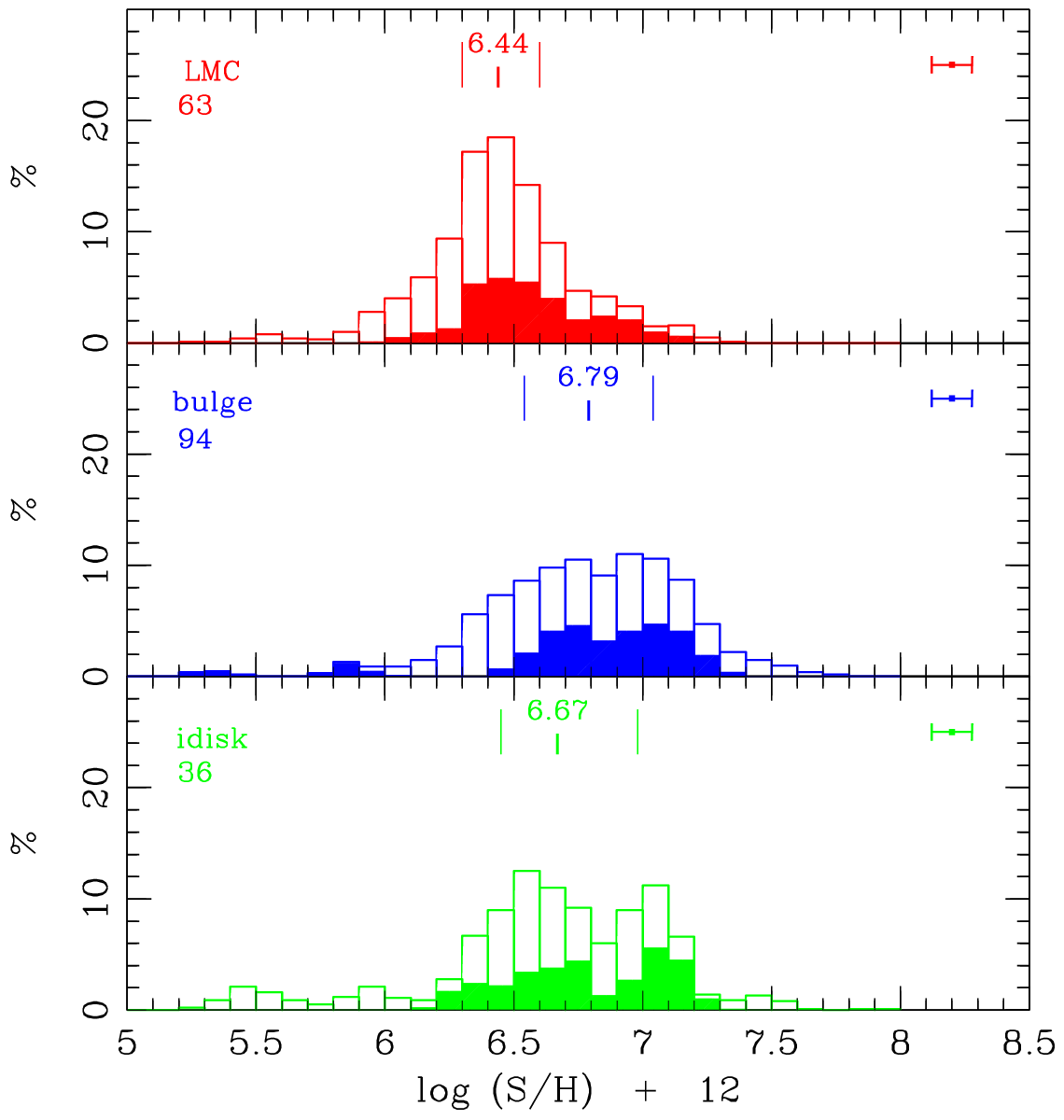}}
\resizebox{0.33\hsize}{!}{\includegraphics{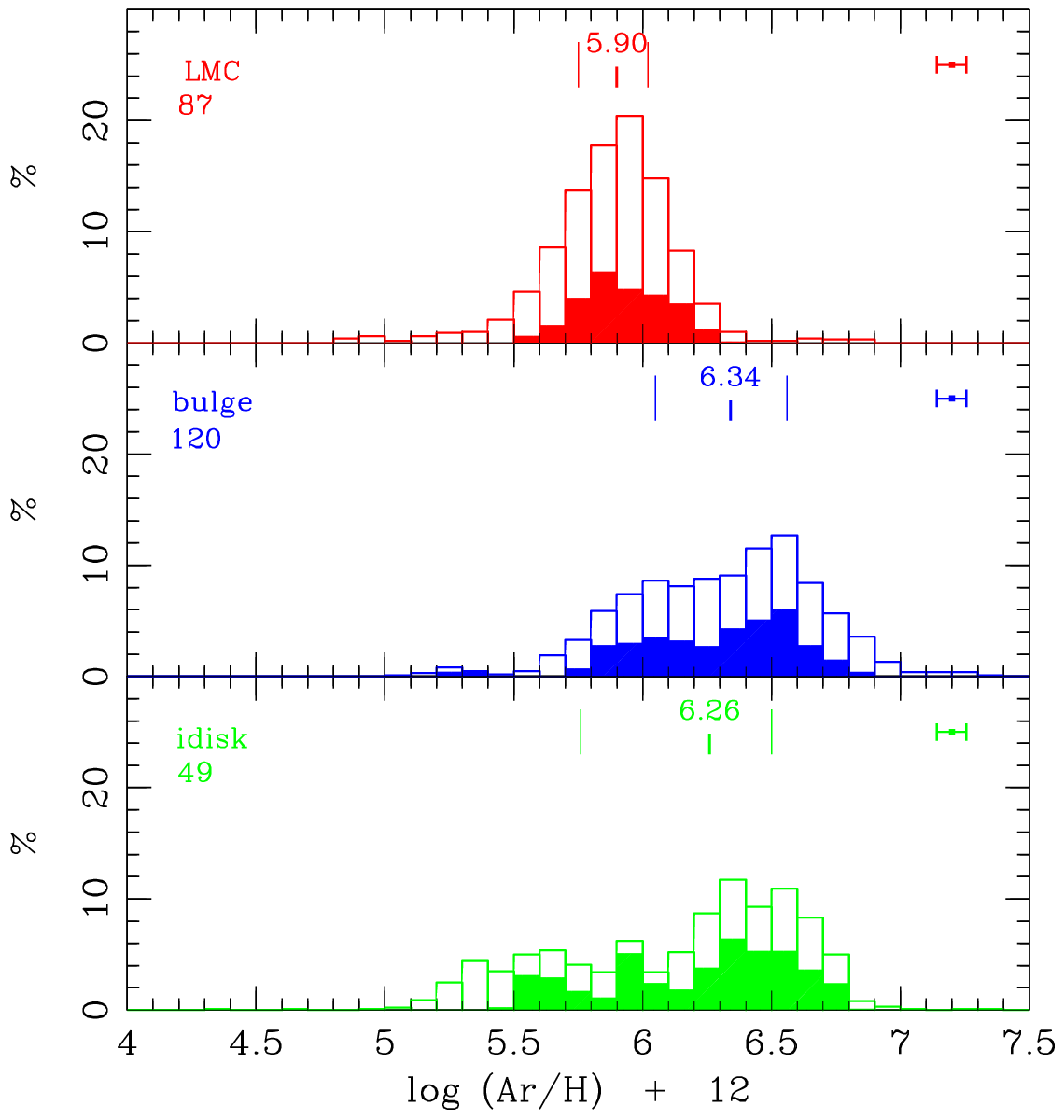}}
\resizebox{0.33\hsize}{!}{\includegraphics{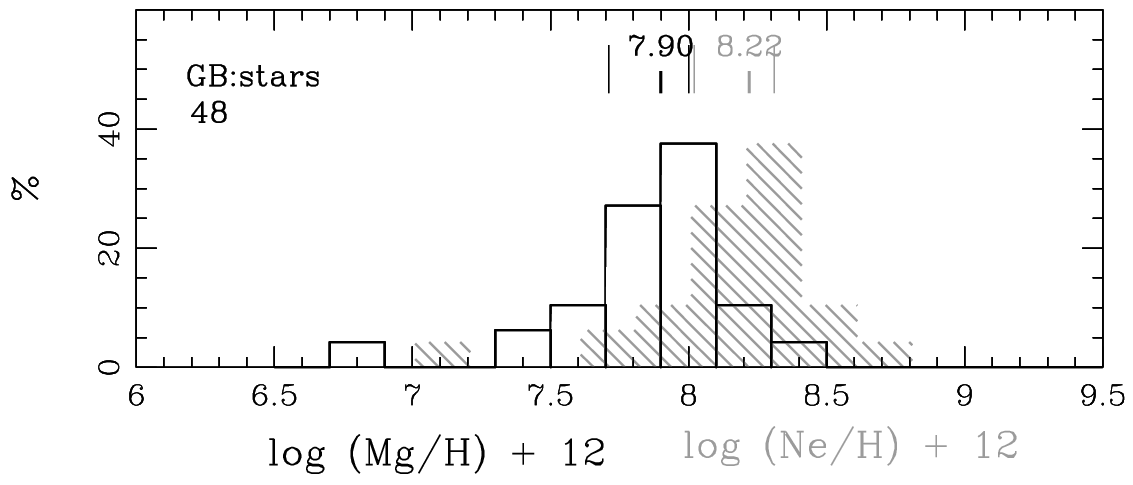}}
\resizebox{0.33\hsize}{!}{\includegraphics{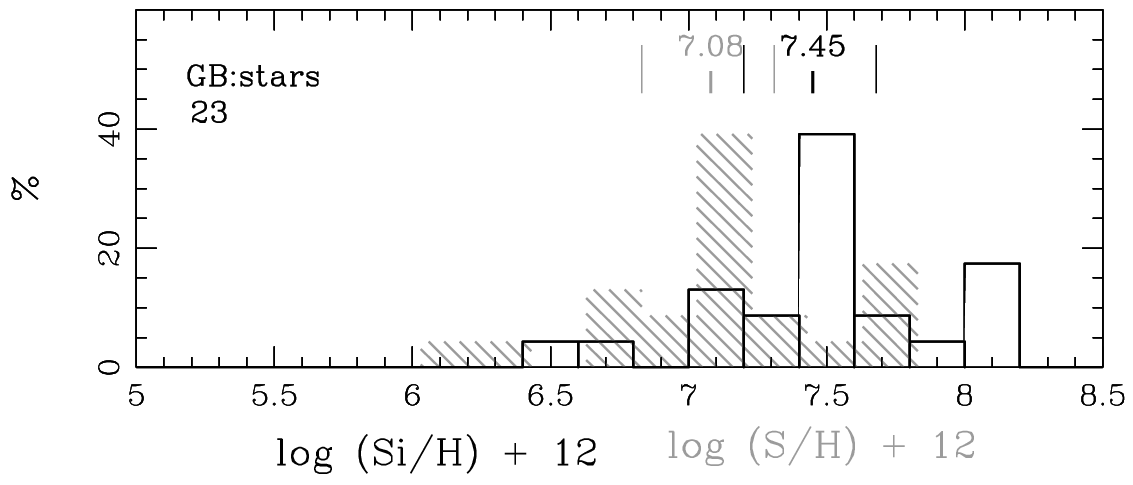}}
\caption[] {
In the first three rows, we present the PN abundance distributions of Ar, Ne, and S (labelled as in Fig.~\ref{Distributions1}). In the last row, we show the abundance distributions of Si and Mg obtained for field-bulge giants. The shaded histograms in this row represent the distributions of Ne/H and S/H obtained from the distributions of Mg/H and Si/H by assuming solar X/O ratios, where X$=$Si, Mg, Ne, and S (see Sect.~\ref{sec:starsindirect}).
}
\label{Distributions2}
\end{figure*}

\begin{figure*}
\resizebox{0.33\hsize}{!}{\includegraphics{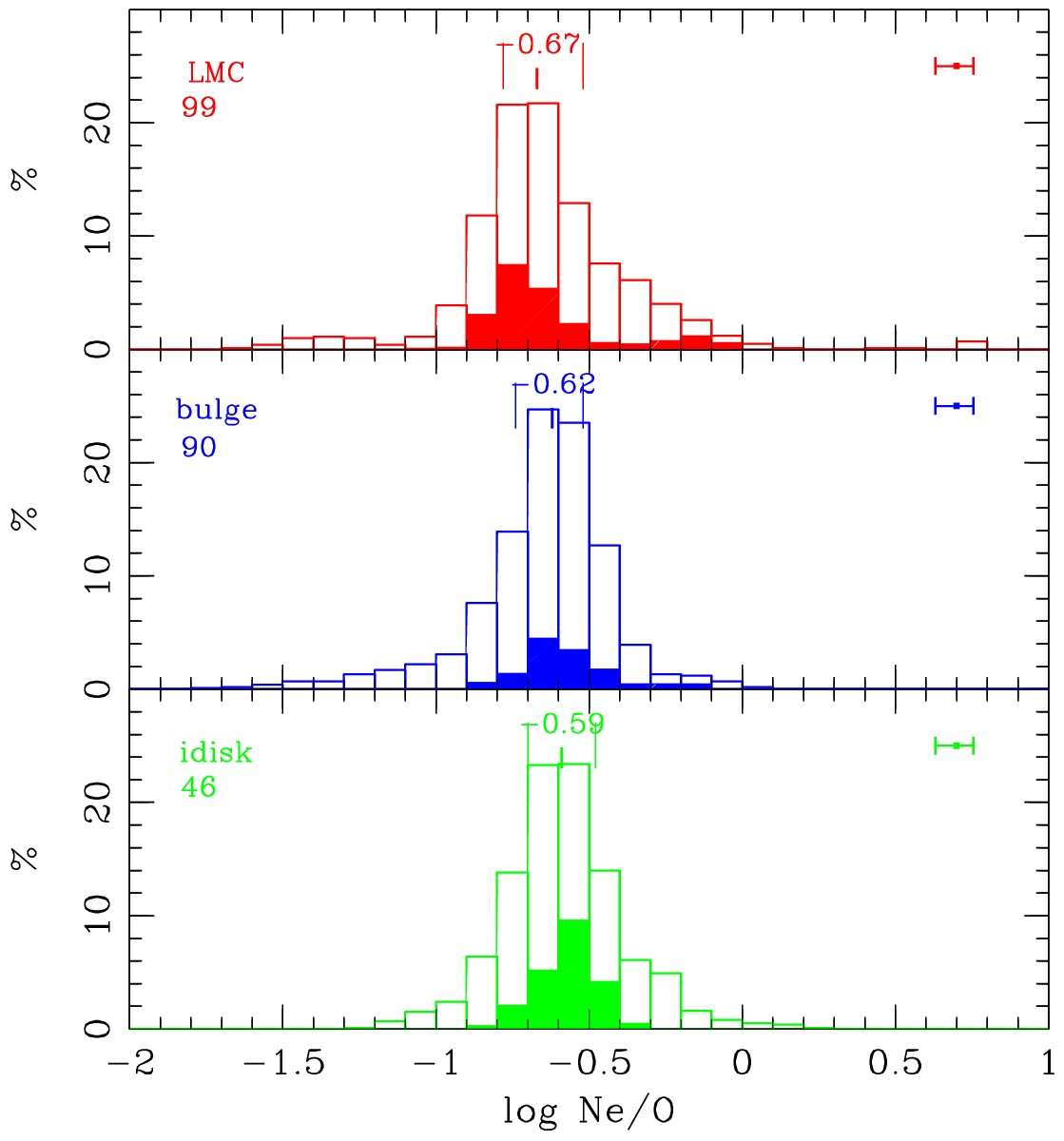}}
\resizebox{0.33\hsize}{!}{\includegraphics{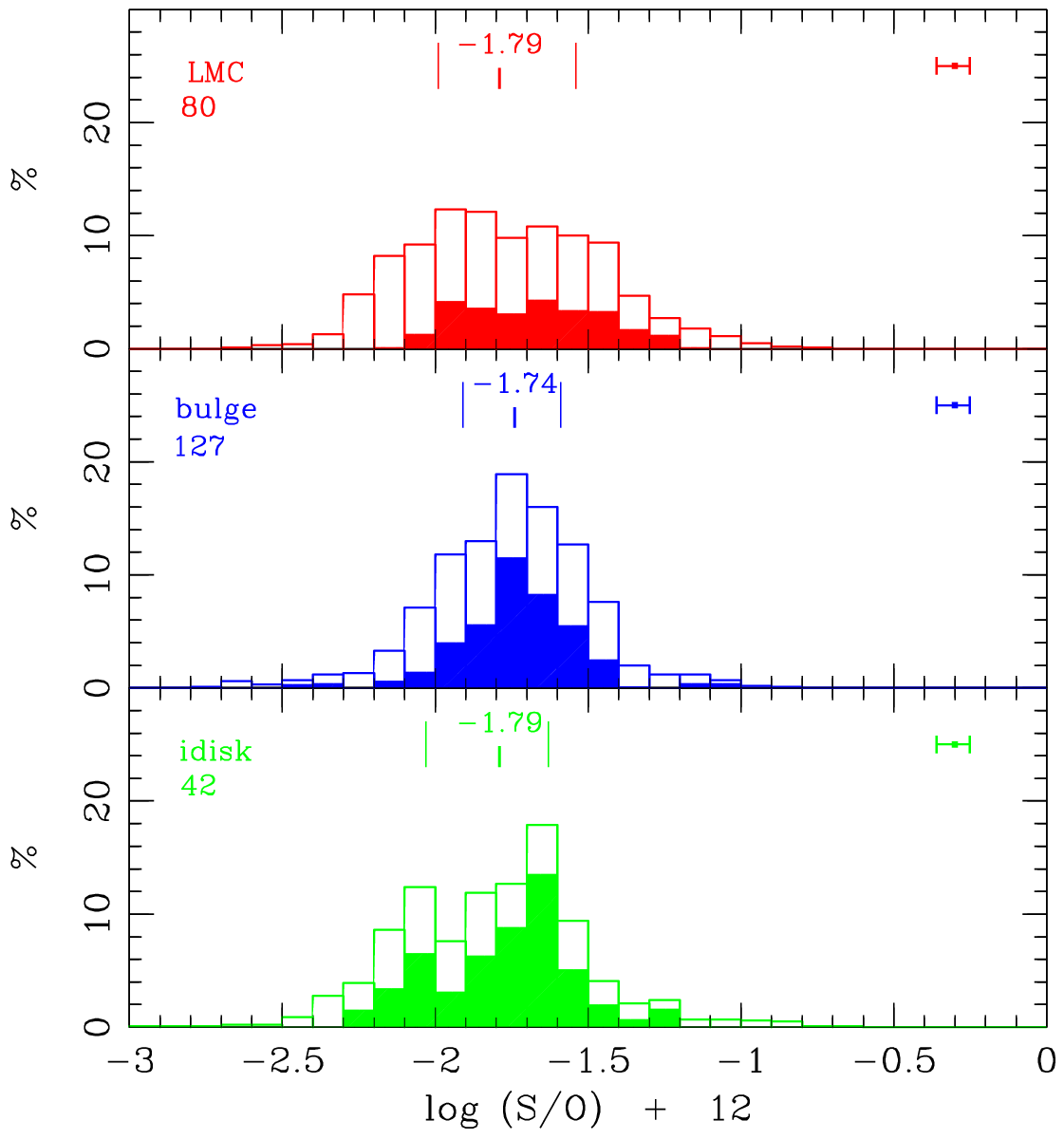}}
\resizebox{0.33\hsize}{!}{\includegraphics{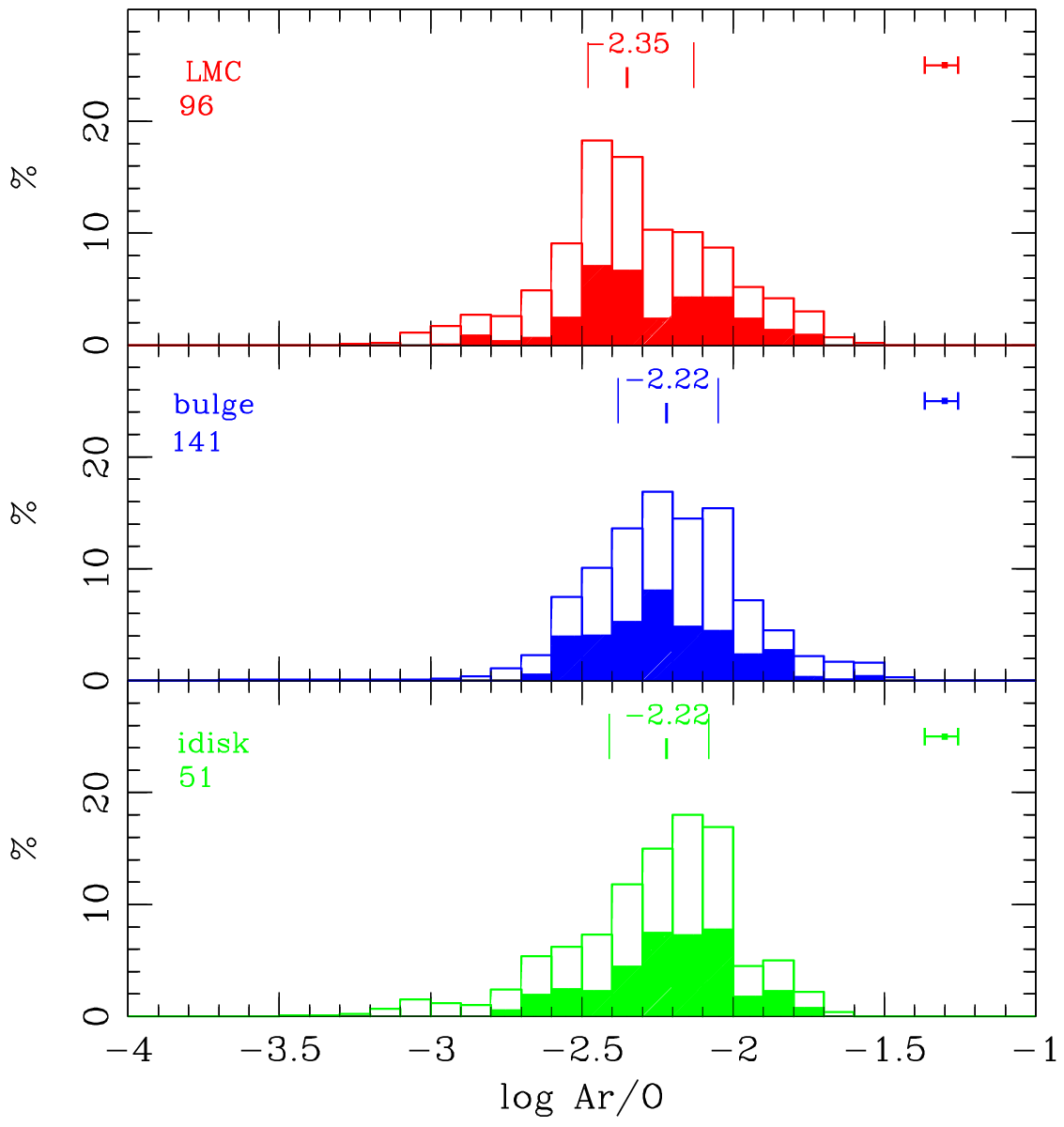}}
\resizebox{0.33\hsize}{!}{\includegraphics{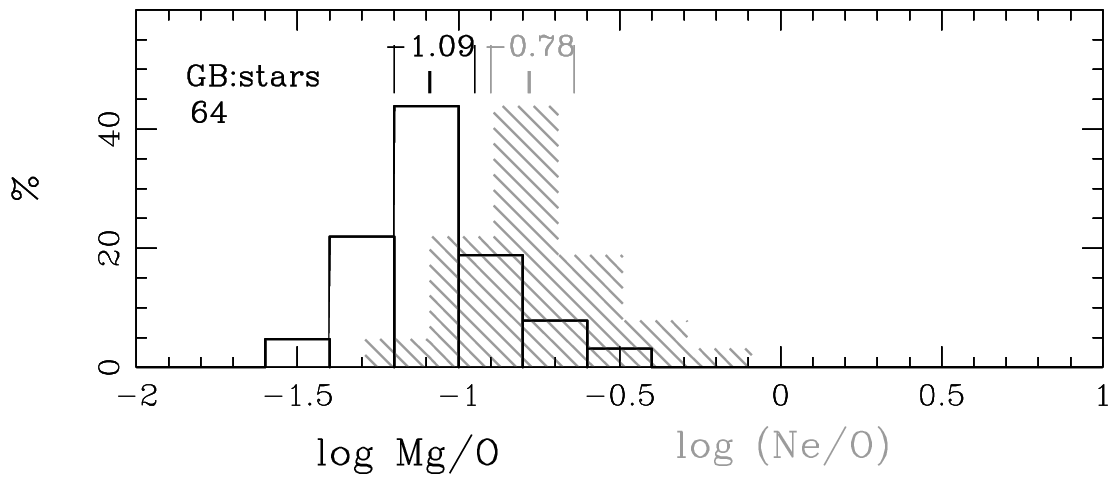}}
\resizebox{0.33\hsize}{!}{\includegraphics{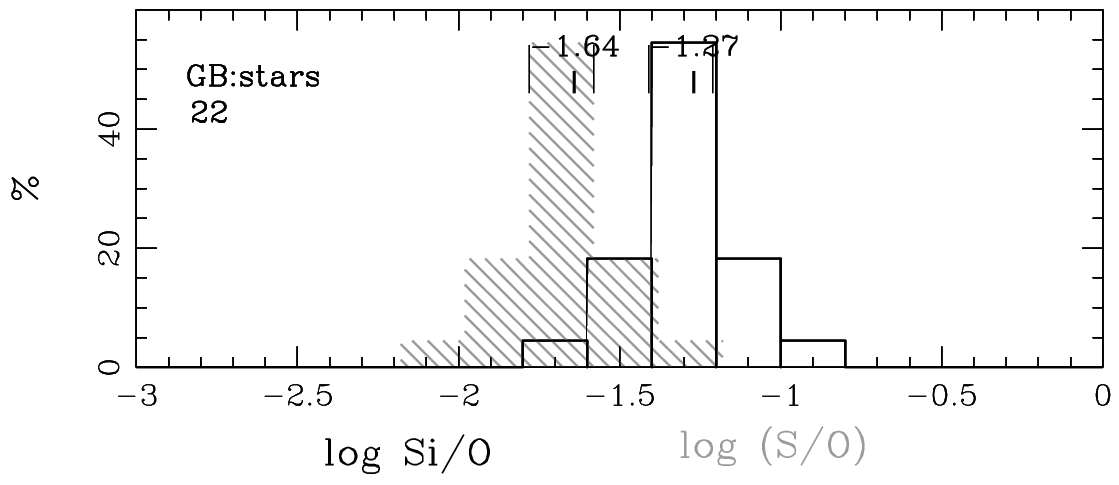}}
\caption[]{
In the first three rows, we present the PN abundance distributions of Ne/O, 
Ar/O, and S/O (labeled as in Fig.~\ref{Distributions1}). In the last row, we show the abundance distributions of Si/O and Mg/O obtained for field bulge giants. The shaded histograms in this row represent the distributions of Ne/O and S/O obtained from the distributions of Mg/O and Si/O by assuming solar X/O ratios, where X$=$Si, Mg, Ne, and S (see Sect.~\ref{sec:starsindirect}).
}
\label{Distributions3}
\end{figure*}

\begin{figure*}
\center
\resizebox{0.3\hsize}{!}{\includegraphics{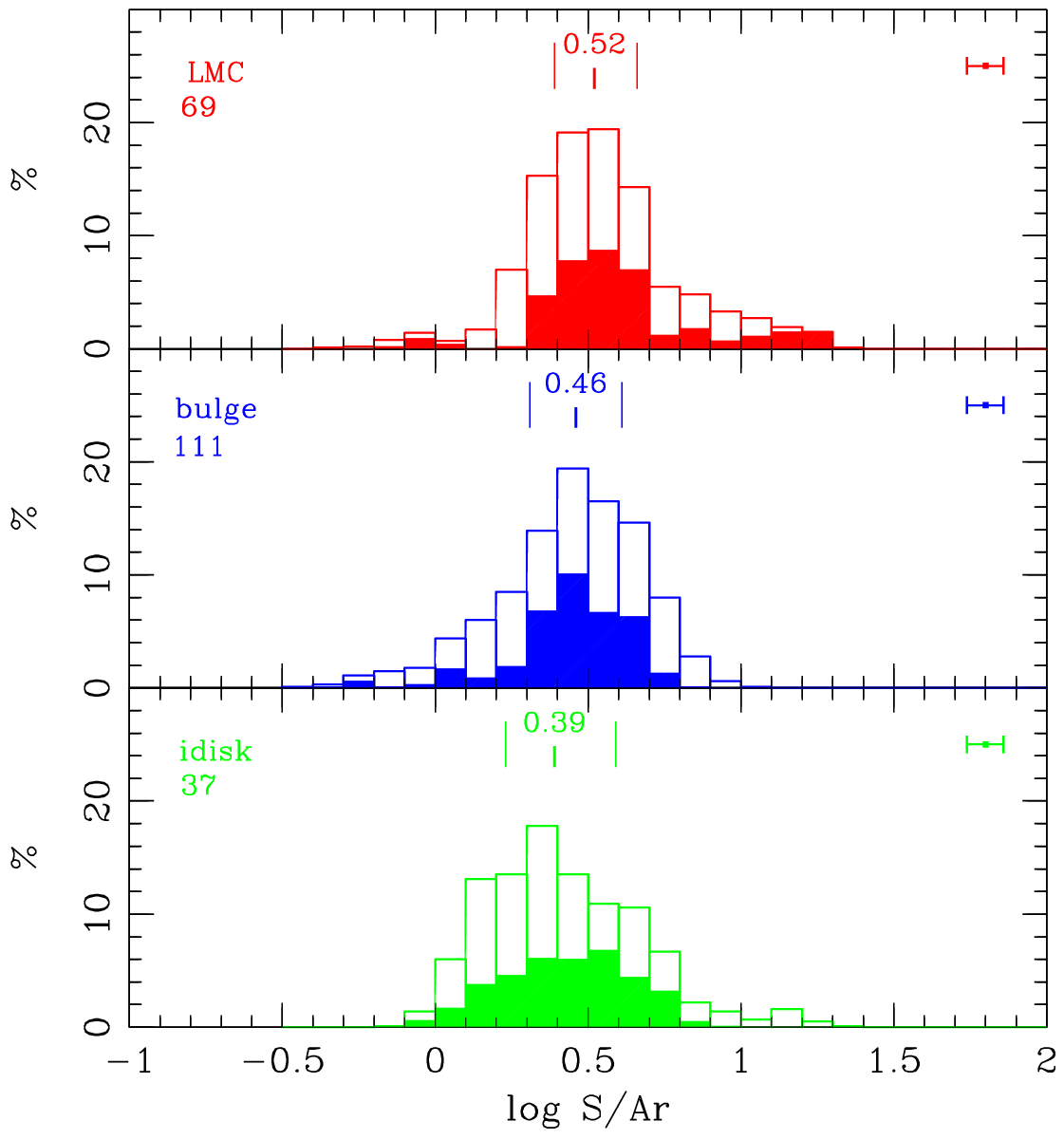}}
\caption[]
   {The S/Ar distribution for PNe in the LMC, bulge, and
   inner-disk (labeled as in Fig.~\ref{Distributions1}).}
\label{SARD}
\end{figure*}

Also shown are the PNe abundance distributions for:
$\log\epsilon$(N), $\log\epsilon$(O), and log(N/O)
(in Fig.~\ref{Distributions1}); $\log\epsilon$(Ne), $\log\epsilon$(S), and $\log\epsilon$(Ar)
(in Fig.~\ref{Distributions2}); log(Ne/O), log(S/O), and log(Ar/O)
(in Fig.~\ref{Distributions3}), and log(Ar/S) (Fig.~\ref{SARD}).
To alleviate the problem of small number statistics and, at
the same time, take into account the errors in the abundance ratios,
we constructed histograms not only for the nominal values of the
abundance ratios, but also the entire set of our Monte Carlo realizations for each object
\footnote{The last rows of Figs.~\ref{Distributions1}, \ref{Distributions2}, and \ref{Distributions3}
show the abundance distributions obtained for bulge field giants, which are discussed in Sects.~\ref{sec:Stars} and \ref{sec:StarsPNeCE}.}.

\begin{table*}
\caption{Median abundance values (number of objects)}\label{table_PN}
\tiny{                        
\begin{tabular}{l @{\hspace{0.10cm}}
                  l @{\hspace{0.15cm}}
                  l @{\hspace{0.5cm}}
                  l @{\hspace{0.15cm}}
                  l @{\hspace{0.5cm}}
                  l @{\hspace{0.15cm}}
                  l @{\hspace{0.5cm}}
                  l @{\hspace{0.15cm}}
                  l @{\hspace{0.5cm}}                 
                  l @{\hspace{0.15cm}}
                  c @{\hspace{0.15cm}} }
\hline\hline
               & Bulge PNe           &       & Inner-disk PNe &
& LMC PNe            &        & Bulge giants          &     &
Sun$^\text{a}$ & HII regions$^\text{b}$ \\
\hline
      $\log\epsilon$(He)     & 11.11 ~[11.05, 11.19] & (144) & 11.08 ~[11.02, 11.14] & (56)
& 11.00 [10.92, 11.07] & (99)  &  --            &            & 10.93
$\pm$ 0.01 & --\\
      $\log\epsilon$(O)      &  ~~8.57 ~~[8.40,   8.72] & (117) &  ~~8.51 ~~[8.28,   8.66] & (44)
&  ~~8.24 ~[8.05,   8.39] & (87)  &  ~8.91 ~[8.82,   9.02] & (42)  & ~8.66
$\pm$ 0.05 & -- \\
      $\log\epsilon$(Ar)     &  ~~6.34 ~~[6.05,   6.56]  & (120) &  ~~6.26 ~~[5.76,   6.50] & (49)
&  ~~5.90 ~[5.75,   6.02] & (87)  &  ~--                  &      & ~6.18
$\pm$ 0.08 & -- \\
      $\log\epsilon$(Ne)     &  ~~7.93 ~~[7.71,   8.17] & (77)  &  ~~7.91 ~~[7.61,   8.09] & (36)
&  ~~7.62 ~[7.42,   7.80] & (85)  &  ~8.22 ~[8.02,   8.31] & (48)  & ~7.84
$\pm$ 0.06 & -- \\
      $\log\epsilon$(S)      &  ~~6.79 ~~[6.54,   7.04] & (94)  &  ~~6.67 ~~[6.45,   6.98]  & (36)
&  ~~6.44 ~[6.30,   6.60] & (63)  &  ~7.08 ~[6.83,   7.31] & (23)  & ~7.14
$\pm$ 0.05 & -- \\
      $\log\epsilon$(N)      &  ~~8.11 ~~[7.76,   8.50] & (123) &  ~~7.99 ~~[7.42,   8.34] & (45)
&  ~~7.75 ~[7.35,   8.15] & (85)  &  ~8.40 ~[8.09,   8.49] & (48)  & ~7.78
$\pm$ 0.06 & -- \\
      $\log\epsilon$(Cl)     &  ~~6.22 ~~[6.00,   6.50] & (47)  &  ~~6.27 ~~[5.97,   6.63] & (19)
&  ~~5.67 ~[--,       --] & (4)   &  ~--                  &      & ~5.50
$\pm$ 0.30 & -- \\
\hline
      log(S/O)      & $-$1.74 [$-$1.91, $-$1.59] & (127) & $-$1.79 [$-$2.03, $-$1.63] & (42)
& $-$1.79 [$-$1.99, $-$1.54] & (80)  & $-$1.64 [$-$1.78, $-$1.58] & (22)  & $-$1.52 &
$-$1.696, $-$1.739\\
      log(Ne/O)     & $-$0.62 [$-$0.74, $-$0.52] & (90)  & $-$0.59 [$-$0.70, $-$0.48] & (46)
& $-$0.67 [$-$0.78, $-$0.52] & (99)  & $-$0.78 [$-$0.90, $-$0.64] & (64)  & $-$0.82 &
$-$0.834, $-$0.688 \\
      log(Ar/O)     & $-$2.22 [$-$2.38, $-$2.05] & (141) & $-$2.22 [$-$2.41, $-$2.08] & (51)
& $-$2.35 [$-$2.48, $-$2.13] & (96)  &  --             &           & $-$2.48 &
$-$2.441, $-$2.376\\
      log(S/Ar)     &  ~~0.46 ~[0.31,  0.61]  & (111) &  ~~0.39 ~[ 0.23,  0.59] & (37)
&  ~~0.52 ~[0.39,  0.66] & (69)  &  --           &             &  ~~0.96 &
0.745, 0.637 \\
      log(N/O)      & $-$0.39 [$-$0.68, $-$0.11] & (136) & $-$0.55 [$-$0.77, $-$0.27] & (50)
& $-$0.51 [-0.89,  ~0.00] & (85)  & $-$0.54 [$-$0.75, $-$0.45] & (42)  & $-$0.88 &
-- \\
      log(N/Ar)     &  ~~1.86 ~[1.60,  2.05] & (138) &  ~~1.81 ~[ 1.51,  1.95]  & (45)
& ~ ~1.86 ~~[1.55,  2.22] & (84)  &  --       &                 &  ~~1.6  &
-- \\
\hline\hline
\end{tabular}\\
$^\text{a}$ Asplund et al. 2005\\
$^\text{b}$ Izotov et al. 2006}
\end{table*}

The filled bar histograms show the location of our highest
quality PNe data, namely those with errors in the abundance ratio smaller than
the median error for the entire PN sample (these objects are represented
by filled symbols in Fig.~\ref{O-FHB}
and in the abundance ratio diagrams discussed in
Sect.~\ref{sec:PNeCE}). This shows that,
in general, accuracy does not depend on the value of the abundance
ratio. In Table~\ref{table_HISTOTESTS}, we show the results of the
Kolmogorov-Smirnov and Wilcoxson tests for pairs of samples, for the
different histograms shown in Figs.~\ref{Distributions1}--\ref{SARD}. 
With both methods, one can test the null hypothesis that the investigated samples originate in identical parent distributions. If, as a result of the test, the probability of the latter hypothesis being 
correct is found to be very low (usually a border value of 1\% is adopted, or at most 5\%), then it can be assumed that
the two distributions or populations under study are truly different. Tests of this kind are crucial for allowing one to discard situations of apparent 
differences, which can result from random effects caused by sample selection and/or uncertainties.

These results are discussed in Sect.~\ref{sec:PNeCE}.

\begin{table}
\caption{Results of Kolmogorov-Smirnov and Wilcoxson statistical tests for abundance
distributions. For each pair of samples, the probabilities of the hypothesis that two given distributions
originate in identical parent distributions is given.}\label{table_HISTOTESTS}
\tiny{
\begin{tabular}{llll}
\hline\hline
Element$^*$      & Bulge/Inner-Disk & Bulge/LMC     & LMC/Inner Disk \\
\hline
$\log\epsilon$(He) &  0.09 / 0.04     &  0.00 / 0.00  &  0.00 / 0.00  \\
$\log\epsilon$(O)  &  0.30 / 0.08     &  0.00 / 0.00  &  0.00 / 0.00  \\
$\log\epsilon$(Ar) &  0.08 / 0.06     &  0.00 / 0.00  &  0.00 / 0.00   \\
$\log\epsilon$(Ne) &  0.17 / 0.36     &  0.00 / 0.00  &  0.00 / 0.00  \\
$\log\epsilon$(S) &  0.30 / 0.12     &  0.00 / 0.00  &  0.00 / 0.00  \\
$\log\epsilon$(N) &  0.09 / 0.03     &  0.00 / 0.00  &  0.11 / 0.13  \\
$\log\epsilon$(Cl) &  0.95 / 0.60     &   --  /  --   &  --   /  --   \\
\hline
log(S/O)     &  0.21 / 0.17     &  0.35 / 0.72  &  0.59 / 0.38  \\
log(Ne/O)    &  0.31 / 0.11     &  0.27 / 0.46  &  0.05 / 0.06  \\
log(Ar/O)    &  0.67 / 0.41     &  0.00 / 0.00  &  0.01 / 0.12  \\
log(S/Ar)    &  0.25 / 0.24     &  0.17 / 0.04  &  0.01 / 0.01  \\
log(N/O)     &  0.05 / 0.01     &  0.09 / 0.41  &  0.10 / 0.28  \\
log(N/Ar)    &  0.21 / 0.13     &  0.02 / 0.42  &  0.00 / 0.08  \\
\hline\hline
\end{tabular}\\
$^*$ $\log\epsilon$(X)~=~$\log$(X/H) + 12}
\end{table}

\section{Stars: samples and abundances}\label{sec:Stars}

Comparing the abundances of bulge planetary nebulae with those of a sample of
bulge stars is not a straightforward task. Abundance measurements for bulge stars
require the use of large telescopes and are only feasible for
giants\footnote{
There have been a few abundance measurements for extremely metal-rich
bulge G dwarfs. This was possible due to the magnification
from microlensing (see Cohen et al. 2008).
Abundances for dwarf bulge stars obtained from
microlensing are promising, however complete samples will not be
feasible before the 30m class telescopes are available.}.
As a consequence the bulge giant samples, for which several chemical
elements are measured, are still small compared to our present bulge PN
sample\footnote{Although this situation is changing quickly -- e.g.
Zoccali et al. (2008) obtained [Fe/H] for $\sim$800 bulge field stars,
and the analysis of some elements measurable at the resolution of
R$\sim$22,000 of the Flames instrument are under way for this large
sample -- at the present moment, abundances are  available only
for a smaller number of objects.}. Hence, for elements other
than iron, the sample selection effects can play
an important role in shaping the
abundance distributions. In addition, whereas
for PNe, the measurements can cover the entire bulge area, for stars the
measurements are made in the low extinction lines of sight (such as the
Baade's Window) and/or require infrared spectra (e.g. Rich and Origlia
2005). Aware of these caveats (see discussion in
Sect.~\ref{sec:Starsdistributions}), we present
here the adopted samples of bulge giants to be compared
with our sample of bulge PNe in Sect.~\ref{sec:StarsPNeCE}. We also
briefly summarize the main steps and uncertainties involved in the abundance
determinations of bulge giants.

\subsection{The adopted samples}\label{sec:Starssample}

In the literature, there are few field\footnote{
   We do not include the abundance measurements
   in bulge stars inside globular clusters, since in this case the abundances
   of oxygen could have been affected by the so-called {\it Na-O anti-correlation}
   typical of globular clusters (e.g. Gratton et al. 2004). In the sample
   of Fulbright et al. (2007) for field stars, two objects showing low
   [O/Fe] ratios for low metallicities are also probably ex-members of
   globular clusters and were not included in the present discussion.}
giant stars for which high quality abundance measurements are available.
Moreover, only oxygen (and in a few cases C and N)
are measured in both bulge PNe and bulge giants, making direct comparison difficult.

For the present work, we chose the Fulbright et
al. (2007, hereinafter F07) and the
Lecureur et al. (2007, hereinafter L07) samples
as our two main reference samples for
bulge giants.
These two samples constitute presently the largest samples of bulge field
giants with abundance measurements for several chemical elements.
These data come from optical
high-resolution spectra taken with 8m-class telescopes.

F07 reported O, Na, Mg, Al, Si, Ca, and Ti abundances for
27 red giant branch stars observed towards the Baade's Window. These
abundances were inferred from high resolution spectra obtained with the Keck
I telescope. Of particular interest to our discussion are the O, Mg, and Si
measurements.

L07 reported abundances (spectra obtained on the
VLT) of O, Na, Mg, and Al for 13 core
He-burning giant stars and 40 red giant branch stars in four 25' fields
spanning the bulge from $-$3$^{\circ}$ to $-$12$^{\circ}$
along the bulge minor axis.
Their stars were selected from bulge colour-magnitude
diagrams, and the expected
number of foreground contaminants in each field was around 15\% in all
fields, with the exception of one at b$=-$12$^{\circ}$ for which this
contamination can reach 45\%. We chose not to use their b$=-$12$^{\circ}$
field (with 5 stars) to avoid significant contamination by disk stars.
L07 also obtained C and N abundances.

We also include the abundance measurements from three other 
smaller samples, namely: Cunha \& Smith (2006),  Rich \& Origlia (2005),
and Mel\'endez et al. (2008).
These abundances were obtained from infrared spectra. Cunha \&
Smith (2006) measured the abundances in 7 K and M red giant members of the
bulge (their target stars are a subset of the K giant sample of F07,
plus two additional M giants, one in common with Rich \& Origlia
2005) from high-resolution infrared spectra obtained with the Phoenix
spectrograph on Gemini South. The elements studied in this case were C, N,
O, Na, Ti, and Fe. For objects in common, the oxygen abundances obtained by
Cunha \& Smith (2006) in the infrared agree with those obtained
by F07 in the optical. Of interest to our discussion
are their abundances for O and N.  Rich \& Origlia (2005) reported
abundances of O, Fe, Si, Mg, Ca, Ti, C, and $^{12}$C/$^{13}$C ratio,
for 14 bulge M giant stars obtained by using the
NIRSPEC spectrograph at the Keck
Telescope. This sample spans a narrow metallicity range (from $-$0.33
$\leq$[Fe/H] $\leq -$0.03). We use their O and Si abundances in
the abundance ratio plots, but not in the histograms (see
Sect.~\ref{sec:Starsdistributions}).
Mel\'endez et al. (2008) derived O, N, and Fe for 19 Baade's
Window giants. The Mel\'endez et al. (2008) stars were taken from the
F07 sample.

The final adopted stellar sample is described in
Sect.~\ref{sec:Starsdistributions}.
Figure~\ref{lb} shows the location of the adopted samples in
a (l, b) diagram. We also report the number of stars studied in each field.

\subsection{Abundance determinations, uncertainties, and biases}
\label{sec:Starsbiases}

The formation of absorption lines is computed by sophisticated
model atmospheres that take into account  the radiative transfer from deeper
to shallower layers (optical depths
 2 $\ge$ log $\tau$ $\ge$ $-$5). Available grids of model atmospheres
assume conditions of Local Thermodynamic
Equilibrium (LTE).
The detailed abundance analysis of cool stars consists
of using a set of FeI and FeII lines to derive
four basic parameters: effective temperature $T_{\rm eff}$,
gravity log $g$, metallicity [Fe/H], and microturbulence velocity
$v_{\rm t}$.
Since these four stellar parameters are interdependent, the analysis
has to consider simultaneously excitation equilibrium, ionization
equilibrium, and an optimal $v_{\rm t}$ value that satisfies a range
of equivalent widths. When the stellar parameters have been decided, the
abundances of different elements are derived line-by-line, either
using their equivalent widths or by fitting line profiles.\\

\noindent
{\it Effective Temperatures}

\noindent
In the optical, the effective temperature is preferentially
derived from the excitation equilibrium of FeI and FeII lines
of different excitation potential.
A change of 100~K in $T_{\rm eff}$  causes a
recognizable trend in the plane FeI abundance versus excitation potential and we can therefore
consider that the uncertainty in $T_{\rm eff}$ is of that order.

In the near-infrared (H, K) bands region, the FeI lines have high
excitation potentials (around 5.5 -- 6.5 eV), with little variation, and
thus have little sensitivity to excitation temperature equilibrium.
Consequently, for bulge stars studied from H and K-band spectra,
temperatures must rely on photometric colours, and given
the rather high and variable reddening in bulge regions,
it is challenging to obtain intrinsic colours with the
accuracy needed for effective temperature derivation
(e.g. Cunha \& Smith 2006).\\

\noindent
{\it Gravity}

\noindent
The surface gravities $\log g$ are in general
derived using the effective temperature (T$_{\rm eff}$) and in some
cases also the parallax as input (in which case
the bolometric magnitude -- M$_{\rm bol}^*$ -- can be derived),
with the classical relation (where M$_*$ is the stellar mass, in solar
masses):

$$log~{g_{*}\over g_{\odot}} = 4~log~{{T_{\rm eff}}\over{T_{\odot}}} +
0.4~(M_{\rm bol}^{*} - M_{\rm bol}^{\odot}) + log~{{M_*}\over{M_{\odot}}} $$

The spectroscopic gravity $\log g$ derived from ionization
equilibrium of FeI and FeII lines shows typical uncertainties
of 0.30~dex.\\

\noindent
{\it Microturbulent velocities}

\noindent
The microturbulent velocities $v_{t}$ are usually determined using FeI lines.
The uncertainty derived from the FeI abundance
versus the equivalent width W$_{\lambda}$
is in general around 0.2~km~s$^{-1}$.\\

\noindent
{\it Atmospheric parameters: impact on oxygen abundance}

\noindent
Assuming typical parameters for a bulge giant
[$T_{\rm eff}$,  $\log g$, [Fe/H], $v_{t}$,] of 4500~K, 2.0, 0.0, and
1.5~km~s$^{-1}$, we estimate the uncertainties in the derivation of
oxygen abundances due to the choice of stellar parameters, by showing in
   Table~\ref{error} the sensitivity of the abundances to variations
 in the temperature, gravity, and microturbulent velocity of
$\Delta$T$_{\rm eff} = $100~K,
$\Delta$log $g =$ 0.30~dex, and
$\Delta$v$_{\rm t} =$ 0.20~km~s$^{-1}$.
The total error is given in the last column as
the quadratic sum of all uncertainties.
We can see that the total uncertainties
in [O/Fe] are about 0.05~dex. The [O/H] abundances
are affected by additional unknown systematic effects,
whereas for abundance ratios such as [O/Fe], part of these 
uncertainties cancel out.

\begin{table*}
\begin{flushleft}
\caption{Sensitivity of abundances to changes of
$\Delta$T$_{\rm eff}$ = 100 K,
$\Delta$log g = +0.3, and $\Delta$v$_{\rm t}$ = 0.2 km s$^{-1}$. In the last
column, the corresponding total error is given. }
\label{error}

\begin{tabular}{lccccc}
\hline\hline
\noalign{\smallskip}
\hbox{Species}  & \hbox{$\Delta$T} & \hbox{$\Delta$$\log$ g} &
\hbox{$\Delta$v$_{t}$} & \hbox{($\sum$x$^{2}$)$^{1/2}$} \\
\hbox{(1)} & \hbox{(2)}& \hbox{(3)} & \hbox{(4)} & \hbox{(5)} \\
\noalign{\smallskip}
\noalign{\smallskip}
\noalign{\vskip 0.1cm}
\noalign{\hrule\vskip 0.1cm}
\noalign{\vskip 0.1cm}
\multicolumn{5}{c}{Original: T$_{\rm eff}$=4500 K, log g=2.0, [Fe/H]=0.0,
   v$_{\rm t}$=1.5 km.s$^{-1}$}\\
\noalign{\vskip 0.1cm}
\noalign{\hrule\vskip 0.1cm}
\hbox{[Fe/H](I)} & $-$0.02 & +0.04   &  $-$0.09 & +0.10 \\
\hbox{[Fe/H](II)}& $-$0.19 & +0.13   &  $-$0.05 & +0.23 \\
\hbox{[O/Fe]}    & +0.01   & +0.05   &  $-$0.01 & +0.05 \\
\hline\hline
\end{tabular}
\end{flushleft}
\end{table*}

Finally, in general, ionized lines are combined with FeII,
and neutral species with FeI. Some ratios are more reliable
when compared with FeII,
such as oxygen derived from [OI]~630~nm, since these lines form in
similar layers. In terms of the effects of non-LTE, the use of FeII is more reliable, since FeII lines are far less sensitive to NLTE than FeI lines.\\

\noindent
{\it Other uncertainties}

\noindent
The adopted oscillator strengths and oxygen abundances in the reference
star may systematically alter the oxygen abundance.
The important issue is the combination of oxygen abundance in the
reference star, and the $\log$~gf value.
For example, Bensby et al. (2004) use for the [OI]630~nm line
a $\log$~gf value of $-$9.819 and a solar oxygen abundance of
$\log\epsilon$(O)=8.69, and the same values were adopted by Zoccali
et al. (2006) and L07, for comparison purpose.
The more usual value of $\log$~gf$= -$9.717 (e.g. Allende Prieto et al. 2001) leads to oxygen abundances that are higher by 0.1~dex.

Concerning the oxygen abundance used in the reference star,
for the Sun, values of $\log\epsilon$(O)~=~8.77 recommended
by Allende Prieto et al. (2001) for 1D model atmosphere calculations
down to $\log\epsilon$(O)~=~8.66 (Asplund et al. 2004) are presently being used.
For Arcturus, when used as the reference star for fitting oscillator
strengths, the stellar parameters and C,N,O abundances also can vary among
different authors.

In the derivation of the bulge giant abundances, Arcturus has been adopted
as the reference star in all cases, except for L07
who adopted $\mu$ Leo as reference star, but they give the parameters
for Arcturus based on those deduced from fitting $\mu$ Leo lines.
The log gf values of FeI and FeII lines (which are the lines
used to derive effective temperature, gravity, metallicity, and
microturbulence velocity) are fitted to Arcturus, and then used to
derive stellar parameters of the sample bulge giants.
For example, [Fe/H]~$=~-$0.50 or $-$0.60
for Arcturus can produce a 0.1~dex difference in log~gf values
of FeI and FeII, which can affect the final abundances by this much.

In Table~\ref{tabteff}, we show the basic parameters and
resulting C,N, and O abundances obtained or adopted by different authors
for Arcturus and the Sun. It is clear that, while for C and N, the abundances are
similar to within 0.15~dex, for oxygen differences of up to 0.27~dex
are seen. This is caused by differences in solar oxygen abundance,
oscillator strengths, and oxygen lines used, as well as small differences
in  stellar parameters for Arcturus.

Another source of error is the use of 1-D model atmospheres, since
the strengths of lines depend on the detailed structure of the atmospheres.
Stellar granulation, including the effects of temperature gradients, atmospheric
inhomogeneities, and velocity fields caused by convection, affect the ratio of
line to continuous absorption (Asplund 2005).
These effects were taken into account by the use of 3-D
time-dependent, hydrodynamical, model atmospheres. Such calculations when 
applied to [OI]630nm and FeII lines by Nissen et al. (2002), demonstrated 
that for metal-poor stars, a correction of 0.2~dex was required. 
On the other hand,
essentially no difference was found for metal-rich stars. Another point
is that 3-D effects compensate for NLTE effects, so that both improvements
must be taken into account at the same time. However, in the case of
[OI]~630nm, the NLTE effects are negligible and the predictions by Nissen et al. (2002) using 3D-models should apply.
For infrared OH lines, NLTE and 3-D corrections
could have some effect, but detailed
calculations are unavailable.

\begin{table*}
\caption[1]{Comparison of results from different recent studies for Arcturus (employed as a reference star in bulge studies) and the Sun. This Table is organized as follows: stellar parameters employed for Arcturus (columns 1--4); solar abundances (columns 5--8); Arcturus abundances (columns 9--12); References$^\text{a}$ (column 13). Note: $\log\epsilon$(X) = log(X/H)+12.}
\begin{flushleft}
\tiny{
\begin{tabular}{llllllllllllll}
\noalign{\smallskip}
\hline\hline
\noalign{\smallskip}
T$_{\rm eff}$&\phantom{-} log g  & \phantom{-}v$_{\rm t}$ & [Fe/H]
& \phantom{-}$\log\epsilon$(O)$_{\odot}$ & \phantom{-}$\log\epsilon$(C)$_{\odot}$ &
 \phantom{-}$\log\epsilon$(N)$_{\odot}$ & \phantom{-}$\log\epsilon$(Fe)$_{\odot}$
& \phantom{-}$\log\epsilon$(O) &
\phantom{-}$\log\epsilon$(C) &
\phantom{-}$\log\epsilon$(N)  &
\phantom{-}$\log\epsilon$(Fe)
 &  Reference & \\
(K) &    & km.s$^{-1}$ &  & &  &  & &  & &  &  &  &  \\
\noalign{\smallskip}
\noalign{\hrule\vskip 0.2cm}
4275 & \phantom{-}1.55 &  1.65 &\phantom{-}-0.54
&8.87 &8.55 &7.97 &7.50 & 8.76 & 7.93 & 7.73
&6.96 & 1 & \\
4300 & \phantom{-}1.7  & 1.6 & \phantom{-}-0.60
&8.77 &8.59 &8.00 &7.50 & 8.49 & 7.92 & 7.60
&6.78 & 2 & \\
4250 & \phantom{-}1.5  & 1.5 & \phantom{-}-0.60
&8.83 &8.52 & 7.92  &7.50 &8.58 & 7.72 &--- &6.90
& 3&  \\
4290 & \phantom{-}1.55 & 1.67 & \phantom{-}-0.50
&8.69 &--- &---   &7.45 & 8.67 & --- & --- &6.95
& 4 & \\
4300 &  \phantom{-}1.5 & 1.5 & \phantom{-}-0.52
&8.77 &8.46 &7.94 &7.50 & 8.70 & 7.96 & 7.74&6.98
& 5& \\
\noalign{\smallskip} \hline\hline \end{tabular}}
\end{flushleft}
\label{tabteff}
$^\text{a}$ 1 - Mel\'endez et al. 2008; 2 - Cunha \& Smith 2006 (based on
Smith et al. 2000); 3 - Rich \& Origlia 2005; 4 - Fulbright et al. 2006, 2007;
5 - Lecureur et al. 2007 (based on Smith et al.
2002).\\
\end{table*}

\subsection{Abundance Distributions}
\label{sec:Starsdistributions}

\begin{figure*}
\resizebox{0.49\hsize}{!}{\includegraphics{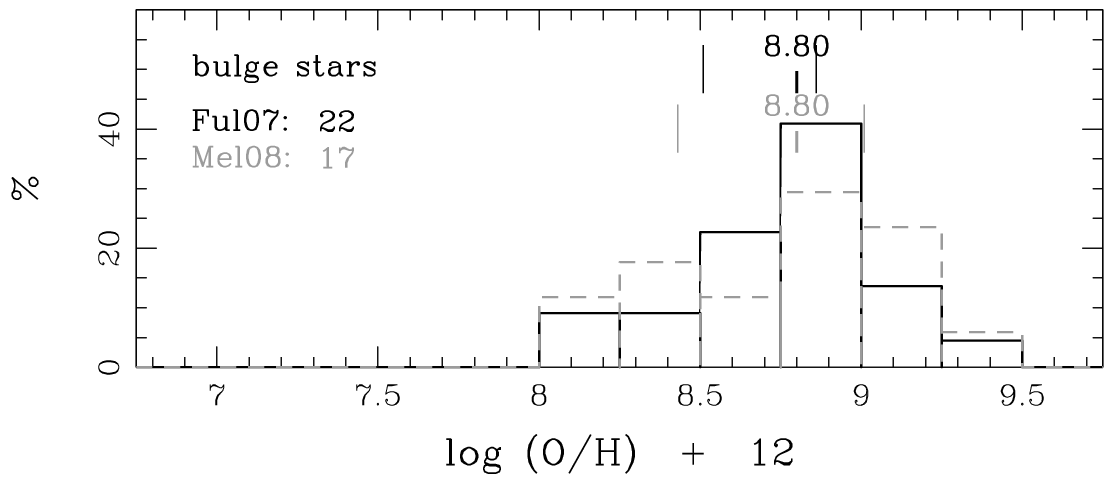}}
\resizebox{0.49\hsize}{!}{\includegraphics{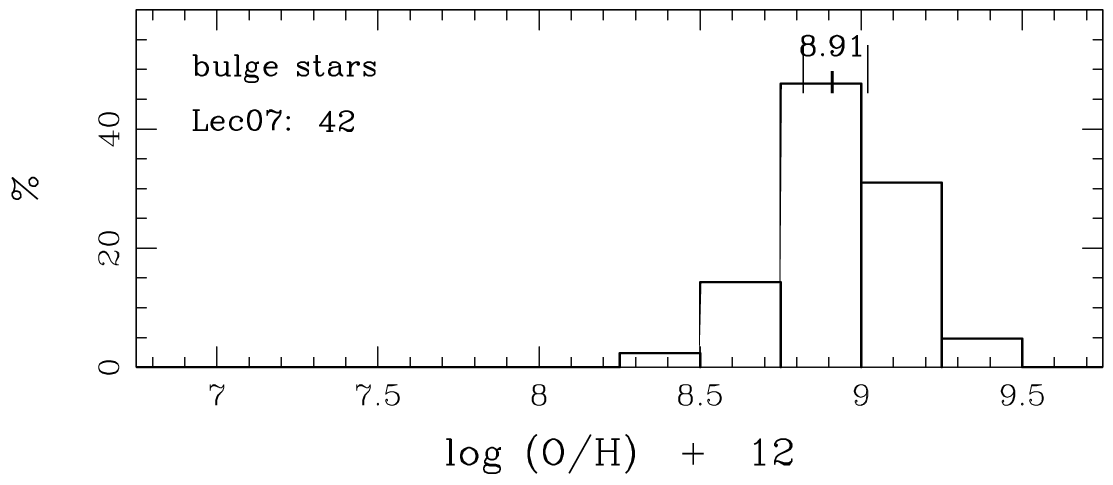}}
\caption[]{
   The distribution of O/H for the F07 (left panel, solid histogram), Mel\'endez et al. (2008) (left panel, dashed histogram), and L07 (right pannel) samples of bulge
   giant stars. The solar values from Asplund et al. (2005) is $\log\epsilon$(O)~$=$~8.66~$\pm$~0.05. The final adopted oxygen distribution
   (see text) can be seen in Fig.~\ref{Distributions1}.
}
\label{Distribution_stars0}
\end{figure*}

Figure~\ref{Distribution_stars0} shows the oxygen distributions for the
L07, F07, and Mel\'endez et
al. (2008) samples (for the other two samples, the number of objects was
too small and we do not show their oxygen histograms). We note the
generally good agreement between the F07 (optical) and Mel\'endez et
al. (2008 -- infrared) oxygen distributions.

The median oxygen abundance of each of the L07 and F07 samples
differ by 0.11 dex, the Lecureur sample
being narrower and shifted towards larger oxygen abundance values.
This difference reflects, in part,
the fact that these samples were selected according to different criteria.
This means that the final abundance distributions may not 
reflect the true bulge distributions. This is certainly the case for
the F07 sample (and of course will also be the case for the 
Mel\'endez et al. 2008 sample). The latter authors selected
a number of stars covering a wide range of metallicities to 
study the [O/Fe] versus [Fe/H] diagram and not the metallicity
distribution. In addition, the oxygen distributions of F07 and L07
can differ due to the combined effect of
basic parameters such as oxygen abundance in the Sun and Arcturus and
oscillator strength differences, as previously
discussed (see Sect.~\ref{sec:Starsbiases}).

The L07 sample is derived from the larger sample of
Zoccali et al. (2008). To avoid strong biases in the
resulting iron-metallicity distribution, Zoccali et al. (2008)
included targets spanning the entire colour range in the red giant
branch at a given magnitude. The raw distribution was corrected
for a possible remaining metallicity bias, but the resulting
distribution turned out to be indistinguishable from the original
one, which implied that their selection criteria was robust (see Zoccali
et al. (2008) for details). For a subset of data from Zoccali et al. (2008),
UVES spectra were obtained, and these data constitute the L07
sample. A comparison between the iron metallicity distribution
calculated with the L07 sample with the more
complete one of Zoccali et al. (2008) confirms that the former still provides 
a good estimate of the bulge metallicity distribution.

For reasons already discussed, the final oxygen abundance
distribution (shown in Fig.~\ref{Distributions1}) was obtained 
using only the L07 sample, and will be compared with
one obtained from bulge PNe in Sect.~\ref{sec:PNeCE}.

\begin{figure*}
\resizebox{0.49\hsize}{!}{\includegraphics{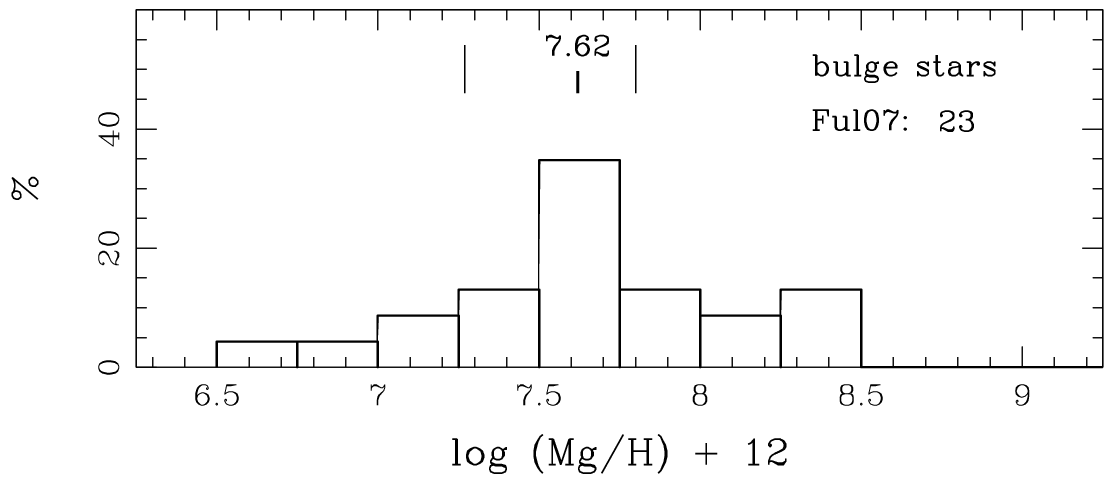}}
\resizebox{0.49\hsize}{!}{\includegraphics{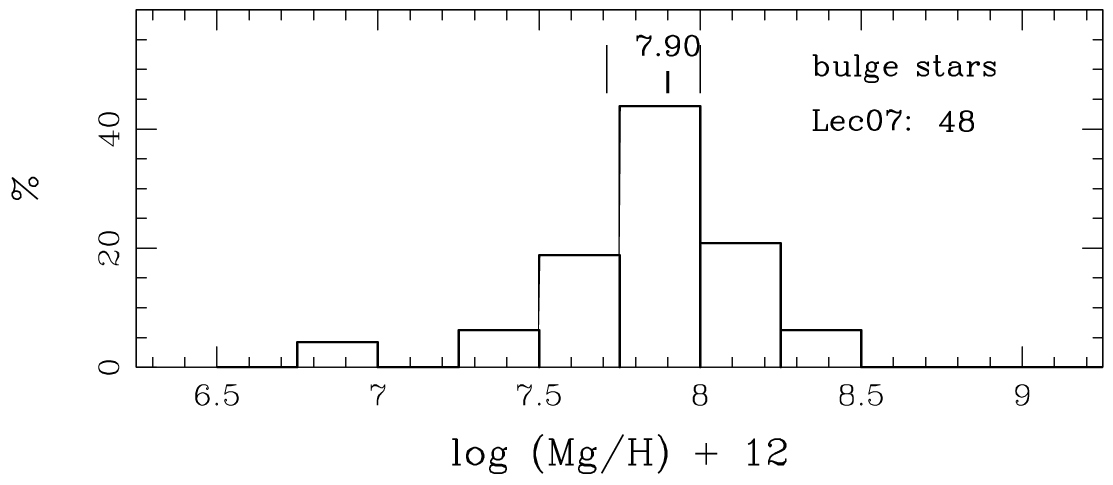}}
\caption[]{The distribution of Mg/H for the L07 (left panel) and F07 (right panel)
  samples of bulge giant stars. The solar values from Asplund et
   al. (2005) is $\log\epsilon$(Mg)~$=$~7.53~$\pm$~0.05. The adopted final magnesium
   distribution (see text) can be seen in Fig.~\ref{Distributions2}.
}
\label{Distribution_starsMg}
\end{figure*}

Figure~\ref{Distribution_starsMg} shows that the Mg distribution obtained
with the L07 sample is narrower than that of F07
and is also shifted towards larger Mg values by almost 0.3~dex. This
again reflects the different sample-selection criteria, as
discussed before. However, in this case, the discrepancy is larger than
for oxygen showing that other systematic effects can also play a role.
In this case, we again use only L07 data in our final Mg/H
distribution (see Fig.~\ref{Distributions2}).

\begin{figure*}
\resizebox{0.49\hsize}{!}{\includegraphics{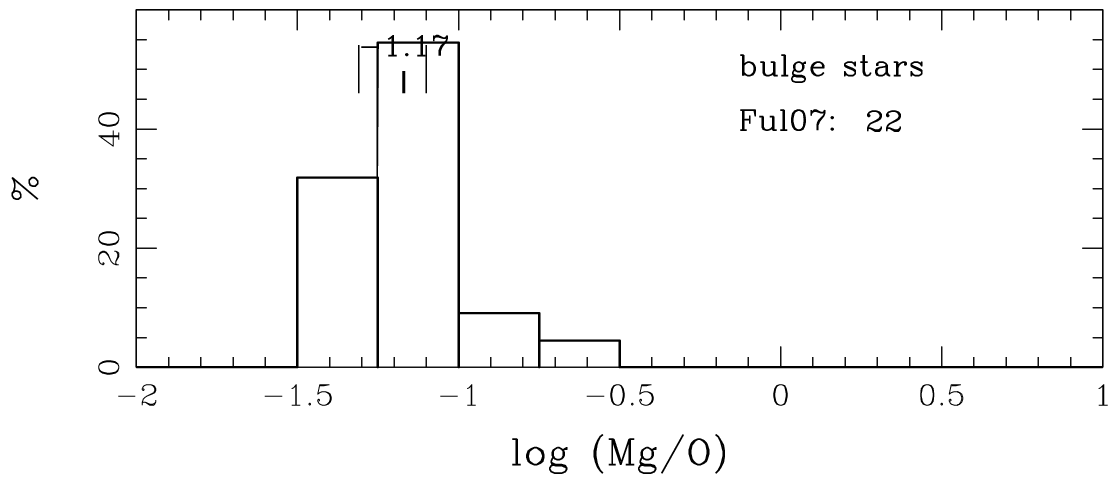}}
\resizebox{0.49\hsize}{!}{\includegraphics{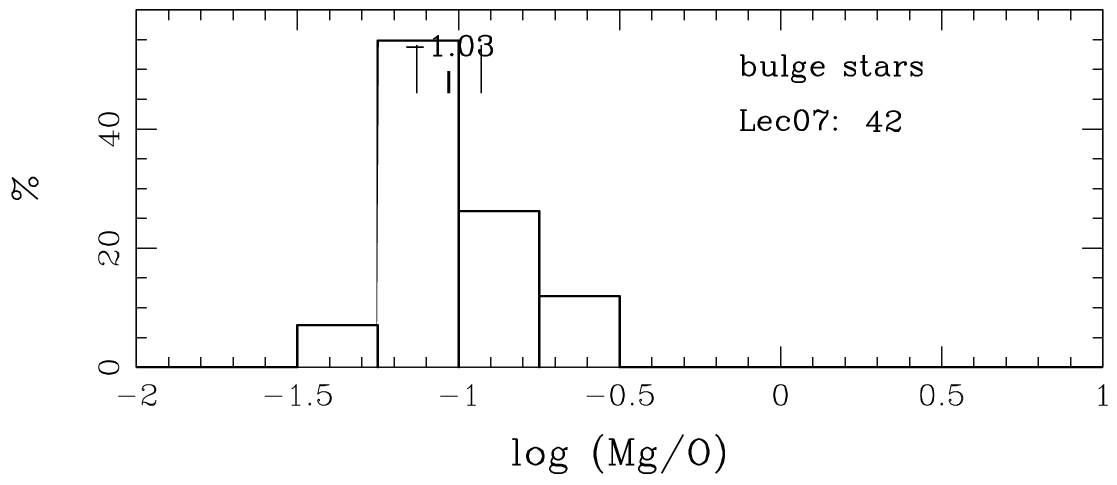}}
\caption[]{The distribution of Mg/O for the F07 (left panel)
    and L07 (right panel) bulge giant samples. The solar value from
    Asplund et al. (2005) is
    log(Mg/O)$= -$1.13. The final distributions for bulge giant
    stars (see text) can be seen in
Fig.~\ref{Distributions3}.
}
\label{Distribution_stars1}
\end{figure*}

\begin{figure*}
\resizebox{0.49\hsize}{!}{\includegraphics{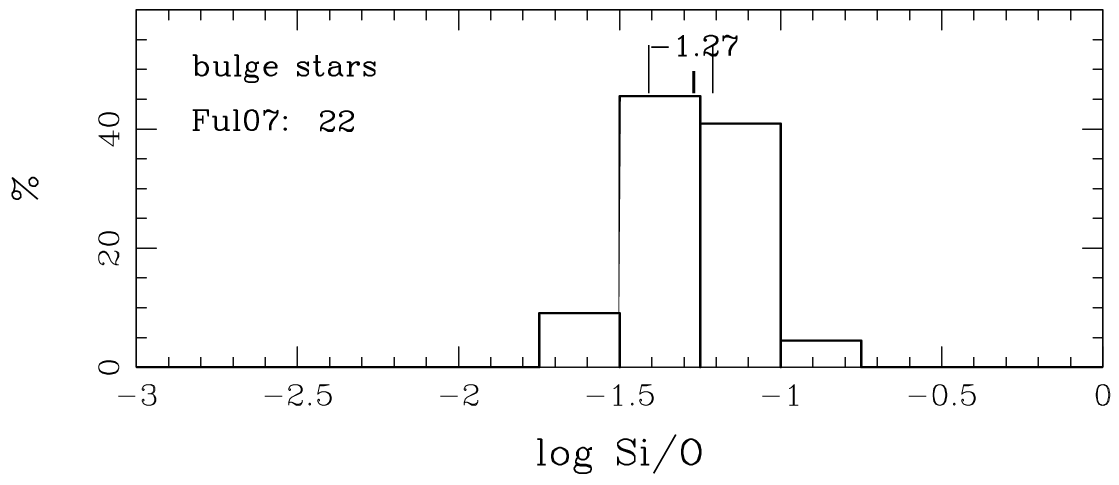}}
\resizebox{0.49\hsize}{!}{\includegraphics{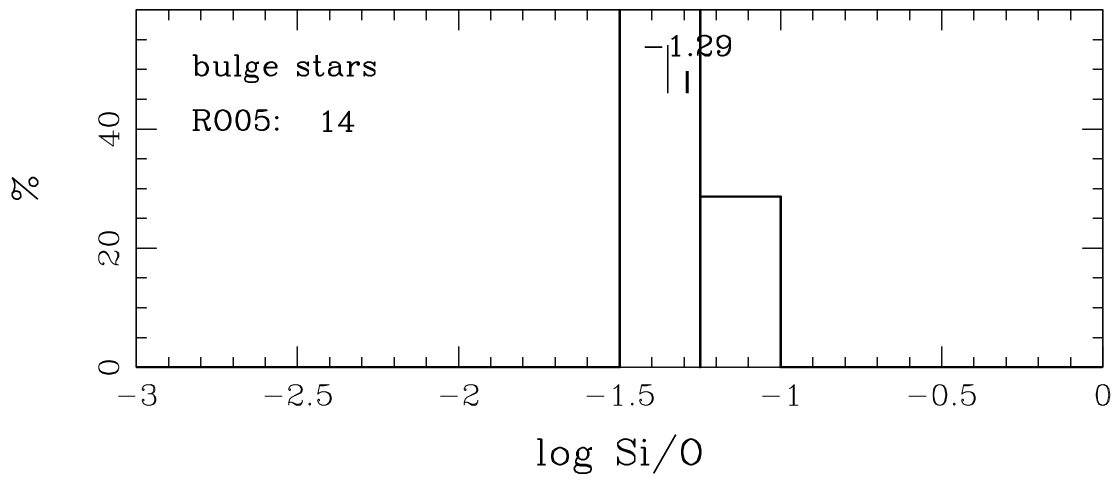}}
\caption[]{The distribution of Si/O for the F07 (left panel) and Rich
    \& Origlia (2005) (right panel) samples. The solar value from
    Asplund et al. (2005) is
    log(Si/O)$= -$1.15. The final S/H and Si/O distributions for bulge giant
    stars can be seen in Figs.~\ref{Distributions2} and \ref{Distributions3}, respectively.
}
\label{Distribution_stars2}
\end{figure*}

Figures~\ref{Distribution_stars1} and \ref{Distribution_stars2} show the
distributions of the Mg/O and Si/O ratios. The Mg/O ratio
measured by both L07 and F07, are close to the solar Mg/O ratio
of Asplund et al. (2005). Similarly, the Si/O ratio measured by F07 was found to be close to the solar Si/O ratio of Asplund et al. (2005). We note that the Si/O ratio from
the Rich \& Origlia (2005) sample, is not solar. In this case, the final
Si/O distribution includes only the data from Fulbright and collaborators.
The final distributions of Si/H, Mg/H, Si/O, and Mg/O for the
bulge giant stars are shown in Figs.~\ref{Distributions2} and \ref{Distributions3}.

\begin{figure*}
\resizebox{0.49\hsize}{!}{\includegraphics{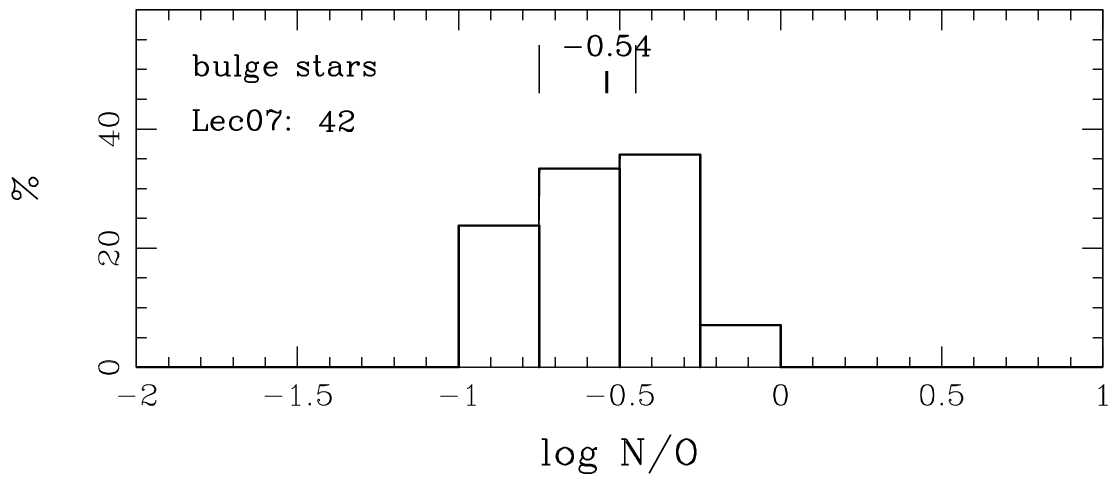}}
\resizebox{0.49\hsize}{!}{\includegraphics{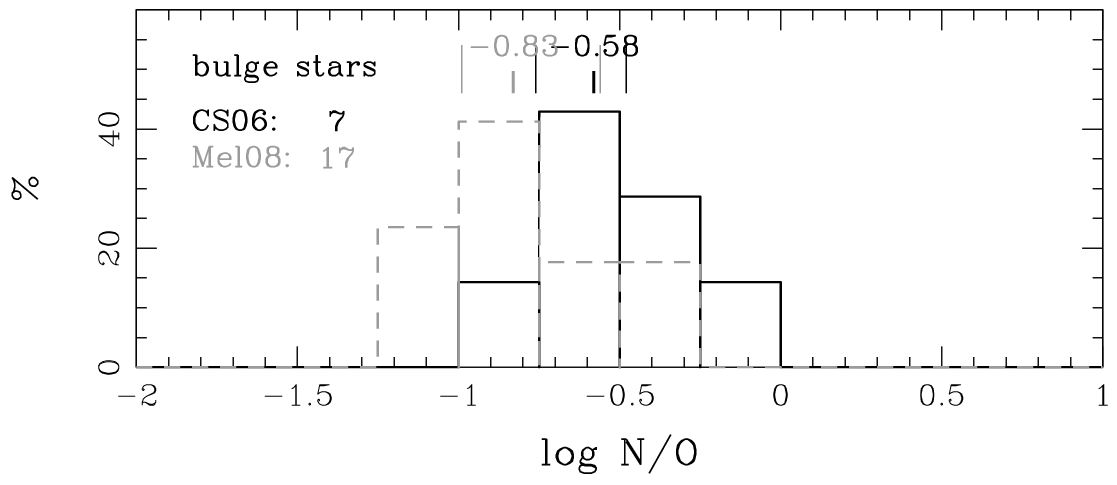}}
\caption[]{The distribution of N/O ratios for the samples of L07 (left panel),
Cunha \& Smith (2006) (right panel, solid histrogram) and Melendez et al. (2008) (right panel, dashed histogram). The final total
   N and N/O distributions for bulge giant
    stars can be seen in Fig.~\ref{Distributions1}.
}
\label{Distribution_stars3}
\end{figure*}

Finally in Fig.~\ref{Distribution_stars3}, the distribution of
N/O for the samples of L07 and Cunha \& Smith (2006) are
shown. Since Cunha \& Smith (2006) also included some stars from F07,
we compare this histogram with that obtained by Mel\'endez
et al. (2008), who measured N in 17 of the F07 stars.
Despite the low number of objects in the Cunha \& Smith (2006) sample,
the median N/O ratio values agree well with those of L07.
The Mel\'endez et al. (2008) N/O distribution
is instead shifted to lower values by $\sim$0.25~dex. Given these differences,
we adopted only that of L07 (see
Fig.~\ref{Distributions1}) as our final N/O bulge giant distribution.

\section{How do mixing processes affect the observed abundances?}
\label{sec:mixing}

To be able to interpret the abundance results described
in Sects.~2 and 3,
we recall some of the main aspects of the evolution of
low- and intermediate-mass stars (for details, see the review
by Siess 2007).

Before ascending the asymptotic giant branch (AGB) phase, the original
surface abundance of low- and
intermediate-mass stars can be modified.
Products of central and shell hydrogen burning are brought to the outer
layers by the first and second dredge-up episodes
(DUPs) taking place during the
red giant branch (RGB) and early AGB phases, respectively. During the first
DUP, the surface abundance of $^4$He is increased, $^{14}$N and
$^{13}$C are enhanced at the expense of $^{12}$C, while $^{16}$O remains
essentially unchanged. The second DUP occurs in stars initially more
massive than 3-5\Ms{} during the early AGB phase, and modifies 
the stellar surface abundances by increasing $^{4}$He and $^{14}$N, 
and decreasing $^{12}$C, $^{13}$C and $^{16}$O 
(an increase in $^{16}$O is predicted for this phase if stars rotate and are metal-poor - see
below).

Products of shell helium burning are also brought to the stellar
surface during the AGB phase, after the third dredge-up episodes (3DUP)
that occur in stars of masses higher than 1.5M$_{\odot}$ of solar
composition, starting at lower masses for lower metallicities. The outcome
is an increase in the $^{12}$C surface abundance and, in stars more massive
than $\simeq$ 3\Ms, a small increase in the $^{16}$O, heavy s-elements,
$^{22}$Ne, $^{25}$Mg, and $^{24}$Mg abundances. Part of the helium-burning
products such as $^{12}$C and $^{16}$O can undergo further H burning if the bottom of the convective
envelope reaches sufficiently high temperatures. This leads to a decrease in
$^{12}$C and $^{16}$O and the further production of $^{14}$N and $^{13}$C (a
process called hot bottom burning -- HBB -- and thought to occur only in stars of
masses above $\simeq$ 4\Ms).
The 3DUP is believed to be more efficient at lower metallicities.
Due to lower mass loss rates, the stellar lifetime is increased and
more thermal pulses can occur, 
producing more N at the expense of C and, in some cases, O
(e.g. van den Hoek \& Groenewegen 1997).

These processes not only depend on the stellar mass and metallicity,
but also on mass loss, stellar rotation, opacity, and overshooting 
(see Charbonnel 2005 for a review). In particular, rotation can have
an important impact on the contribution of low- and intermediate-mass
stars to the production of
$^{16}$O and $^{14}$N. During central He burning, primary
$^{16}$O diffuses from the core and is later carried to the surface
in  large quantities during the 2nd DUP. This process can be
extremely efficient 
(especially at low metallicities and for the high mass end of
intermediate-mass stars) to the point that such an enhancement cannot be
erased by HBB later on during the AGB phase (in other words, these processes
could affect the predictions for the N and O yields in intermediate mass
stars). The quantitative estimates of this effect are still uncertain
since it depends on many unconstrained model parameters, the
most important being
overshooting, rotational velocity, and mass loss (which in turn affects the
stellar lifetime). 

Calculations by Decressin (2007) and Decressin et
al. (in preparation) show that the masses of ejected oxygen and neon, for a
5\Ms{} star of initial metallicity of Z$=$5$\times$10$^{-4}$, increase by
a factor of $\sim$2 upon the inclusion of rotation (for a rotational velocity
of 300\kms). It is interesting to note that in this case, despite the
increase in O and Ne due to rotation, the PNe would have the
same Ne/O ratio as if their surface abundance had not been modified.
However, for a 3\Ms{} star, the opposite result is obtained, that models with
rotation predict a factor of $\sim$2 decrease in O and almost no change
in Ne. The differences seen between the 5\Ms{} and 3\Ms{} cases
are due to the fact
that the second DUP is more efficient in the former, and hence brings
more helium-burning material to the surface.

The effects of both the first and second DUPs are expected to affect the
surface abundances of PNe and giants in particular for elements produced by 
the CN-cycle. Carbon deficiencies
and nitrogen excesses are observed in the bulge giants (see discussion
in L07). For PNe, the effects of the 3DUP should also be evident:
during the AGB phase, recurrent occurrences of the 3DUP
enrich the stellar surface with freshly synthesized nuclides, which
are then ejected into the ISM by strong winds. For stars of masses
higher than 4\Ms, the 3DUP signatures may be modified by HBB.
The 3DUP and HBB are expected to produce the most significant
changes in CNO and He abundances. It has been common practice to assume
that oxygen would reflect the composition of the ISM from which PNe formed,
and that the quantities of oxygen brought to the stellar surface during the
3DUP or consumed during HBB were negligible in stars of masses
lower than 4\Ms (e.g. Maciel \& K\"oppen 1994, Henry 1989).
However, in the presence of stellar rotation,
the surface abundance of oxygen could increase considerably
and it would no longer represent the {\it pristine} value (hereinafter defined as 
the element abundance in the ISM when the PN progenitor formed). This effect
should be stronger in low metallicity stars and could be negligible in old
metal-rich populations.

In summary, the C and N abundances measured in PNe
provide important information about the efficiency of the 3DUP
compared to that of the HBB, as well as the impact of rotation. 
On the other hand, He
abundance probes the cumulative effect of the 1st, 2nd, and 3rd DUPs, and the 
HBB. Oxygen also contains information about the efficiency of the HBB and 
the effects of rotation. In constrast, the abundance of Ne is related to the
efficiency of the 3DUP and HBB (Charbonnel 2005; Marigo et al.
2003).

\section{Planetary Nebulae: disentangling mixing and chemical evolution}
\label{sec:PNeCE}

We compare our results for our  bulge PN sample with those for the
LMC and inner-disk ones.

We expect that the LMC PN population should differ
from that of the bulge for the following main reasons.
The first is that, due to the almost continuous and
ongoing star formation in the LMC (Cioni et al. (2006); Hill et
al. (2000); Pagel \& Tautvaisiene (1998)), the LMC PN progenitors
should cover the full low- and intermediate-mass range (the same is
expected for the inner-disk sample). This is not
expected to be the case for the bulge, where star formation is
believed to have stopped long ago (see Sect.~\ref{sec:starsCE})
and PNe produced by the most massive AGBs will have already disappeared\footnote{
However, if a fraction of low-mass stars coalesce, producing objects
of higher mass, the difference between the LMC and bulge PN progenitor
mass distributions could be reduced. This suggestion has
been made to explain the observational fact that the bright
end of the PN luminosity function (PNLF) is the same in star-forming
and old systems (see Ciardullo et al (2005) and references therein).}.

The second reason is that, due to the lower metallicity of the LMC, its PN
progenitors must have undergone more important mixing processes
than the bulge and inner-disk ones\footnote{There is one caveat: our PN samples
could be biased against the highest metallicity
objects (especially the inner-disk and bulge
samples). Indeed, at high metallicities,
high mass-loss rates can prevent the star from reaching
the upper AGB and the PN phases,
leading to the formation of the so-called AGB manqu\'e (e.g. O'Connel
(1999), see discussion in Sect.~\ref{sec:direct}).} (see Sect.~\ref{sec:mixing}). In the case of the inner
disk sample, we would expect the metallicity
to be larger than in the solar vicinity, although by how much is still
an open question (e.g. Perinotto \& Morbidelli 2006; Cescutti et al. 2007).
In addition, there is a transition region between the
disk and the bulge with a stellar ring,
which can be described by a Gaussian
centered around 3.5 kpc from the Galactic centre with $\sigma$=0.5kpc
(Bertelli et al. 1995). If our inner-disk objects come from this region
(which is probably the case), they cannot
be considered as part of the thin disk
 and should be treated as separate entities (see Smartt et
al. 2001 for a detailed discussion\footnote{The interpretation of inner-disk
abundances is complicated by the fact that different chemical elements appear
to provide different answers in this region. Smartt et al. (2001) found a solar oxygen
abundance for 6 B stars located in the inner-disk, whereas Mg and Si
were found to have abundances higher than solar and consistent with that 
expected according to abundance gradients measured by Rolleston et al.
(2000) also from B stars. An alternative way of interpreting these results
is that the stars in the inner-disk belong to the thin disk and not to the inner stellar ring, but
that the yield of oxygen decreases at high metallicities (see Maeder 1992).}
).  

In summary, the first two reasons for the LMC PN
population to differ from the bulge and inner-disk ones
are related to the PN progenitor's evolution and its
dependency on mass and metallicity and can be
expressed as follows: one expects a larger N/O for PNe in
star-forming metal-poor systems (LMC) than in
old (non-star forming), metal-rich ones (bulge).

In addition to the above mentioned facts
related to the PN progenitor's evolution, we also expect
the {\it pristine} N/O ratios (i.e.  the N/O ratio in the ISM when the PN progenitor
formed) of the different systems to differ. Indeed, the N/O ratio will
increase slowly in metal poor systems (due to essentialy primary
contribution from low- and intermediate-mass stars), and much faster
in high metallicity systems where the secondary production of N will
be more important (e.g. Henry et al. 2000, Chiappini et al. 2003).

Finally, another reason for differences among the PNe in the different
samples could exist in the case of oxygen, if the amount of this element
trapped by dust is metallicity dependent. This amount is difficult to estimate and for now we will
assume that it is not strongly metallicity dependent (but see discussion in
Sect.~\ref{sec:comparing} and Gutenkunst et al. 2008), and has no impact on the
relative comparisons we pursue in the following paragraphs.

In what follows we attempt to estimate which are the 
dominant effects and check if the expectations decribed here 
are confirmed by the data. In this way, we propose to clarify
which PN abundances can be used as chemical-evolution 
tracers in the Galactic bulge. We now discuss the abundance
distributions and the
abundance ratio plots of PNe in the different samples. We begin
with N (which is clearly modified during the PN progenitor's
evolution) and O (which can be  modified in particular cases). We then
discuss Ne, Ar, and S, which are more likely to reflect the abundances
in the {\it pristine} ISM.

\subsection{Nitrogen and Oxygen}
\label{sec:PNeoxygen}

In the first three rows of Fig.~\ref{Distributions1}, the distributions of
N/H (left), O/H (middle), and N/O (right) are shown for our three PN
samples. Focusing first on oxygen, we observe that the bulge PN O/H
distribution is shifted to higher values with respect to that of the LMC
by $\sim$0.3~dex (see Table~\ref{table_PN}).

Recent chemical evolution models suggest that the IMF in the
Galactic bulge is flatter (i.e. with a larger
fraction of massive stars) than in the solar vicinity (Ballero et al. 2007a). 
There is no evidence for
a variation in the IMF along the Galactic thin disk (e.g. Chiappini et
al. 2000) and hence a standard IMF (Romano et al. 2005) is
expected for the inner-thin disk. Should we expect
the oxygen histograms of bulge and inner-disk to differ due to
a difference in their IMFs? As shown in Fig.~\ref{Distributions1}, the
inner-disk PN distribution is similar to that of the bulge, 
which is also confirmed by the statistical tests in
Table~\ref{table_HISTOTESTS}. To understand this result,
one should keep an important fact in mind\footnote{Note that our
inner-disk sample probably contains PNe
from both the thin and thick disk populations possibly also
contributing to the similarity of both distributions (Mel\'endez et
al. 2008).}: it is possible that the net result of the oppositely-acting 
effects of the IMF and chemical evolution history
of the different systems are similar metallicity distributions. 
Indeed, the metallicity distribution of the bulge
(Zoccali et al. 2008) and thin disk in the solar vicinity (e.g. Rocha-Pinto
\& Maciel 1995) are similar despite known differences in their
star formation histories, star formation efficiencies, and probably 
their IMFs, which can be inferred from an [O/Fe] versus [Fe/H] plot 
(see Sect.~\ref{sec:starsCE}).

We now turn to our results for the N/O
distributions in the different PN samples.
Richer (2006) compared PNe from
different galaxy types (dwarf spheroidal -
old; LMC - young; dwarf irregular - young; M31 bulge - old; M31 disk -
young) and emphasized that all show a similar range of nitrogen
enrichment (i.e. N/O ratios). In the present work,
we find the median abundance values to be similar among our three PN
samples (within 0.1~dex, see Table~\ref{table_PN}), which 
extends Richer's result to the bulge and inner-disk PN samples.
On the other hand, due to the large number of objects in 
each of our samples, we detect differences in the N/O distributions (see
Fig.~\,\ref{Distributions1}, right column), such as:
a) the N/O distribution that we obtain for LMC PNe appear 
broader than in the two other cases;
b) the highest values of the N/O ratio are observed in the LMC diagram, whereas
the bulge and inner-disk PNe show very few objects with log(N/O)$>$0
(the same is found for bulge giants that will be discussed in
Sect.~\ref{sec:comparing}); and c) there is marginal indication
that the LMC sample distribution shows an excess of objects with low
N/O ratios, with respect to the bulge sample. The inner-disk also 
appears to contain the same excess with respect to the bulge, but this
result is even less clear due to the small number of objects.

Assuming that the above differences are real\footnote{Although, for the
N/O distributions, the statistical tests shown
in Table~\ref{table_HISTOTESTS} are inconclusive.},
could we use them to disentangle the
effects of stellar evolution and Galactic chemical evolution
summarized at the beginning of this Section?\\

{\it{Different pristine N/O ratios}}\\

Both the LMC and inner-disk samples have experienced continuous star
formation. The effects of the different primary/secondary
N contributions should be evident because the inner-disk is more metal-rich
than the LMC (as confirmed by the oxygen metallicity distribution).
To estimate the {\it pristine} N/O ratios, we can analyze the HII
regions in both systems. The difference between the mean LMC
value of N/O obtained from HII regions and the solar vicinity value is
estimated by Russell \& Dopita (1992) to be around 0.08~dex. Assuming that
there is no N/O variation across the Galactic
disk from the solar vicinity to the
inner regions (e.g. Chiappini 2005), this could partially explain the larger number 
of PNe with low N/O in the LMC with respect to those in the
inner-disk or bulge (although the effect is marginal).
For the bulge, we expect an important secondary nitrogen
contribution from massive stars and a small contribution from low- and
intermediate-mass stars (e.g. Ballero et al. 2007a), although this cannot
be empirically estimated since there are no HII regions with which to compare).\\

{\it{Different amounts of mixing in the PN progenitors}}\\

The fact that in the LMC (and also SMC -- see Leisy \& Dennefeld 2006) there
is a larger fraction of PNe with log(N/O) $>$ 0 than in
the bulge is consistent with the idea that the stellar evolution
processes that increase N and consume O are less efficient in
bulge PNe. Moreover, since the {\it pristine} N/O
ratios are lower in the LMC, the fact that for
this system several PNe show log(N/O) $>$ 0 suggests a significant
contribution from mixing processes during the PN progenitor's life.

This can be understood as a: a) {\it mass effect} -- the bulge PNe originate 
in less massive progenitors (as expected if the bulge formed 
rapidly and is not currently forming stars) than the LMC ones; 
and/or b) {\it metallicity effect} -- bulge
PNe have a systematic higher metallicity than the LMC PNe 
(confirmed by their distinct oxygen distributions, discussed above). 
Does one of these effects dominate?

We can address this question by comparing the N/O metallicity
distributions of bulge and inner-disk PNe. If the {\it mass effect} plays
a dominant role, we would expect the inner-disk
PNe to also show objects with log(N/O)~
$>$~0, since in this case the PN progenitors should span the full
mass range, due to the ongoing star formation in the inner-disk 
(as in the LMC). However, this is not observed. 
Objects with log(N/O)~$>$~0 are almost absent in the
inner-disk sample. It therefore appears that {\it metallicity} 
plays a dominant role in the observed differences of 
the N/O metallicity distribution between bulge and LMC PNe
\footnote{Interestingly, this interpretation is consistent with the idea that part of
the PN progenitor mass difference expected for bulge vs. LMC objects could
be {\it washed out} if stellar mergers occur (especially if the fraction of binaries
is higher in the bulge - Gutenkunst et al. 2008). Stellar mergers in the bulge have been 
invoked to explain the
fact that the upper part of the PN luminosity function is the same in spiral and elliptical
galaxies (Ciardullo, 2005).}.

\begin{figure}
\resizebox{0.98\hsize}{!}{\includegraphics{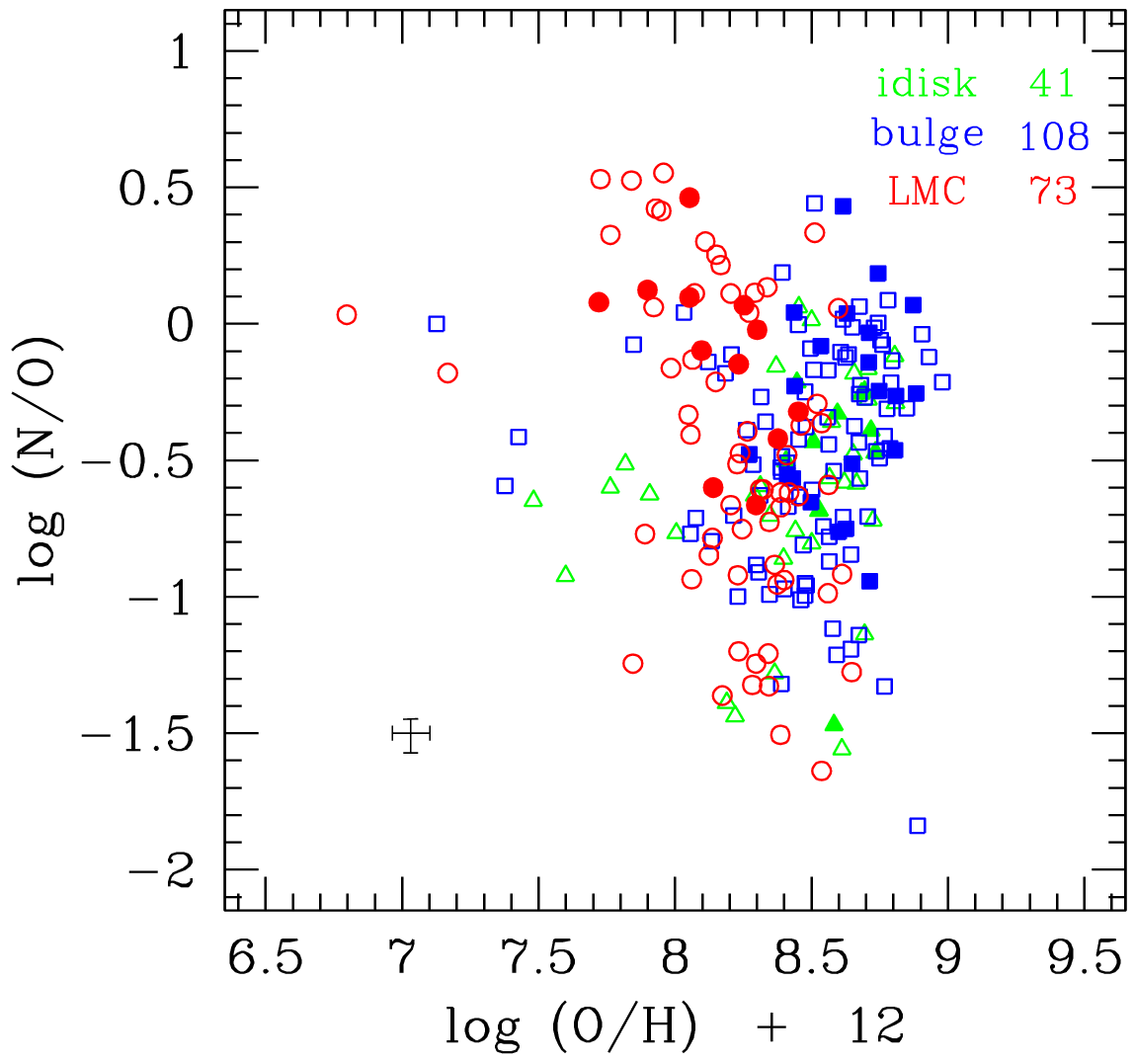}}
\caption[]
   {Left panel:
   The nebular abundance ratios N/O versus O/H for our bulge (squares),
LMC (circles) and inner-disk (triangles) PN samples. The filled symbols represent
the more accurate abundances (see
Sect.~\ref{sec:PNedistributions}).
}
\label{ONO}
\end{figure}

In Fig.\,\ref{ONO}, we plot O/H versus N/O abundances for
PN samples.
This figure presents a consistent picture: for the LMC (red filled circles),
log(N/O) is anti-correlated with oxygen as
expected if the cycle O-N had occurred converting part of the oxygen into
N.
This again indicates that in the LMC, more objects are prone to this
effect, and thus in this galaxy, oxygen cannot be assumed to be a metallicity
tracer that can be compared with chemical evolution predictions. 
On the other hand, the correlation is not present for the bulge, which 
suggests that, in this case, the oxygen abundance can be safely 
used as a chemical evolution tracer.

\subsection{Ne, S, and Ar}
\label{sec:PNeNeSAr}
While Ar and S are produced in massive stars during explosive
burning, Ne, like O, is
produced during quiescent burning (e.g. Limongi
\& Chieffi 2007). It is thus expected that Ne
should trace O more closely than Ar and S.
On the other hand, the possibility is not excluded
that Ne, as O, can be modified during the evolution of the more
massive PN progenitors (see Sect.~\ref{sec:mixing}, Leisy \&
Dennefeld 2006).

Figure~\ref{Distributions2} shows the distributions of Ar/H, Ne/H, and
S/H. As in
the case of oxygen, the distribution of these elements in bulge PNe are
shifted to higher values that those in the LMC. This is
expected since the bulge had a more rapid and effective 
chemical enrichment than the LMC (see Sect.~\ref{sec:starsCE}). 
Thus, the shift in the abundance distributions of elements 
produced essentially by Type II SNe can be understood on chemical
evolution grounds.

A second point to notice in Fig.~\ref{Distributions2} is
that the abundance distributions of Ne/H, Ar/H, and S/H in the LMC are
narrower than those of the bulge and inner disk samples (as confirmed by the
statistical tests in Table~\ref{table_HISTOTESTS}). This is also expected from
chemical evolution: in the bulge and inner-disk, the rapid chemical enrichment
leads to a broader metallicity distribution (a proportionally larger
number of metal-poor stars). In contrast, in the LMC, less
efficient star formation and longer infall timescales produce a
narrower distribution \footnote{Indeed, in the solar vicinity, one can solve
the so-called {\it G-dwarf problem} -- i.e. the fact that we observe
fewer metal-poor objects than predicted by the Simple Model --
by assuming that the disk formed by slow gas accretion, or infall
(e.g Chiappini et al. 1997).}.

Figure~\,\ref{Distributions3} shows the distributions of the abundances of 
Ar, S, and Ne with respect to the oxygen abundance. 
The Ne/O distributions are narrower and more
symmetric than those of Ar/O and S/O because of: a) the smaller
uncertainties involved in the Ne/O ratios, and b) the similarity of the 
nucleosynthetic sites of O and Ne (see above). Within the
uncertainties, we consider that for each element X$=$Ar, Ne, and S, the
X/O distributions of LMC, bulge, and inner-disk reach their maxima at 
similar values. The bulge, inner-disk, and LMC also exhibit similar 
median values of S/Ar (see Table~\ref{table_PN}). 
The mean PN values are close to the
values of Izotov et al. (2006) for solar metallicities
\footnote{The log(X/O) (X= Ne, Ar, S) ratios obtained by Izotov
et al. (2006) at low metallicities are the most representative
of the ISM value, because they are the least affected by dust depletion.
Their values are close to the solar ratios of Asplund et al. (2005),
with the exception of S/O. In this case, the Izotov et al. (2006) value at
low metallicities is $\sim$0.18~dex below the solar one.
This implies that either the
solar S value of Asplund and co-authors is overestimated, or that the
sulphur abundances of Izotov et al. (2006) 
are underestimated. Most probably, the second alternative is true (the
S photospheric abundance given in Asplund et al. (2005) is in excellent 
agreement with the meteoritic value). }. This indicates
that the PN X/O ratios are affected by dust in a similar way as 
HII regions of the highest metallicities in the Izotov et al. (2006) 
sample and that,
globally, the PN abundances of Ne, Ar, S, and O are not modified significantly 
inside PN progenitors (but see below).

A constant value of each of the Ne/O, Ar/O, S/O, and
S/Ar (see Fig.~\ref{SARD}) ratios, is expected, independently of the specific 
chemical evolution of a galaxy or a particular IMF, if these elements: 
a) originated only in Type II SNe; b) have stellar yields that are not 
strongly metallicity dependent\footnote{In fact, although it has been
suggested (Maeder 1992) that the yields of
oxygen should decrease strongly with increasing metallicity, the
more recent calculations found this effect to be
reduced and (Meynet \& Maeder 2002)
confined only to the most massive stars (but see
Sect.~\ref{sec:starsCE}).} and c) have
not been modified during the PN
progenitor evolution. As discussed above, these conditions appear to
hold globally when the mean values are considered.

However,
some differences are seen in the form of the
distributions, indicating that some of the above
conditions might not apply to all objects.
In the case of the LMC, there is an excess of objects with larger X/O
ratios with respect to that observed for the bulge and inner-disk
samples. We interpret this as an indication that
in the more massive LMC PN progenitors, the ON cycle took place
(as already indicated by our N/O results
previously discussed), which increased their X/O
ratios. This effect is observed most clearly in the Ne/O distribution
since, as discussed before, Ne and O undergo similar nucleosynthetic processes.

\begin{figure*}
\resizebox{0.33\hsize}{!}{\includegraphics{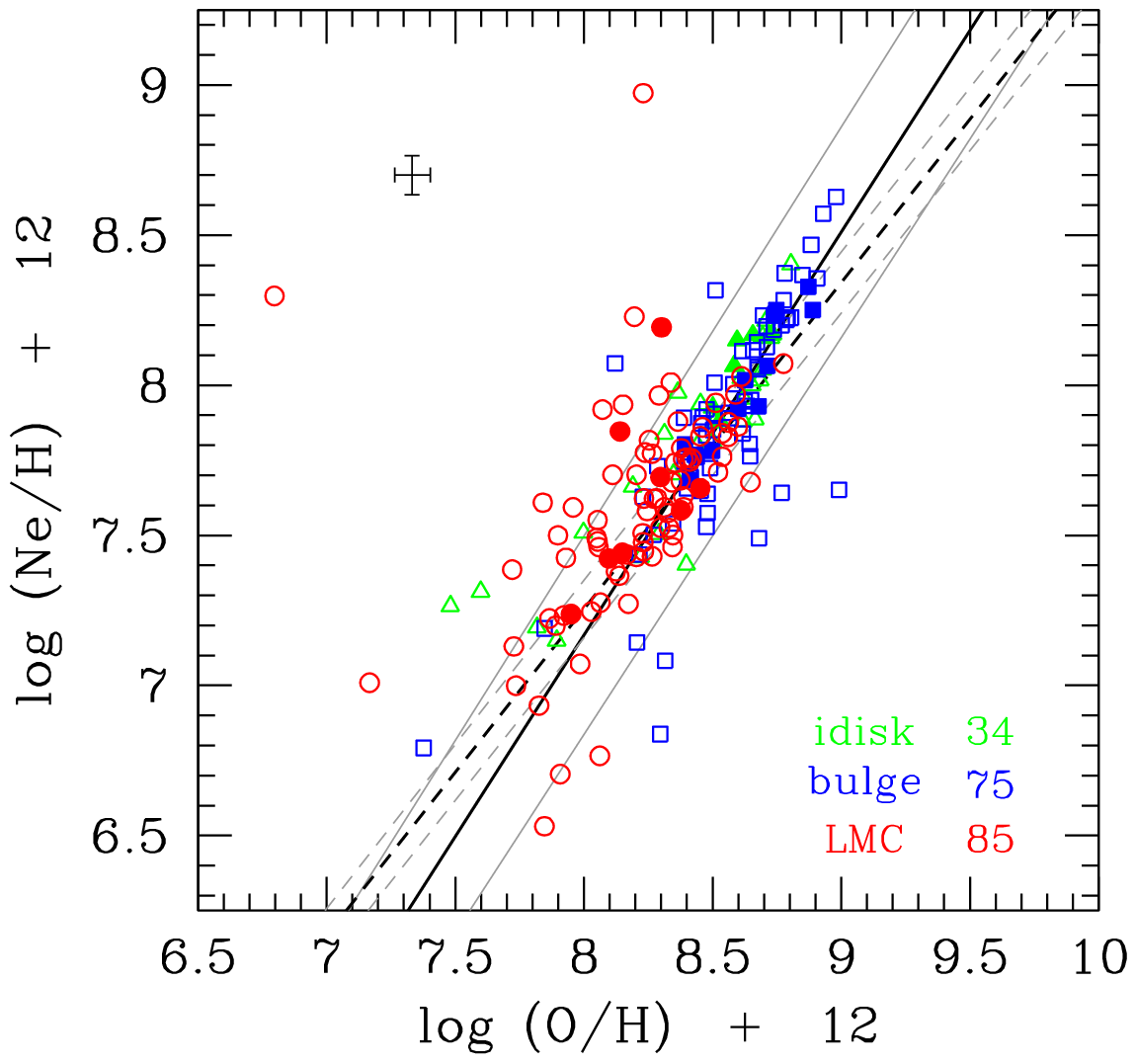}}
\resizebox{0.33\hsize}{!}{\includegraphics{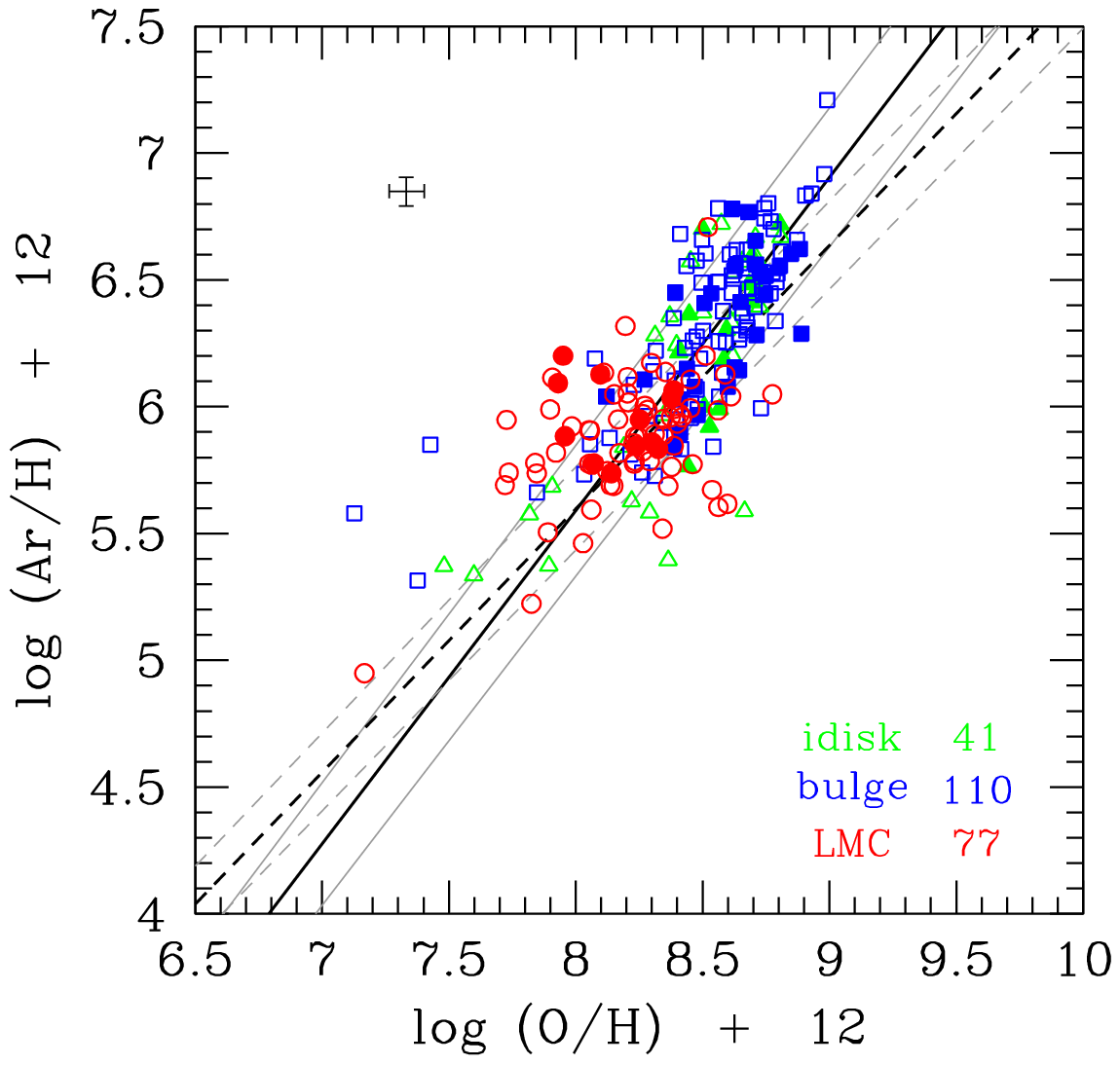}}
\resizebox{0.33\hsize}{!}{\includegraphics{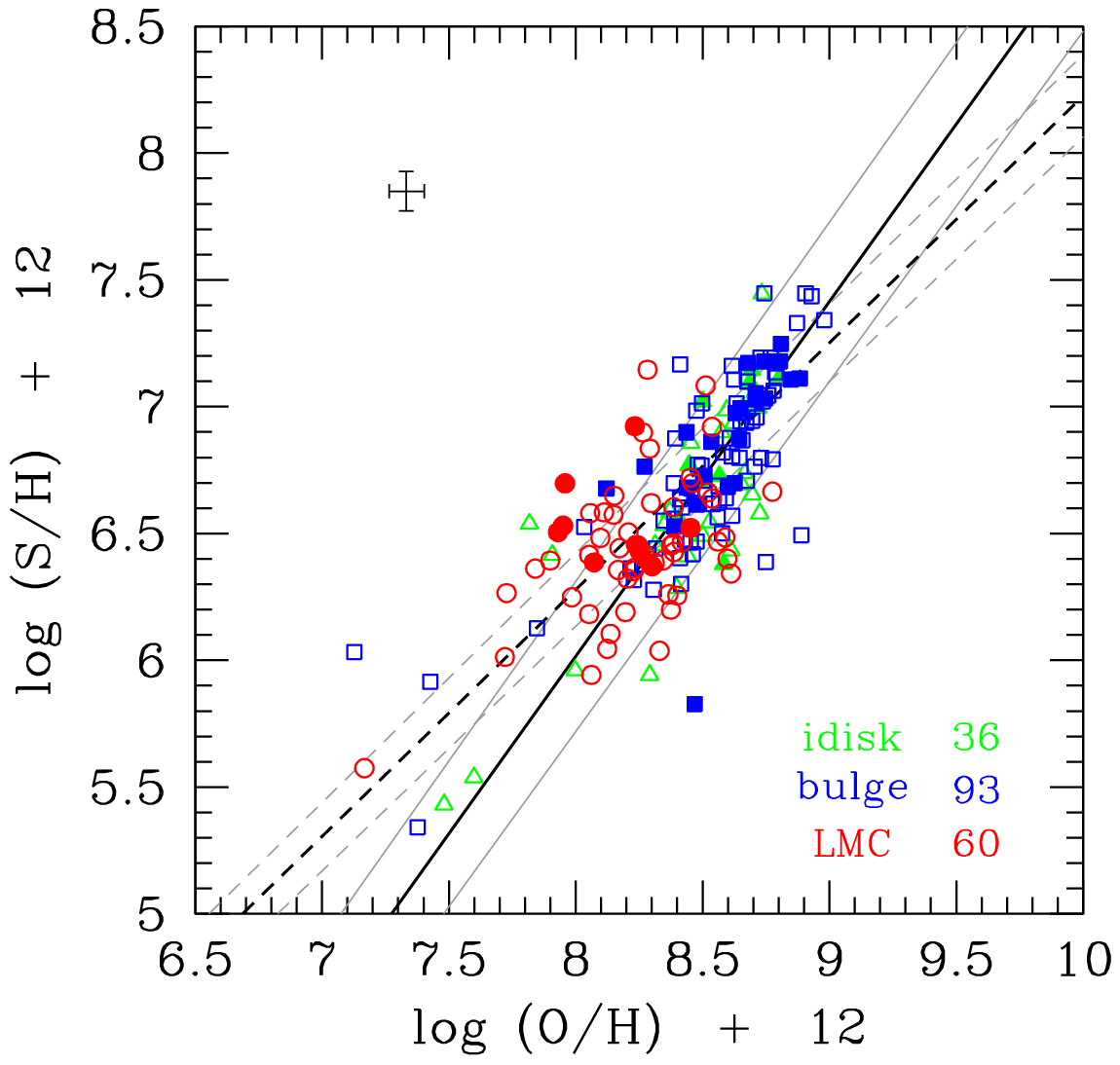}}
\caption[]
    {The variation in Ne, Ar, and S with oxygen for our PN
    sample (bulge: blue squares; LMC: red
circles; and inner-disk: green triangles). The filled
    symbols represent the more accurate abundance measurements. 
The solid line is
    the fit obtained by using only the bulge PNe. This is compared to
    the relations derived by Izotov et al. (2006 - dashed lines) for 
a sample of blue compact galaxies and star-forming galaxies
    from the SDSS survey. Grey lines show the one-sigma ranges obtained by maximizing the combined
    uncertainties in the fit parameters ``a'' and ``b'' (see Table 6).
}
\label{OArNe}
\end{figure*}

Figure\,\ref{OArNe} shows the abundances of Ne, Ar, and S versus log(O/H)+12.
In each panel, the dashed line shows the result obtained by Izotov et al.
(2006) for a sample of blue compact galaxies, while the solid line shows
the fit\footnote{Here, we used the routine
\emph{fitexy} from Numerical Recipes, which performs a linear least-squares 
approximation in one-dimension considering errors both in x and y. 
For the giant stars, we adopted an error of 0.1~dex for all data points.} to our bulge PN sample (the grey lines indicate the one-sigma uncertainty levels in the fits. See also Table~\ref{FITS}). For Ne and Ar, the PN relations agree well with those derived by Izotov et al.(2006) (to within 20\% for the slope and with almost overlapping one-sigma ranges as shown in the left and middle panels of Fig.\,\ref{OArNe}), especially if we consider that blue compact galaxies and bulge PNe span different metallicity intervals (while the median $\log$(O/H)+12 is 8.6 for bulge PNe, it is always below 8.5 for the blue compact galaxies studied by Izotov et al. 2006). These results again suggest that O and Ar in bulge PNe reflect the {\it pristine} ISM composition. A larger difference is seen in the case of
sulphur (at least 30\% in the slope and no overlapping fit ranges -- see right panel of Fig.\,\ref{OArNe}), similar to that reported by Henry et al. (2006). This ``sulfur anomaly'' is probably due to the use of inappropriate ICFs, especially in the case of PNe. For this reason, all of the following results
that are based on sulphur should be taken only as indicative.

\begin{table*}
\caption{Linear least-square fit between oxygen and other elements in bulge
PNe, bulge giants and in the blue compact galaxies studied by Izotov et al. (2006): y=ax+b (and one-sigma uncertainties). 
}\label{FITS}
\begin{tabular}{l @{\hspace{0.50cm}}
                  l @{\hspace{0.50cm}}
                  l @{\hspace{0.50cm}}
                  l @{\hspace{0.50cm}}
                  l @{\hspace{0.50cm}} }
\hline\hline

   & a & b & $\chi2$ & q \\

\hline

Bulge PNe\\

\hline

log(Ne/H) vs. log(O/H)  & 1.34 $\pm$ 0.02 &  -3.58 $\pm$ 0.17 & 208 &
5.79 $\times$ 10$^{-15}$ \\

log(Ar/H) vs. log(O/H) & 1.31 $\pm$ 0.02 &  -4.93 $\pm$ 0.13 & 554 &
4.11 $\times$ 10$^{-61}$\\

log(S/H) vs. log(O/H)  & 1.40 $\pm$ 0.02 &  -5.18 $\pm$ 0.15 & 425 &
1.36 $\times$ 10$^{-44}$\\

log(N/H) vs. log(O/H)  & 2.27 $\pm$ 0.09 & -11.23 $\pm$ 0.27 & 836  &
4.37 $\times$ 10$^{-115}$\\

\hline

Stars & & & & \\

\hline

log(N/H) vs. log(O/H) & 1.32            & -3.49             & 34.4  & -- \\

log(Si/H) vs. log(O/H) & 1.22            & -3.24             & 6.71  & -- \\

log(Mg/H) vs. log(O/H) & 1.02            & -1.26             & 102.  & -- \\

\hline

H II Regions (Izotov et al. 2006) & & & & \\

\hline

log(Ne/H) vs. log(O/H) & 1.09 $\pm$ 0.01 & -1.45 $\pm$ 0.05 & --  & -- \\

log(Ar/H) vs. log(O/H) & 1.04 $\pm$ 0.01 & -2.71 $\pm$ 0.08 & -- & -- \\

log(S/H) vs. log(O/H) & 0.97 $\pm$  0.01 & -1.51 $\pm$ 0.07 & -- & --\\

\hline\hline

\end{tabular}
\end{table*}

The fact that the LMC PNe follow the same O-Ne relation found for bulge PNe, despite their {\it pristine} oxygen having been modified (as previously shown)
suggests that the mixing processes occurring in the PN progenitors
at low metallicities are such that O and Ne are modified
 by similar amounts, leaving the Ne/O ratio essentially unchanged.
However, stellar evolution models discussed in
Sect.~\ref{sec:mixing} suggest that this is not always the case because the
mixing processes depend on several parameters (e.g. mass and
metallicity). This explains why in Fig.~\ref{OArNe}
the LMC sample exhibits a larger scatter than that of the
bulge. On the other hand, Ar is not expected to be modified, and
hence an even larger scatter should be seen in the O-Ar diagram for LMC
PNe\footnote{
The same thing should be seen also in the O-S relation. However, here
a further complication arises. Sulphur can be produced in non-negligible 
amounts in Type Ia SNe (Iwamoto et al. 1999).}. This appears to be
the case in Figs.~\ref{Distributions3} and \ref{OArNe}. 

Finally, Fig.~\ref{SARD} shows the S/Ar distributions.
It can be seen that: a) the median value of S/Ar in the LMC sample is slightly
larger than that in the bulge and inner-disk samples, and b) in the case of
the LMC distribution, there is a clear excess of objects with high 
S/Ar ratios compared with the bulge PNe. 
This can be explained as follows: in the LMC, the chemical
enrichment has proceeded on a longer timescale than that of 
the Galactic bulge. Hence, the contribution of type Ia SNe to the S abundance in
the LMC has been important, whereas this has not been the case in the bulge
(as confirmed by the abundance ratios; Hill et al. 2000). We therefore 
expect to find objects with high S/Ar ratios in the LMC and not
in the bulge (the inner-disk being an intermediate case)
as in Fig.~\ref{SARD}.

\subsection{Summary}

We summarize the main results of this Section.

\begin{itemize}
\item The bulge and inner-disk PN distributions of O, Ne, S, and Ar
  are systematically shifted to higher values compared with those of the LMC. 
This clearly
  indicates that both the bulge and inner-disk are more metal-rich
  than the LMC, a result consistent with those of other abundance tracers (e.g. stars, and HII regions).
\item S and Ar can be used as chemical-evolution tracers because their
  abundances are not modified by the PN progenitor's evolution.
  An important contribution of SNIa to the ISM enrichment in the LMC
is clearly seen by the excess of PNe
  with high S/Ar ratios in the LMC sample compared with that of the bulge.
Although this conclusion has been reached before from stellar (giant) abundance
studies (Hill et al. 2000), it is the first time that
this has been shown for a sample of PNe. The caveat here is that the
S abundance in PNe are still affected by large uncertainties, as discussed before.
\item The oxygen and neon abundances of bulge PNe are close to their ISM
values at the time of PN progenitor formation, and
  hence can also be used as tracers of the bulge chemical
    evolution. This view is supported by: a) the narrow Ne/O
    distribution of bulge PNe; b) the mean values of log(X/O) for bulge PNe 
(where X=Ne, Ar, and S), which are similar to that found in HII regions;  c) the O versus Ne and O versus Ar relations for bulge PNe,
similar to those derived by Izotov et al. (2006) for HII regions; d) the lack of anti-correlation
    between log(N/O) and log(O/H) for bulge PNe, indicating that no/negligible
amounts of oxygen were converted into N via the ON cycle;
e) the lack of objects with log(N/O) $>$ 0 among bulge PNe.
\item In the LMC PNe, both oxygen and neon have been modified by the
  evolution of the PN progenitor (see also Leisy \& Dennefeld 2006).
This is shown by: a) the large number
of objects with log(N/O) $>$ 0 in the case of LMC PNe; b) the clear
anti-correlation between N/O and O/H measured for a large number of
LMC PNe; c) the excess of objects with high log(X/O) (where X=Ne, Ar, 
and S) in the LMC sample compared with that of the bulge. This last result
is a clear signature of the oxygen decrease due to the ON cycle,
and d) the larger scatter in the O versus Ne and O versus Ar relations.
\item Mixing processes responsible for the increase in N inside PNe are
  confirmed to be far more effective in metal-poor PNe (LMC) than in
  more metal-rich objects (bulge and inner-disk PNe). Differences in
metallicity appear to be more important than differences in
the mass range of PNe progenitors.
\end{itemize}

\section{Stars and Planetary Nebulae: do their abundances convey the same story?}
\label{sec:StarsPNeCE}

We compare the abundances of our bulge
PN sample with the stellar abundances of bulge giants obtained using 
high-resolution spectroscopy.
We start  by briefly summarizing the main ideas 
about the bulge chemical evolution arising from the
consideration of stellar samples.
We then compare our bulge PN
abundances with those obtained from giants.

\subsection{Results from bulge stars}
\label{sec:starsCE}
A consensus exists about some of the main properties of our Galactic bulge
from the analysis of the bulge colour-magnitude diagram and the abundances of
both field and globular-cluster giant stars. It has been shown that the
bulge formed on a short timescale
 and, hence, consists of an old population with a rather
small age dispersion (Ortolani et al. 1995; Zoccali et al. 2003).
The bulge stars exhibit a large metallicity dispersion (Minniti et
al. 1995; McWilliam \& Rich 1994; Fulbright et al. 2007) with a peak around
[Fe/H] $\sim -$0.2. Based on a larger amount of spectroscopic
data, Zoccali et al. (2008) demonstrated that the bulge iron distribution
peaks at solar metallicity and is slightly narrower than inferred by 
previous works.

\begin{figure}[b]
\resizebox{\hsize}{!}{\includegraphics{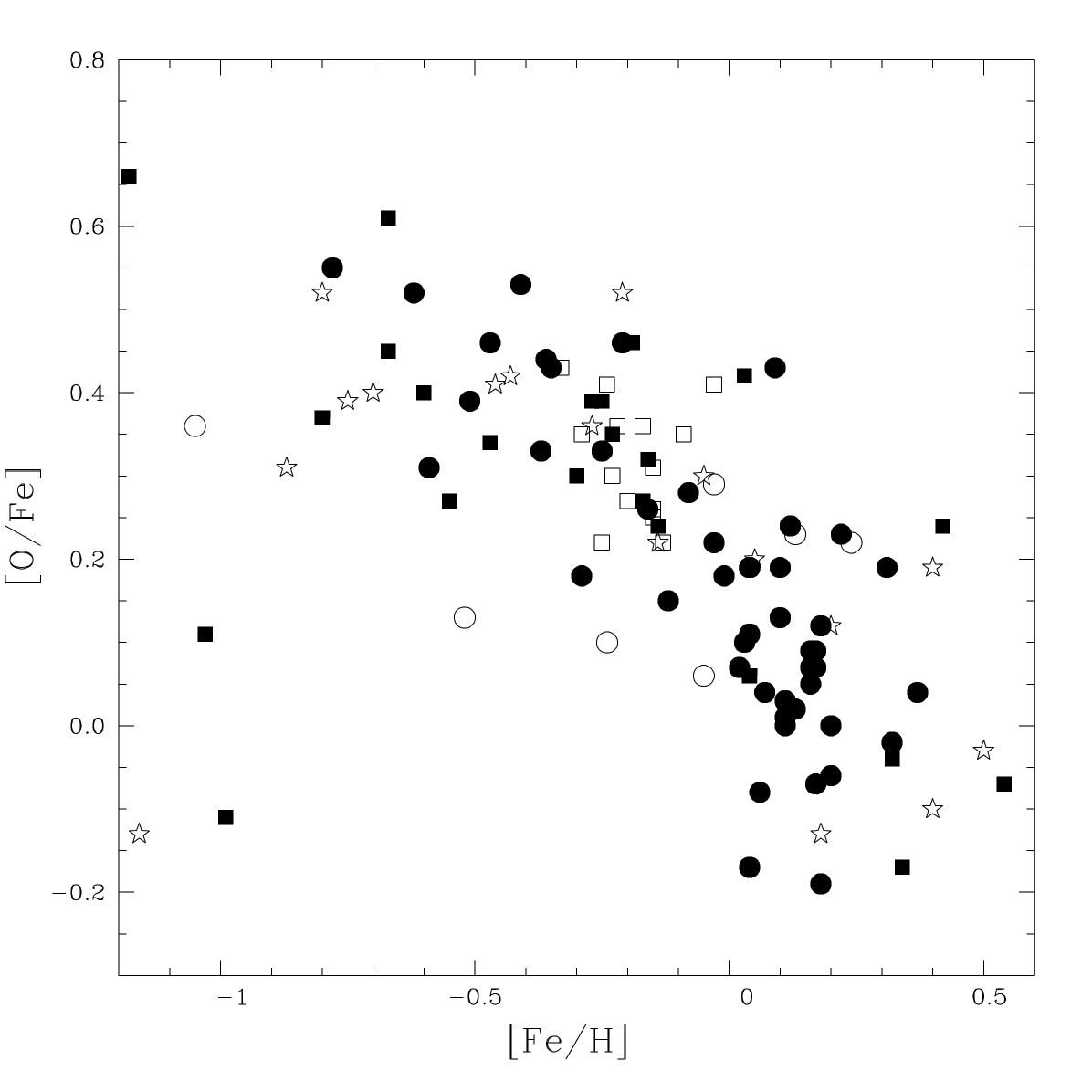}}
\caption{[O/Fe] vs. [Fe/H] in red giants of the Galactic bulge
from five groups, namely: Rich \& Origlia (2005 - open squares), Cunha \& Smith (2006 - open circles), Fulbright et al. (2007 - filled squares), Lecureur et al. (2007 - filled circles) and Mel\'endez et al. (2008 - stars).}
\label{figbulge}
\end{figure}

From abundance-ratio studies of both bulge globular clusters and bulge 
field-giant stars (obtained from high-resolution spectroscopy), it became clear that
massive stars were the main contributors to the bulge chemical enrichment.
The ISM became enriched more rapidly in elements
produced by short-lived stars (i.e. massive stars) and more slowly in those
elements produced by type Ia SNe and low- and intermediate-mass stars. Given
this fact, the ratio of two elements - such as oxygen and iron - that are
returned to the ISM on different timescales can be used as a {\it clock}
when comparing with a metallicity indicator such as [Fe/H] or [O/H].
Indeed, [O/Fe] is higher for bulge-field giants than for thin-disk stars 
(see for instance F07 and L07). This was clearly demonstrated by five studies: 
Cunha \& Smith (2006), Rich \& Origlia (2005), and Mel\'endez et al. (2008)
used the infrared OH lines, and Zoccali et al. (2006) and
F07 who used the forbidden [OI]630~nm line.
In Fig.~\ref{figbulge}, these five data sets are plotted
in an oxygen-to-iron diagram.
The fact that all five groups are in agreement, even if two different
sets of lines are used, indicates that this result is rather robust. 
The high [O/Fe] ratios measured for bulge stars suggest that most 
formed before Type Ia SNe had time to contribute to the ISM enrichment, 
evidence that the bulge was formed on a short timescale (Matteucci \&
Brocato (1990), Ballero et al. (2007a)).

Mel\'endez et al. (2008) concluded that the overenhancements
of oxygen with respect to iron abundance in the bulge and thick disk are 
the same, whereas L07
and F07 argue that it is overenhanced relative to
the thick disk, as found by Bensby et al. (2004).

Abundance measurements of high accuracy are available for a range of 
different elements, other than oxygen and iron. The [$\alpha$/Fe] ratios
versus [Fe/H] detailed behavior depends on the particular $\alpha$ element
chosen. One example is the difference between [O/Fe] and [Mg/Fe]
ratios. F07 found that [Mg/Fe]
ratio declines more slowly with [Fe/H] than O, Si, Ca, and Ti over Fe.
This could be interpreted as being due to the 
oxygen yield dependency on metallicity (McWilliam et al. 2008). Some
metallicity dependency of the stellar yield of oxygen cannot be excluded, 
especially in stars more massive than $\sim$40\Ms{} due to mass loss (see
Hirschi 2007). However, as discussed before, current stellar-evolution models
predict this effect to be less strong than previously thought (Maeder 1992),
and its impact on chemical evolution models is still unclear.
L07 also measured an increase in the Mg/O ratio with increasing metallicity, 
although less pronounced than reported by F07.

\subsection{Planetary Nebulae versus Stars}
\label{sec:comparing}
We compare the bulge-giant sample with our bulge PN
sample. We start
with a discussion of O and N for which a direct comparison between PNe and
giants can be carried out, since these elements have been 
measured in both cases. However,
whereas O, both in PNe (Section~\ref{sec:PNeCE}) and giants, traces
the chemical evolution of
these systems, N is modified during the evolution of both the giant star and 
the PN progenitor by the processes described in
Sect.~\ref{sec:mixing}. In the second part of this Section, we also present indirect comparisons, relating S (PNe) with Si (stars), and Ne (PNe)
with Mg (stars).

\subsubsection{Direct comparison of O and N}
\label{sec:direct}

In Fig.\,\ref{Distributions1}, the distributions of O, N, and N/O for the
sample of bulge giants are shown (last row) and can be compared with those 
found for bulge PNe (second row). There are striking differences that have
to be understood.

The most important result is that for the oxygen distribution of bulge
PNe and bulge giants. The bulge giant distribution is shifted to higher 
oxygen values by $\sim$0.3~dex (see Table~\ref{table_PN}).
Although both F07 and L07 agreed that some stars in the bulge have 
log(O/H)+12~$>$~9 (the same is also found by Mel\'endez et al. 2008), 
our measured PN oxygen abundances are all below this value. 
Systematic effects on the
abundance determinations of both PNe and giants certainly account
for part of this discrepancy.

For the present paper, we adopted PN abundances derived from collisionally 
excited lines. However, it is known that abundances from recombination lines 
are systematically higher. Could this result explain this oxygen discrepancy?
This seems unlikely given that recombination line abundances do not appear 
to represent those of the bulk of the nebula, as described in Wang and Liu 
(2007) and references therein. In some PNe, 
recombination lines lead to oxygen abundances 
that can be higher than those obtained from
collisionally excited lines by a factor of 10 or more. 
It is difficult to imagine these such extreme
values represent the true abundances in PNe.

On the other hand, stellar abundances are given relative to a
reference element and the determinations are carried out
differentially with respect to a reference star with well-known
stellar parameters (e.g. Sun, Arcturus). However,
the comparison with PNe requires the conversion
to an absolute scale. This is not a trivial
matter and involves an important uncertainty, as
can be seen in Table~\ref{tabteff} for both the Sun and
Arcturus. 

One must consider the possibility of two
further biases in our PN sample. The first 
could be due to the fact that we removed some PNe
from our original samples because we could not determine their
abundances or considered the abundance estimates to be too
uncertain. However, among the 53 PNe with
uncertain oxygen abundances (i.e. with an
estimated error larger than 0.3~dex), only 6 objects have
an estimated $\log\epsilon$(O)~$>$~ 8.7 (all of
them have $\log\epsilon$(O)~$<$~8.9). There are
8 objects for which the temperature could not be
derived. Even assuming that all of them have
a $\log\epsilon$(O)~$>$~8.9, this is a small
proportion (6\%) in comparison with the 50\%
of giant stars that have $\log\epsilon$(O)~$>$~8.9.
Therefore, the bias that may be introduced
in our sample of bulge PNe due to the fact that
some objects were rejected does not explain
the important discrepancy we find between the O/H
distribution of bulge PNe and giants.

A second possibility is that our bulge PN sample
is biased against the highest metallicity
objects. As pointed out in Sect.~\ref{sec:PNeCE}, at high metallicities,
high mass-loss rates can prevent the stars from reaching
the upper AGB and the PN phases,
leading to the formation of the so-called AGB manqu\'e (O'Connell,
1999). However, in this case, one expects only the
metal-rich part of the distribution to be affected. Instead, a shift in the 
entire PN distribution is observed towards
lower metallicities. Hence, this effect does not appear to 
account fully for the observed discrepancy.

Apart from systematic effects and biases, are there additional 
reasons why oxygen abundance distributions in bulge stars and PNe differ?

One possibility is the presence of dust in PNe.
If in PNe part of the oxygen is trapped by dust, the
observed oxygen abundance  distribution  would be shifted to lower values. 
However, the maximum oxygen depletion
into refractory material can be estimated using Eq.~24 of Dwek (1998)
and amounts to 27\% in oxygen corresponding to
0.1~dex. Therefore, dust alone cannot
explain the discrepancy between the oxygen abundance distribution of PNe
and giants in the bulge.

Another possibility is that the samples of PNe and giants do not
trace the same population. One way to check this possibility is to monitor 
the behavior of nitrogen. In both cases, stellar evolution affects
the nitrogen abundance\footnote{
In the case of bulge giants, L07 searched for a
C-N anticorrelation as a probe of internal mixing in their stars.
It is well known that the C and N abundances evolve along the
red giant branch. L07 concluded that within the
uncertainties involved, no anti-correlation
was found, but that some other mixing signatures were present for
stars located above the RGB bump. The
C$_2$ and CN lines used in the optical are
weak, whereas a more robust derivation of carbon and nitrogen abundances
can be obtained from OH, CO, and CN lines in the near-infrared. Such
data are currently under analysis for the L07 sample.
Ryde et al. (2007) presented preliminary
abundances of bulge giants computed from IR spectra
obtained with Crires on the VLT, and found the 
giant stars studied to be depleted in C and enriched in N, whereas 
the abundance of oxygen was unchanged 
(a typical sign of matter exposed to the CN cycle,
which conserves the sum of C and N nuclei).}. In fact, the median N/O values of
our bulge PNe and giant samples
peak at similar values, to within 0.1~dex, and are above solar (see
Table~\ref{table_PN}). However, Fig.~\ref{Distributions1} shows that both the N/H and N/O bulge
distributions are broader for PNe than for giants.
Does this means that the two samples
probe slightly different mass/metallicity ranges?

The bulge giants studied here
are truly old objects (consistent with having initial masses 
of about 0.8~M$_{\odot}$, see Zoccali et al. 2003). 
PNe also originate in old objects, but as they are already in the PN phase, 
they were either born earlier and/or had slightly more massive progenitors
than the giants we observe today.
If the PN progenitors had the same mass
as the giants, but were born earlier, and thus more oxygen-poor, we 
would expect them to exhibit more N enrichment since the mixing processes 
are more efficient at lower metallicities (see
Sect.~\ref{sec:mixing}). If instead the PN population originate in 
more massive progenitors than giants, again we would
expect more N enrichment. Both situations could broaden the 
N/H and N/O metallicity distributions of PNe with respect to those of 
giant stars,
and only in the first case would we expect the PNe distribution to be shifted 
to lower oxygen abundances than the giants. However,
we expect the differences
in both age and masses between PNe and bulge giants to be
small (because the dominant bulge population seems to be old - see
Sect.~\ref{sec:starsCE}) and hence
not to account for the observed 0.3~dex difference in oxygen. 
In addition, given the observational uncertainties
and low number of giants with measured N/O abundances, the observed differences
could also be due to systematics effects\footnote{The C and
    N determinations of L07 have uncertainties of the order of 0.2~dex and
    their N abundance is strongly dependent on the derived carbon
    abundance, since it is determined from the strength of the CN
    molecular band. In the spectral region studied, the most prominent
C$_2$ bandhead at 563.5nm is often weak, and only upper limits
can be inferred for the carbon abundance. The result is that
if [C/Fe] is lower than the upper limit derived, the nitrogen abundance
should be higher.} (see Sects.~\ref{sec:biases} and
\ref{sec:Starsbiases}).

Therefore, there is the possibility that
the two samples represent different populations, although 
this is unlikely to explain the oxygen discrepancy.

\subsubsection{Indirect comparisons}
\label{sec:starsindirect}

We are unable to measure Mg and Si abundances using optical spectra of PNe
(which would in all cases be trapped into dust
grains), whereas no measurement of S and Ne are available in bulge giant
stars. However, we can still complete an indirect comparison by
converting the bulge giant Mg and Si into Ne and S abundances,
respectively, if we assume that Mg/O and Si/O ratios are
solar and constant with metallicity (as assumed for Ne/O and S/O
ratios). This assumption seems to be robust because the median values 
of Mg/O and Si/O for the bulge giant star sample are close to the solar 
ratios of Asplund et al. (2005) (see Figs. \ref{Distributions3},
\ref{Distribution_stars1} and \ref{Distribution_stars2}).

Figure~\ref{Distributions2} (bottom row) shows the distributions of Ne/H and
S/H for bulge giants obtained after the transformation explained above
(shaded histograms). The resulting distributions are shifted
to higher values than those of the PNe by 0.29 dex both for Ne/H and
S/H, similar to that found for oxygen.
Although for S/H there is the possibility that
this element is underestimated in PNe (according to the ``sulfur anomaly'' 
described by Henry et al. 2004), this is not expected to be the case for Ne (although see Gutenkunst et al. 2008).
Confirmation that S in giants is systematically higher than in PNe will 
have to await measurements of S in stars. The S lines
are faint and their measurement requires data of very high signal-to-noise. 
On the other hand, no Ne or Ar lines are present in the spectra of cool stars. 

\section{Discussion and Conclusions}
\label{sec:conclusions}
Zoccali et al. (2008) compared the metallicity distribution of about 400
clump and giant stars in Baade's window, with that of about 200 giants at b$=-
$6$^\circ$ and b=$-$12$^{\circ}$.
The metallicity distributions exhibited a gradient in stellar populations 
on the metal-rich side, such that in Baade's window there is a metal-rich 
component at [Fe/H]$\sim$+0.3, which
becomes less evident at b$=-$6$^\circ$. This result is based on iron
abundances, and for oxygen the differences could be smaller (see
below). In a preliminary kinematical study, Gom\'ez et al. (in
preparation) find a higher velocity dispersion for the
metal-rich component by $\sim$~20km/s, which could be interpreted as an 
indication of a different stellar population towards
the inner regions.

Given that the Baade's window field seems to be more contaminated by the
[Fe/H]$\approx$+0.3 component mentioned above
than the field at b=$-$6, and given that
our bulge PN sample are projected across a wider area (see
Fig.~\ref{lb}), it is unclear in what proportion
this new component is present in our PN sample.
If we attempt to explain the 0.3~dex difference in oxygen
(and probably in Ne and S as well) in terms of the properties of the stellar
populations, we may conclude that our PN sample is essentially free of this 
metal-rich component.
However, since [O/Fe] decreases with [Fe/H], the oxygen content in this metal-rich
population will probably resemble that of the more
metal-poor population.

To confirm that the metal-rich 
population in the Baade's window follows the same [O/Fe] relation as shown in
Fig.~\ref{figbulge}, it would be necessary to measure oxygen in the
same stars. This would require very-high resolution
(R=40000) data, such as that acquired for the UVES sample of 
Zoccali et al. (2006) for all of the metal-rich population found by Zoccali et al. (2008) in the Baade's window.

Checking for any difference in the
V$_{lsr}$ velocity dispersion of our bulge PNe with b $<$ 4$^{\circ}$ (88
objects) and b $>$ 4$^{\circ}$ (78 objects), we measured 121~km/s and
101~km/s, respectively. Interestingly, the difference of 20~km/s is similar
to that found for bulge giants in Baade's window with
respect to other fields. However, for PNe this difference is not significant 
according to our statistical tests.

Finally, we note that significant samples of bulge objects are
now available for which the
metallicity distribution can be studied (RGB stars, red clump giants --
Zoccali et al. 2008 -- and in PNe -- this work).
This opens the possibility of looking for additional constraints to stellar
evolution models of low- and intermediate-mass stars, which would enhance 
our understanding of how metallicity and
mass loss affect the metallicity distributions at different
stellar evolution stages. It is also important to
understand another discrepancy measured in the bulge: 
the metallicities of lensed turnoff stars (i.e. turnoff stars observed due to the amplification of their brightness by gravitational microlensing) appear to be systematically higher than giants, according to Cohen et al. (2008).\\

Our main conclusions are summarized below:\\

We have compared the properties of PNe in different systems 
(bulge, inner-disk, and LMC) by using
the largest homogeneous sample of PN abundances presently available. 
We find that:

\begin{itemize}
\item The Galactic bulge and inner-disk PN distributions of O, Ne, S, and Ar
  are shifted to higher values than those of the LMC, indicating
  that the bulge and inner-disk are more metal rich
  than the LMC (a result already known from other stellar studies). 
\item Oxygen and neon in bulge PNe are close to their ISM
values at the time of the PN progenitor formation, and
  hence can also be used as tracers of the bulge chemical
    evolution.
\item In LMC PNe, both oxygen and neon have been modified during the
  evolution of the PN progenitor.
\item Differences in
metallicity appear to play a more dominant role in the mixing processes occurring in low- and intermediate-mass stars than the differences in
the mass range of PN progenitors.
\end{itemize}

After identifying reliably the bulge PN abundances that can be used as
  tracers the bulge chemical evolution (O, Ne,
Ar, S), we compared these abundances with those measured in 
giant stars. We found that:

\begin{itemize}
\item The oxygen abundance distribution of bulge giant stars is shifted to
  higher values by 0.3~dex with respect to that of PNe.
\item A similar shift appears to exist for Ne and S (after converting 
the Mg and Si abundances of giant stars to those of Ne and S by
  adopting the solar values of Asplund et al. (2005), although this
  method is rather uncertain).
\item We discussed many reasons for the discrepancy between the abundances 
of PNe and giant stars, and concluded that the oxygen abundances in PNe 
(distributed over the entire bulge) and in giants (most of them in the Baade's 
window) do not convey the same evolutionary story.
\end{itemize}

After a thorough analysis, we conclude that the observed
discrepancy between PN and giant star abundance distributions is probably due to systematic errors
in the abundance derivations of either PNe or giant stars, or both.
 Our results constitute at least an important warning 
against a careless use of absolute abundances.

\begin{acknowledgements}
C. C. thanks CTIO and ESO staff in Chile and acknowledges
partial support from Pronex-Brazil and the Swiss National Science Foundation (SNF). 
C. C. also thanks R. Walterbos and F. Cuisinier for interesting discussions. We also thank
the referee for a careful reading of this manuscript.
S.K.G. and G.S. wish to thank support by the European Associated Laboratory
``Astronomy Poland-France''. BB acknowledges support from CNPq and Fapesp.

\end{acknowledgements}

\clearpage

\Online

\vspace*{1cm}
{\large\bf Table 1a. Plasma parameters and chemical abundances (Galactic
bulge sample)}

\hspace*{-10cm}
\resizebox{1.8\hsize}{!}{\includegraphics{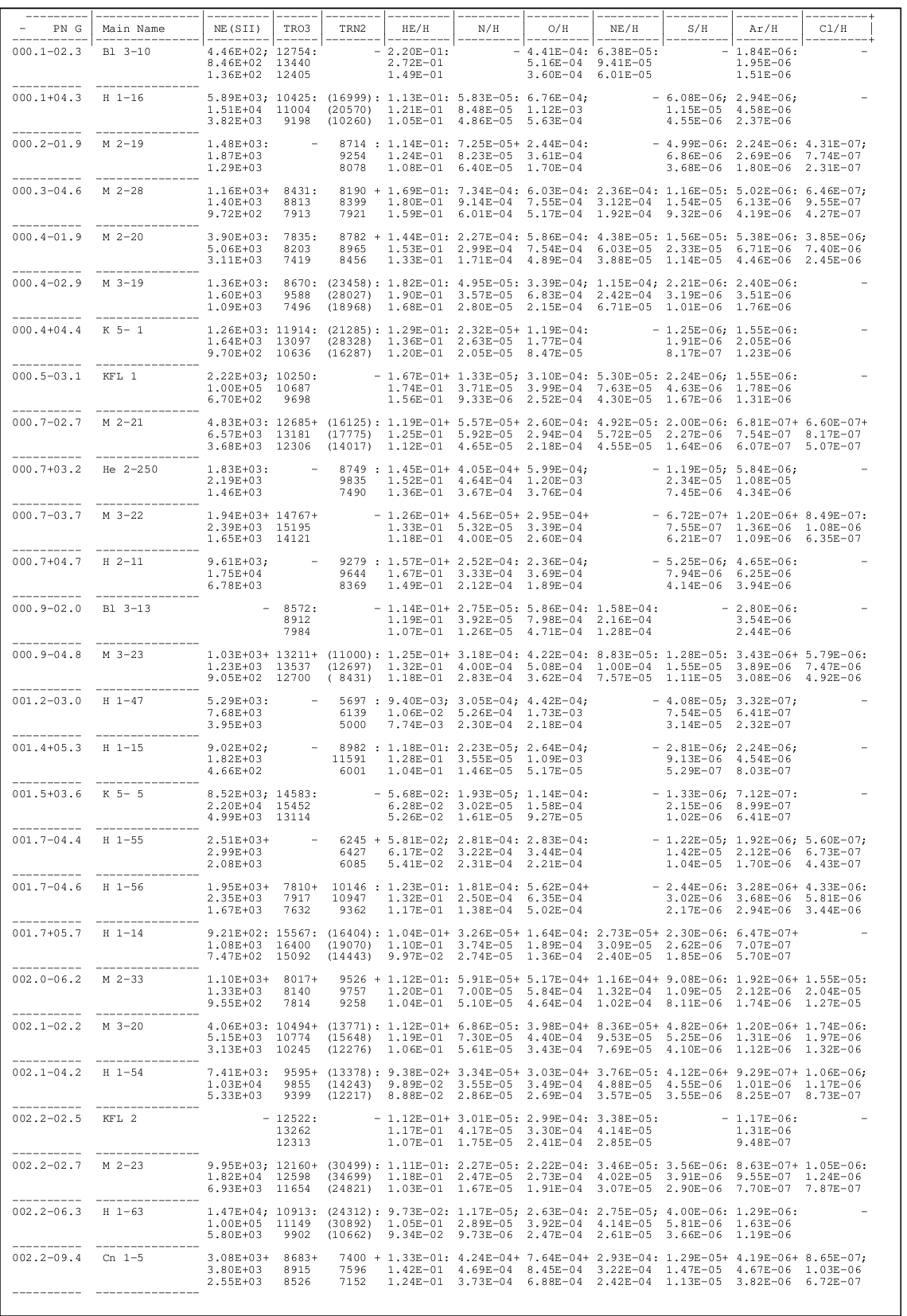}}

\clearpage

\vspace*{1cm}
{\large\bf Table 1a. continued}

\hspace*{-10cm}
\resizebox{1.8\hsize}{!}{\includegraphics{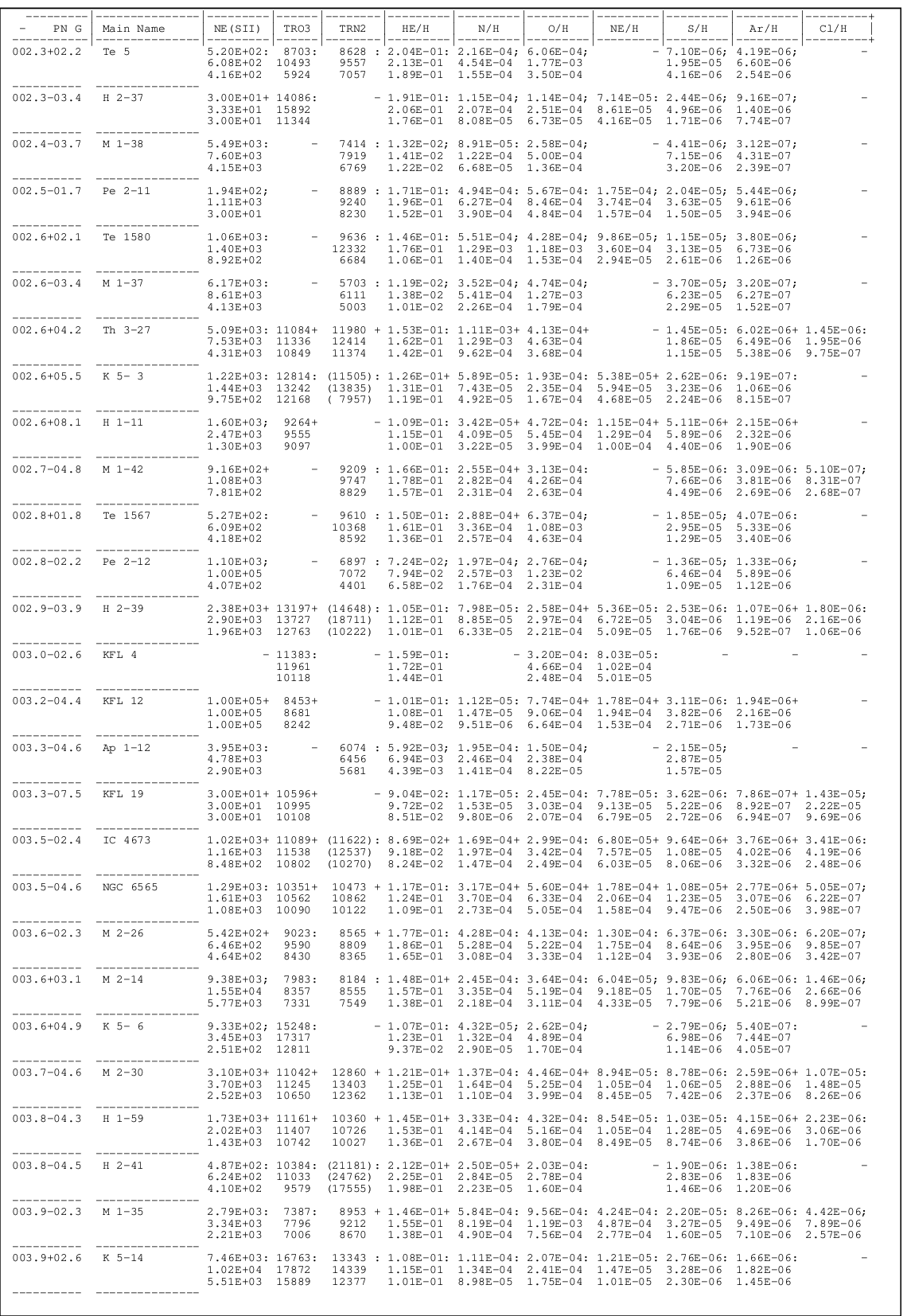}}

\clearpage

\vspace*{1cm}
{\large\bf Table 1a. continued}

\hspace*{-10cm}
\resizebox{1.8\hsize}{!}{\includegraphics{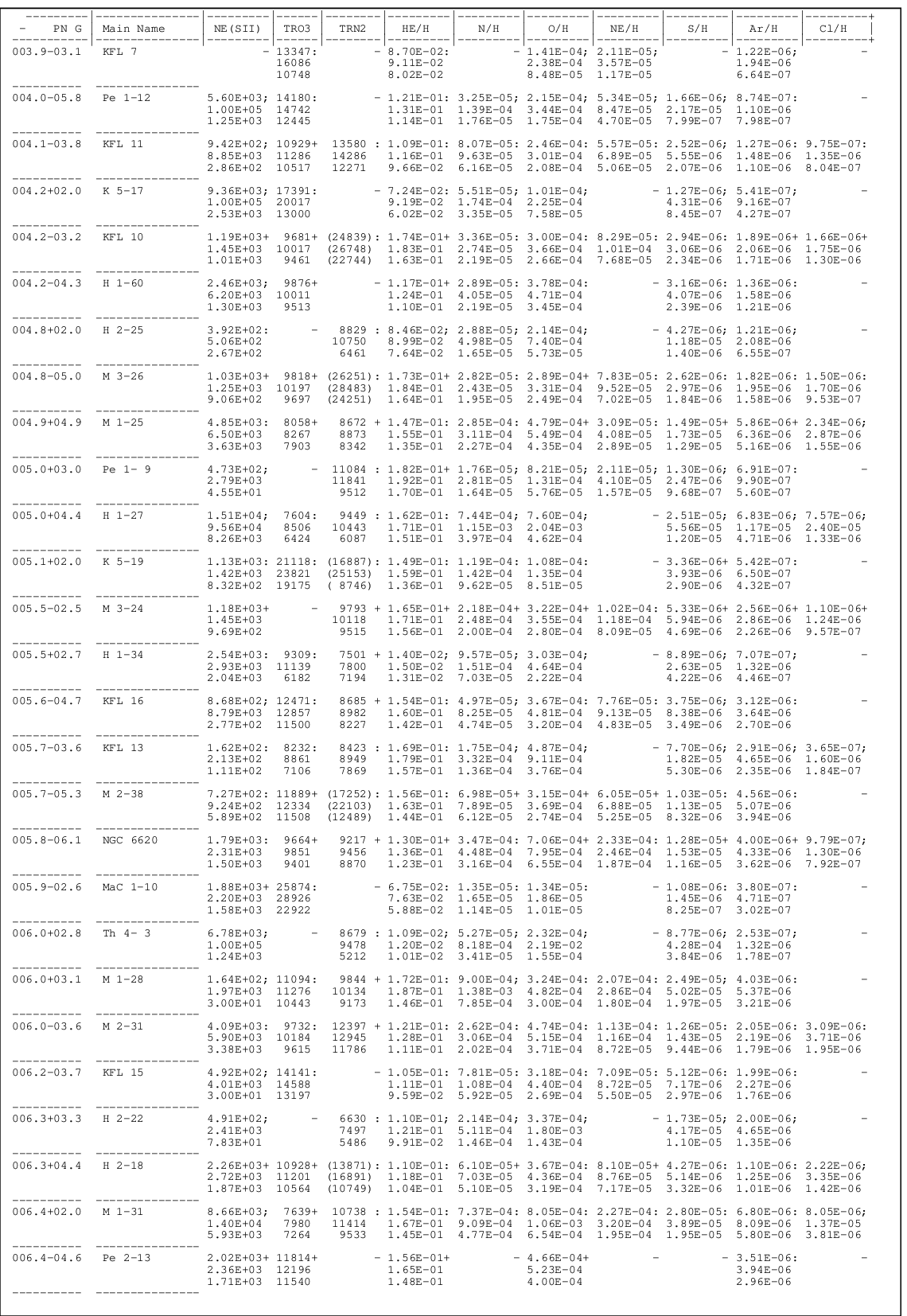}}

\clearpage

\vspace*{1cm}
{\large\bf Table 1a. continued}

\hspace*{-10cm}
\resizebox{1.8\hsize}{!}{\includegraphics{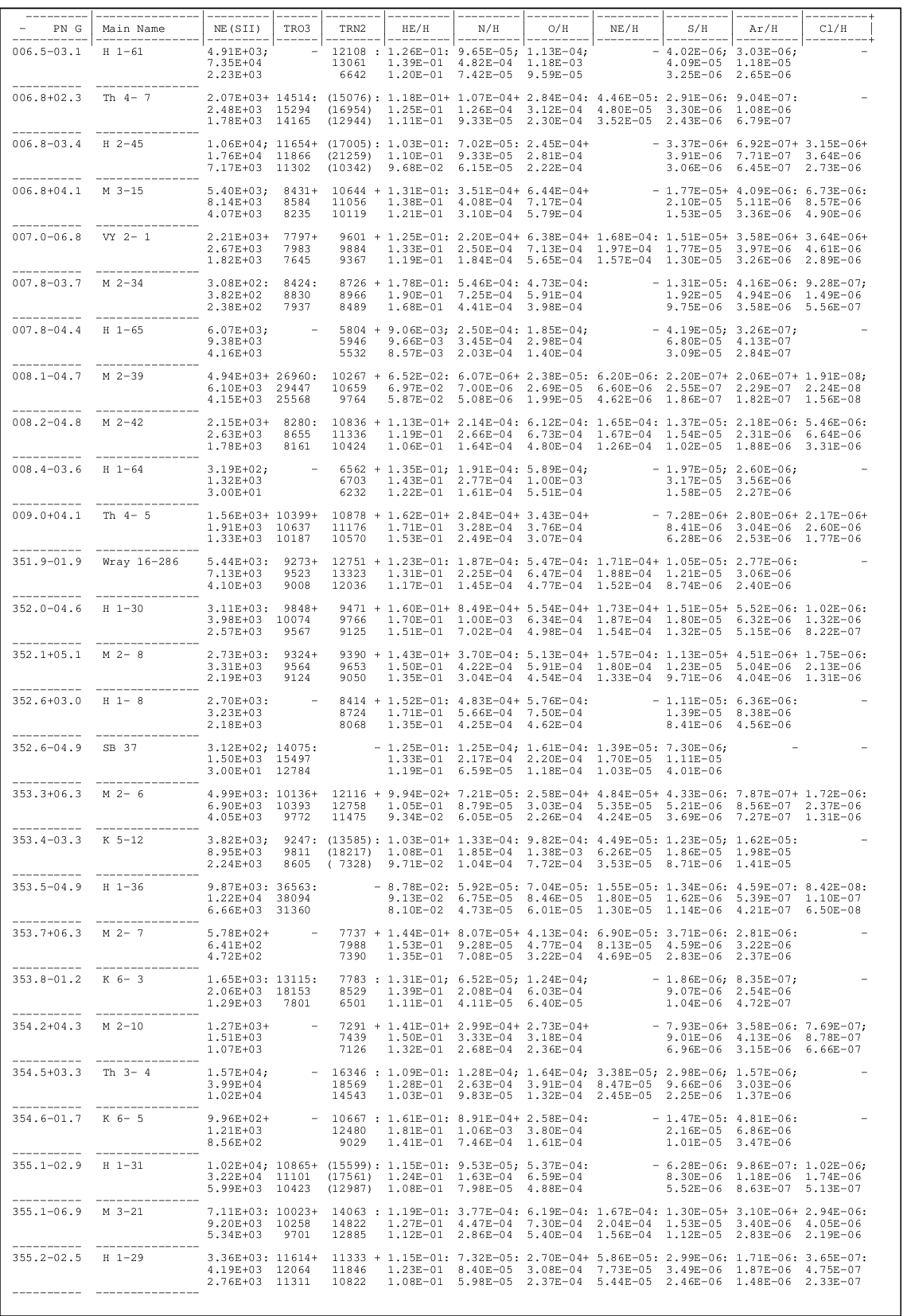}}

\clearpage

\vspace*{1cm}
{\large\bf Table 1a. continued}

\hspace*{-10cm}
\resizebox{1.8\hsize}{!}{\includegraphics{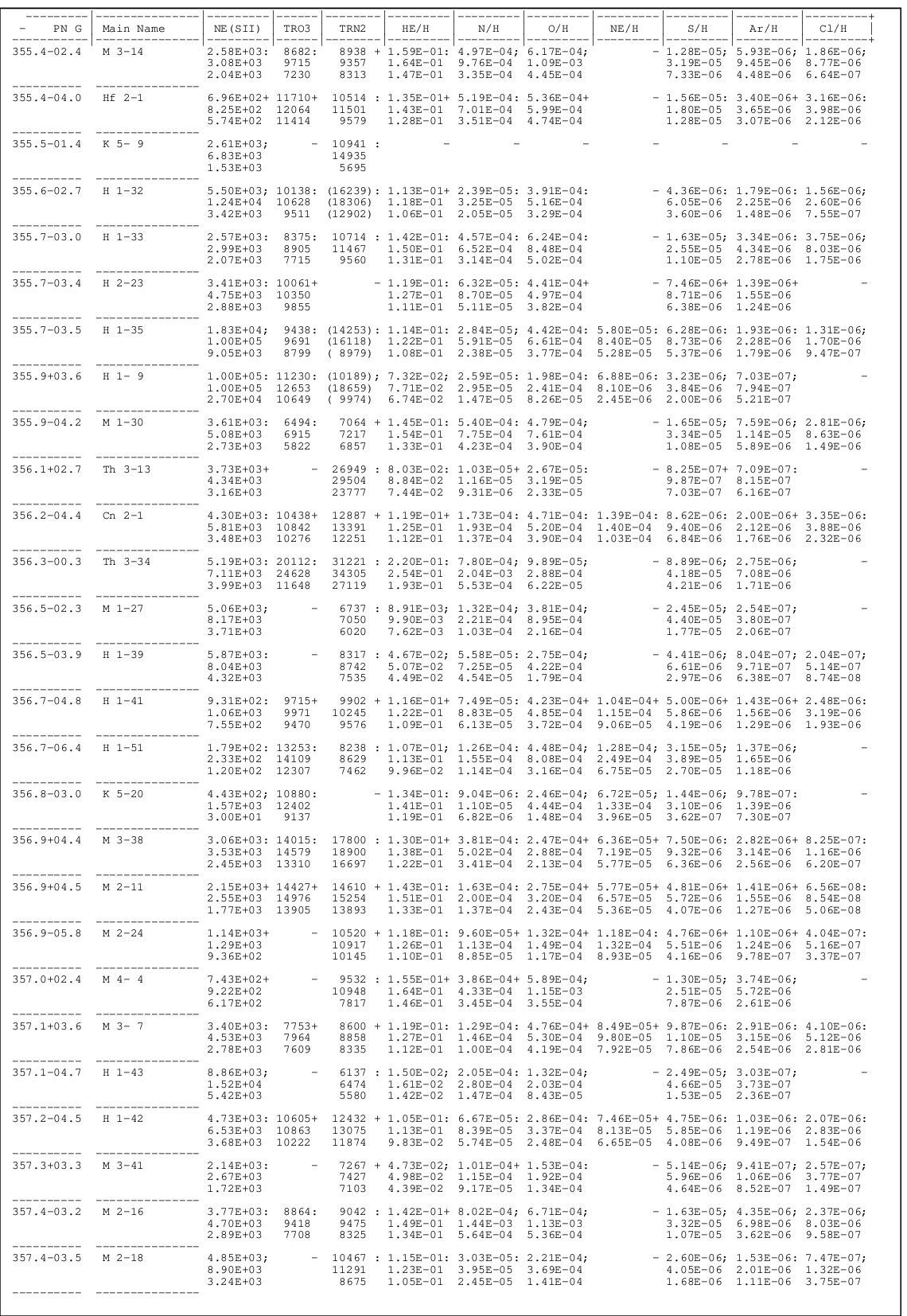}}

\clearpage

\vspace*{1cm}
{\large\bf Table 1a. continued}

\hspace*{-10cm}
\resizebox{1.8\hsize}{!}{\includegraphics{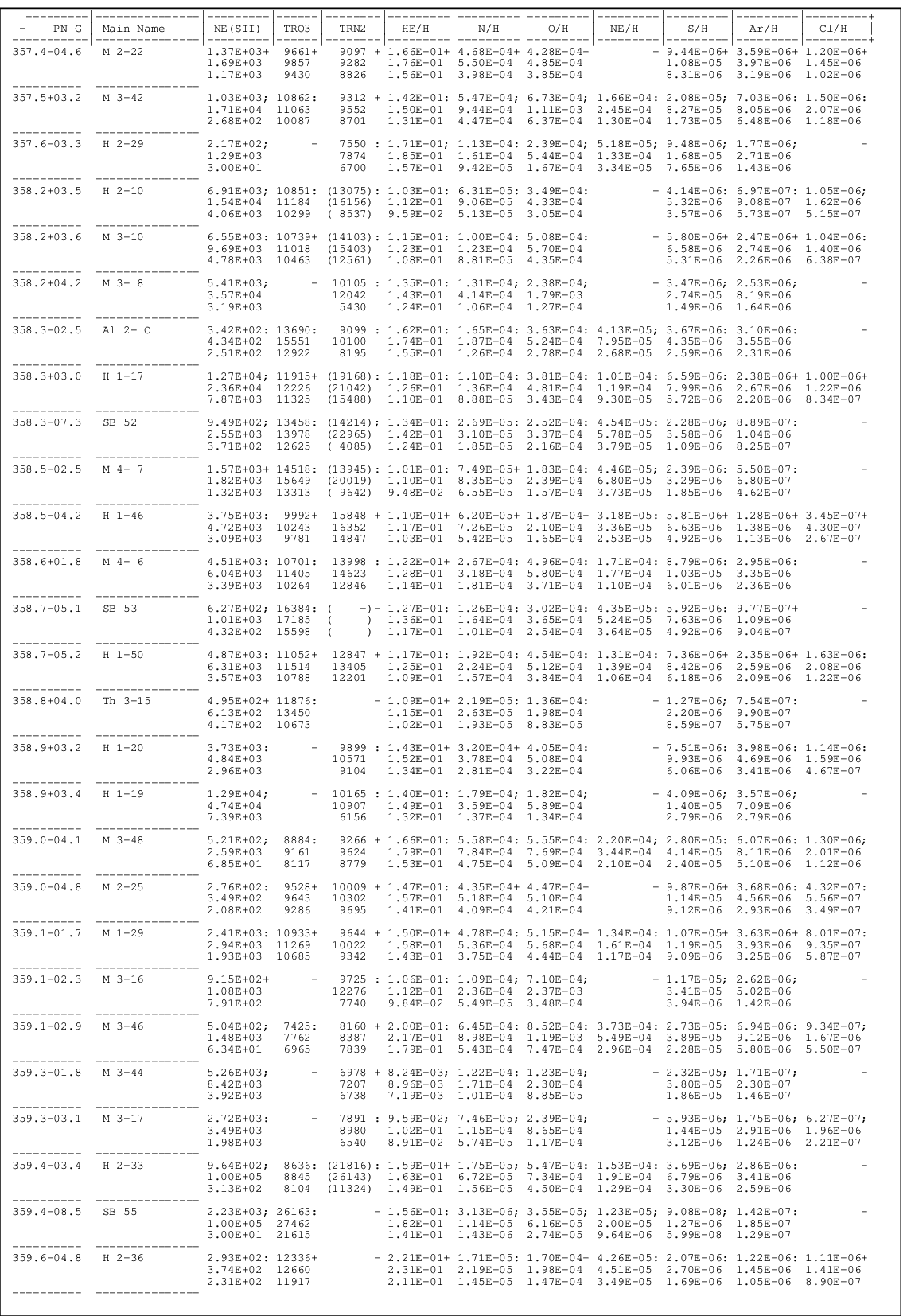}}

\clearpage

\vspace*{1cm}
{\large\bf Table 1a. continued}

\hspace*{-10cm}
\resizebox{1.8\hsize}{!}{\includegraphics{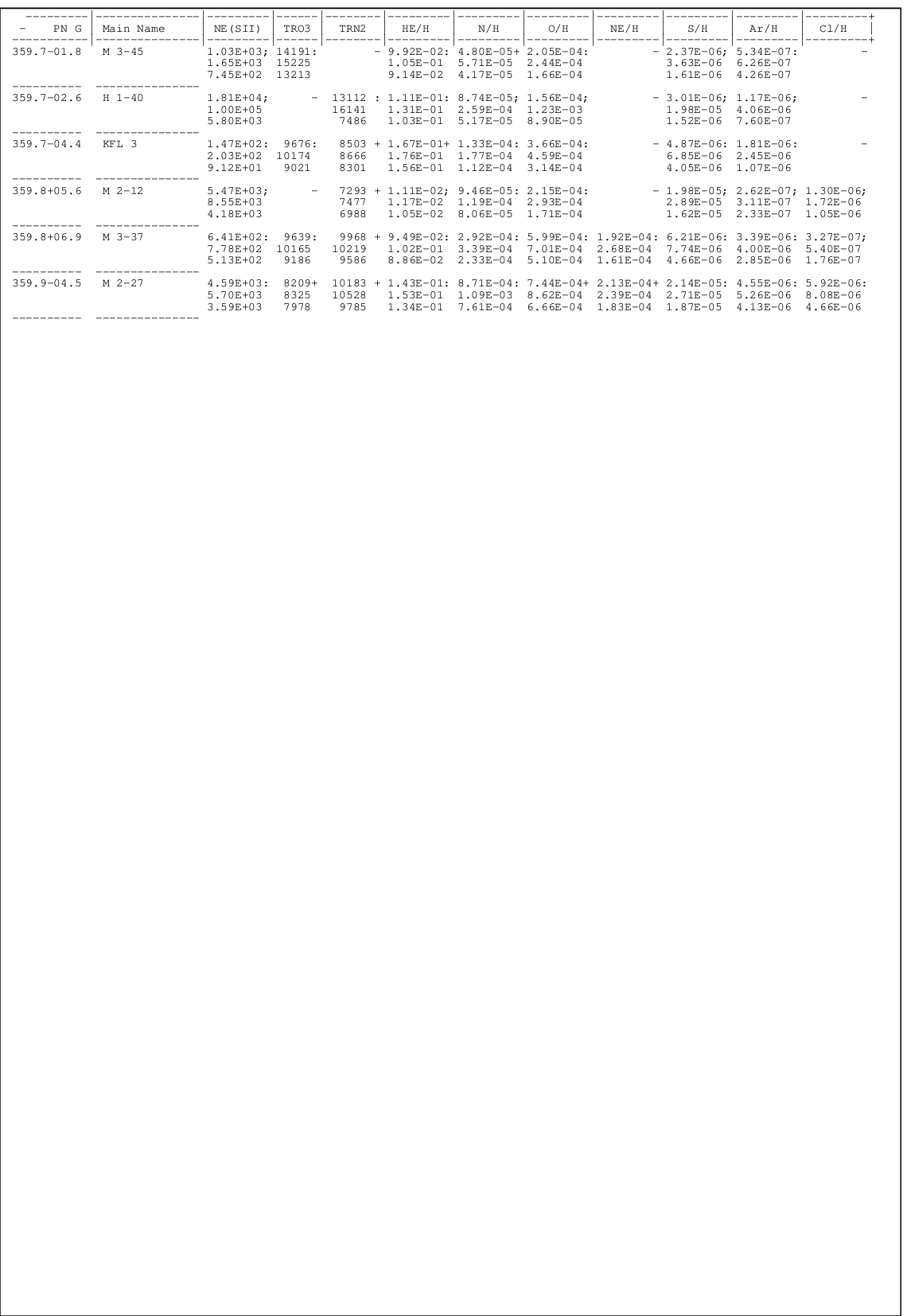}}

\clearpage


\vspace*{1cm}
{\large\bf Table 1b. Plasma parameters and chemical abundances (Galactic
inner-disk sample)}

\hspace*{-10cm}
\resizebox{1.8\hsize}{!}{\includegraphics{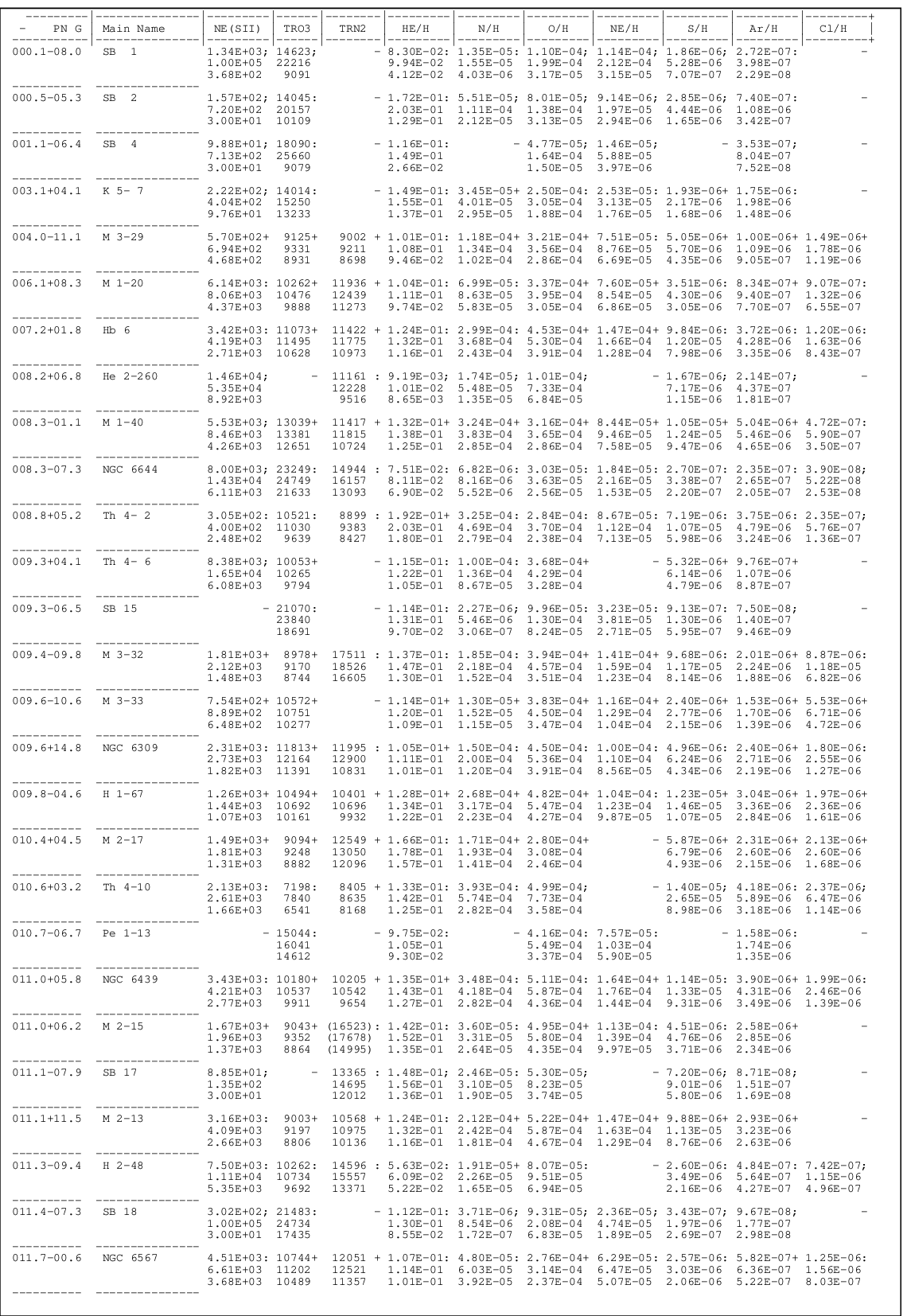}}

\clearpage

\vspace*{1cm}
{\large\bf Table 1b. continued}

\hspace*{-10cm}
\resizebox{1.8\hsize}{!}{\includegraphics{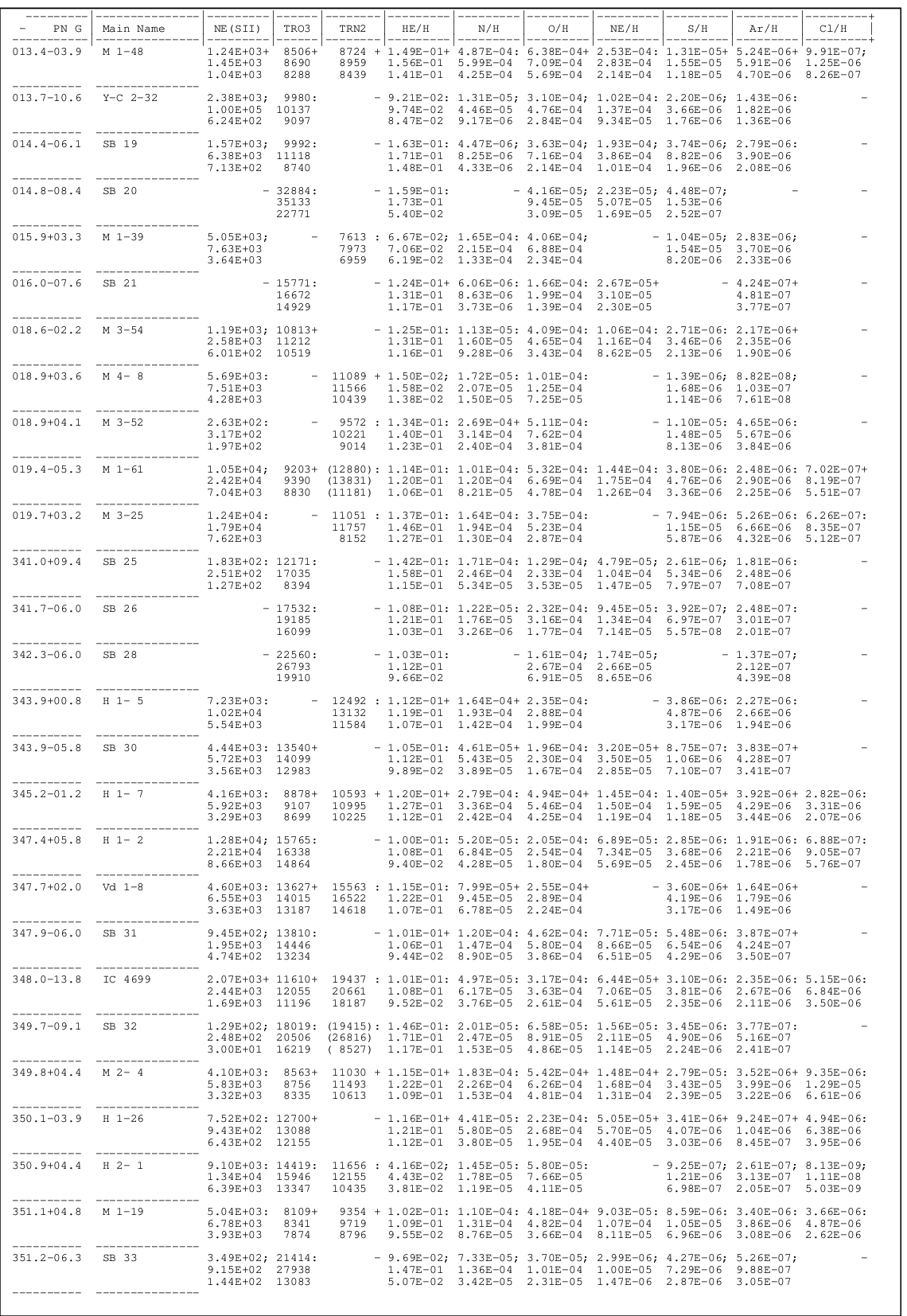}}

\clearpage

\clearpage

\vspace*{1cm}
{\large\bf Table 1b. continued}

\hspace*{-10cm}
\resizebox{1.8\hsize}{!}{\includegraphics{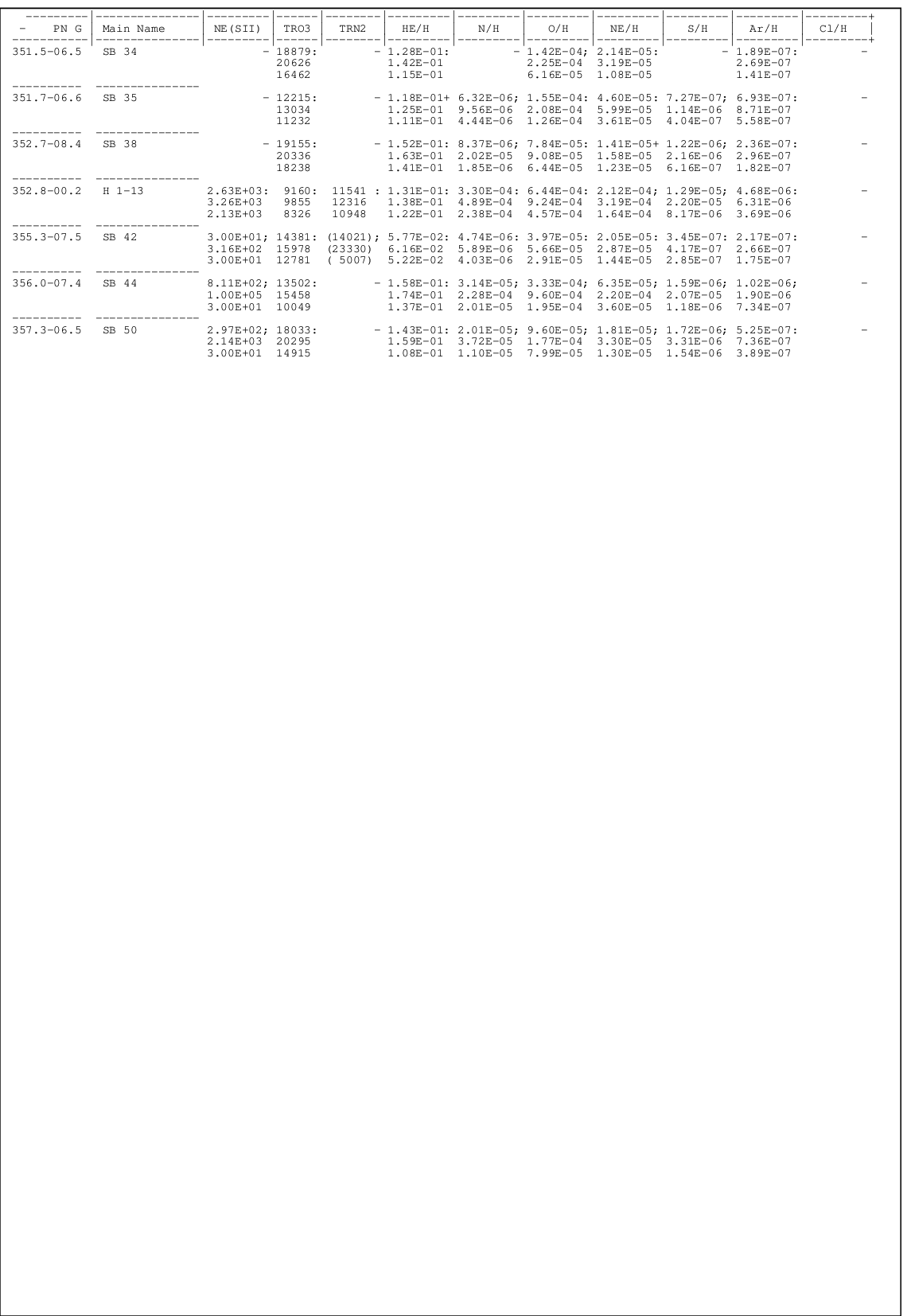}}

\clearpage


\vspace*{1cm}
{\large\bf Table 1c. Plasma parameters and chemical abundances (LMC sample)}

\hspace*{-10cm}
\resizebox{1.8\hsize}{!}{\includegraphics{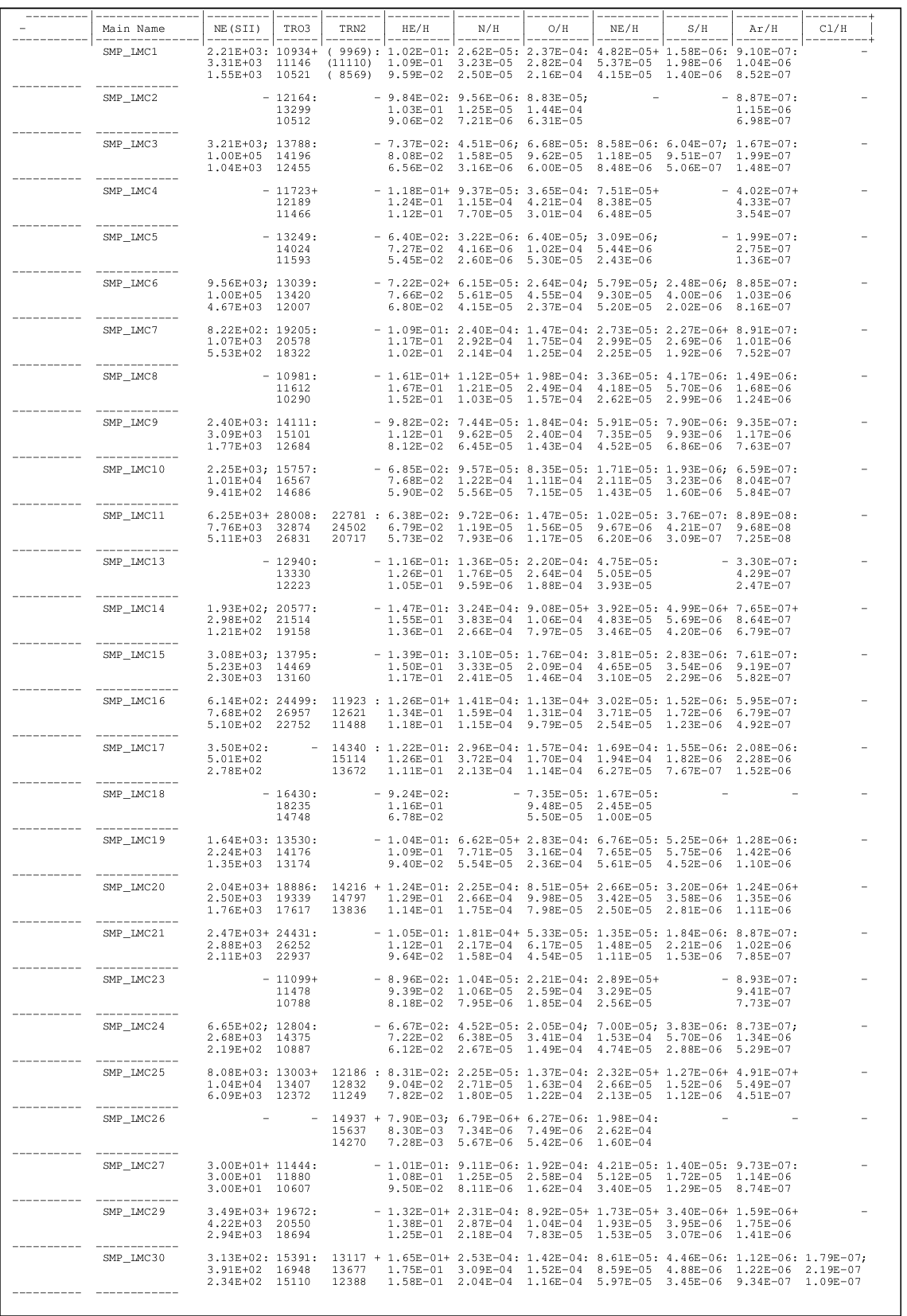}}

\clearpage

\vspace*{1cm}
{\large\bf Table 1c. continued}

\hspace*{-10cm}
\resizebox{1.8\hsize}{!}{\includegraphics{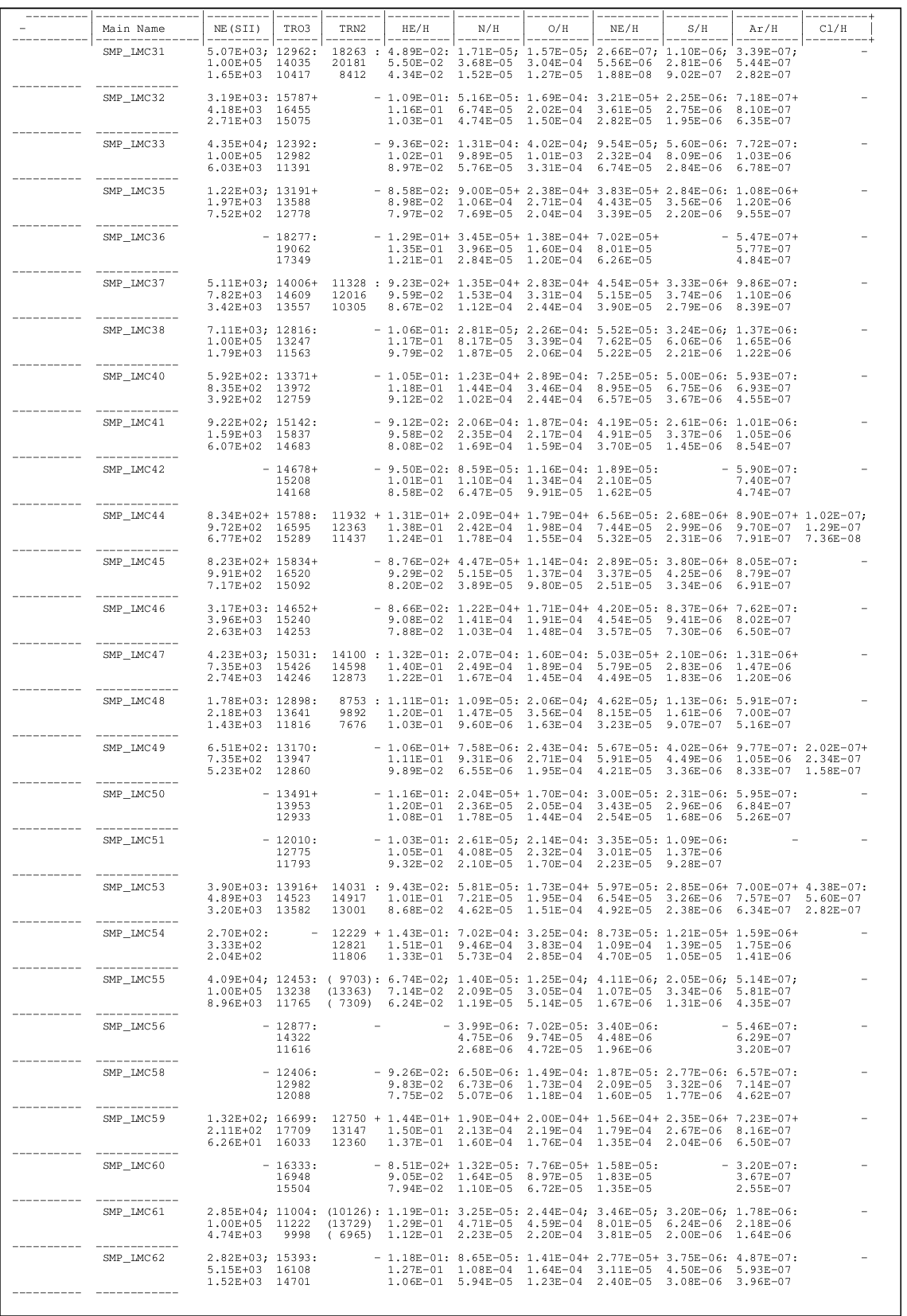}}

\clearpage

\vspace*{1cm}
{\large\bf Table 1c. continued}

\hspace*{-10cm}
\resizebox{1.8\hsize}{!}{\includegraphics{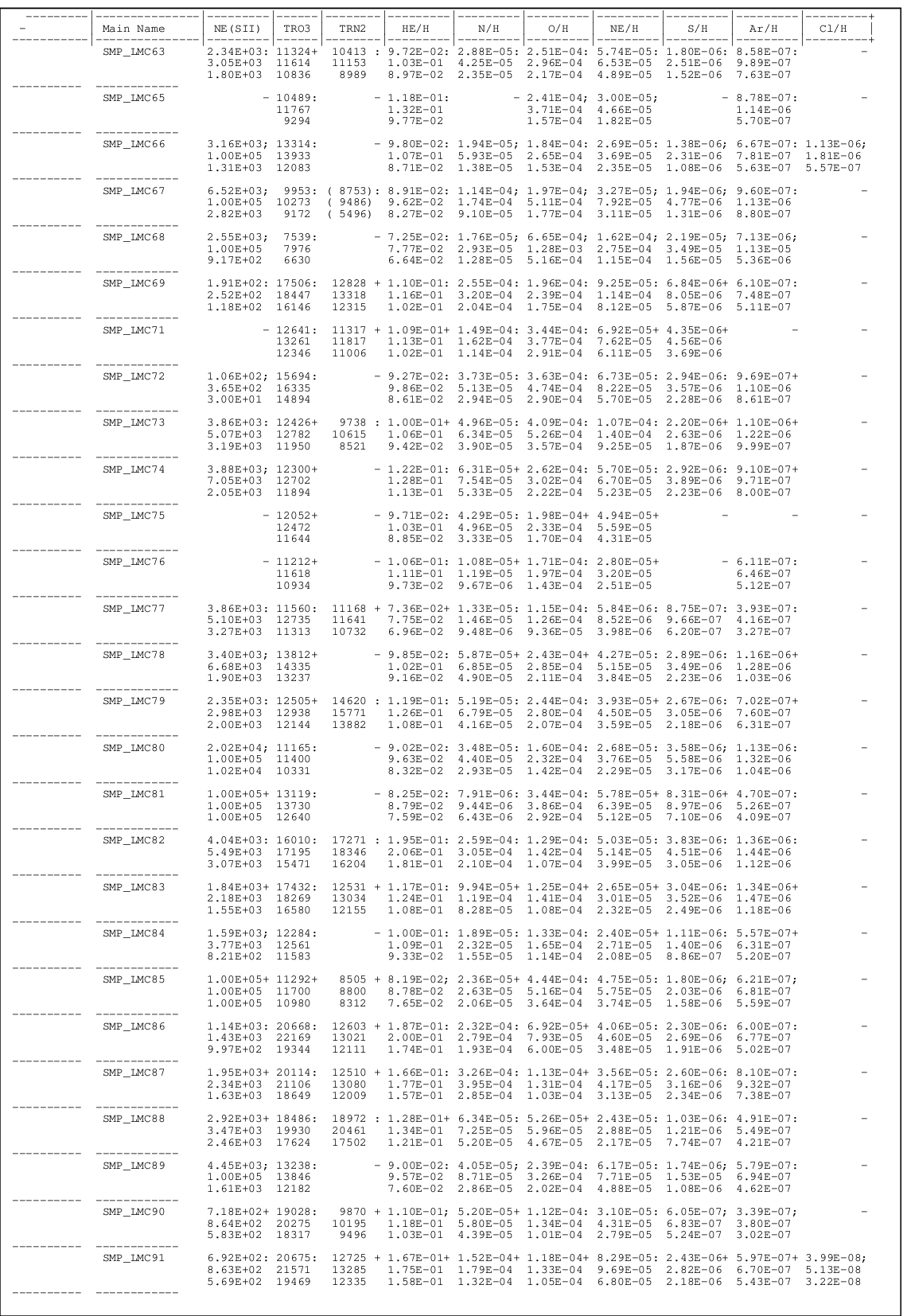}}

\clearpage

\vspace*{1cm}
{\large\bf Table 1c. continued}

\hspace*{-10cm}
\resizebox{1.8\hsize}{!}{\includegraphics{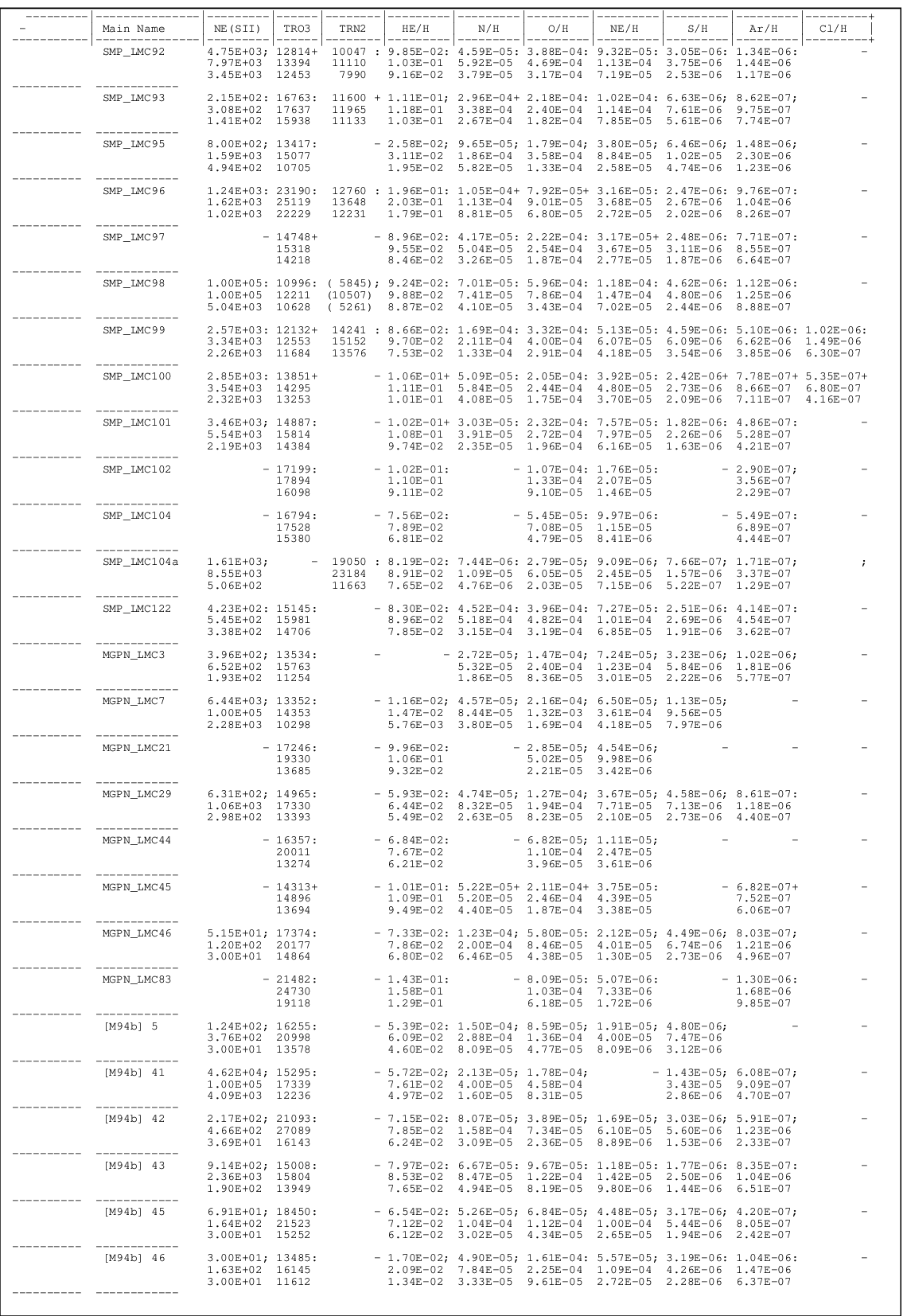}}

\clearpage

\vspace*{1cm}
{\large\bf Table 1c. continued}

\hspace*{-10cm}
\resizebox{1.8\hsize}{!}{\includegraphics{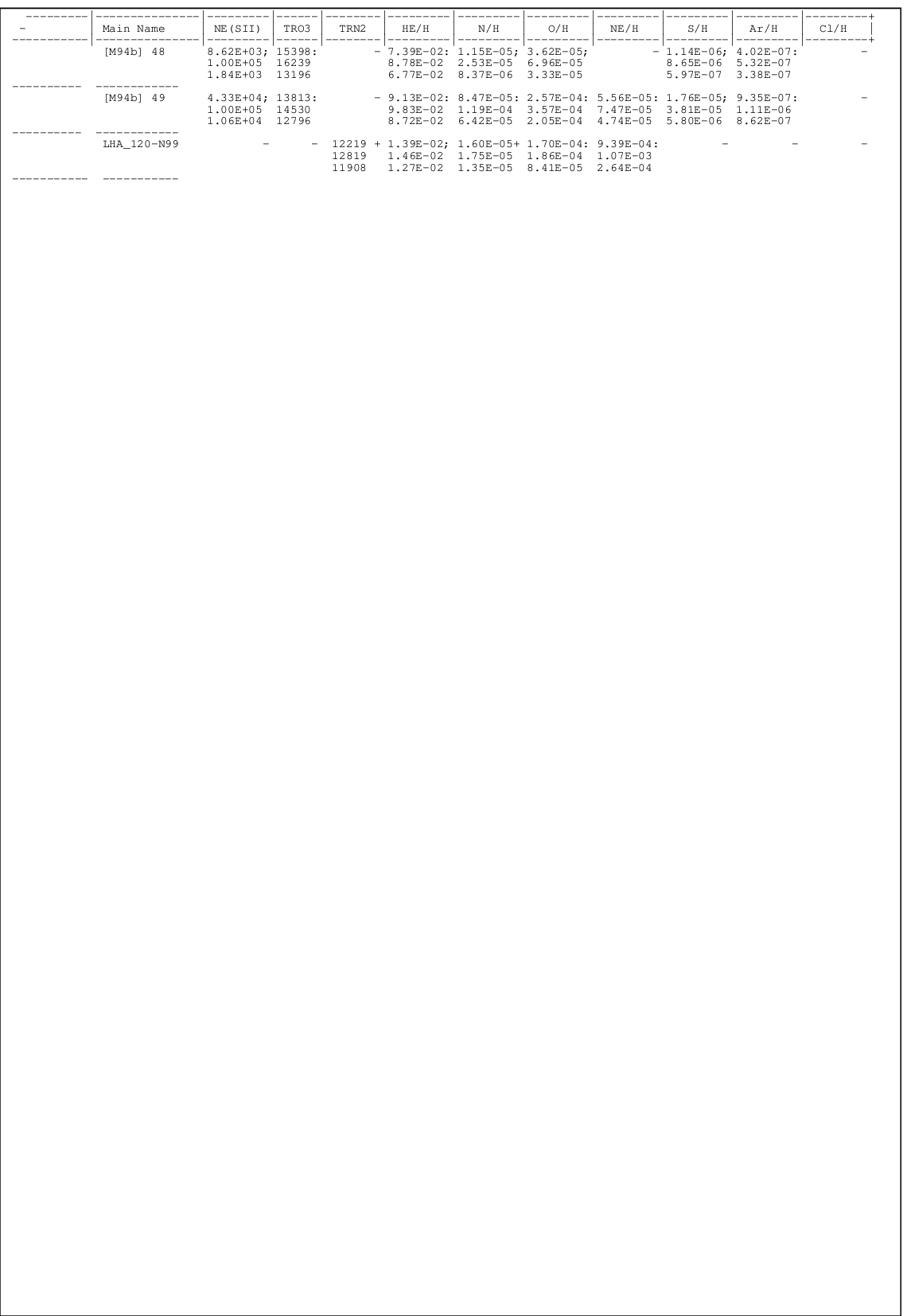}}

\typeout{get arXiv to do 4 passes: Label(s) may have changed. Rerun}

\end{document}